\documentclass[pra,twocolumn,amsmath,amssymb,superscriptaddress,showpacs,floatfix]{revtex4-1}
\usepackage[latin9]{inputenc}

\usepackage{graphicx}
\usepackage{amssymb}
\usepackage{amsmath}
\usepackage{bbm}

\usepackage{bm}
\usepackage{braket}
\usepackage{enumitem}
\usepackage{xfrac}

\usepackage{color}

\usepackage{cancel} 

\usepackage{mathrsfs}

\usepackage[hidelinks=true,pdfborder={0 0 0},colorlinks=true,linkcolor=blue,urlcolor=blue,citecolor=blue]{hyperref}

\usepackage[normalem]{ulem}
\usepackage{array}
\usepackage{xcolor}
\newcommand{\etal}{\textit{et al.~}}

\newcommand{\veck}{\mathbf k}
\newcommand{\vecR}{\mathbf R}

\newcommand{\vecp}{\mathbf p}

\newcommand{\vecr}{\mathbf r}
\newcommand{\abg}{a_\text{bg}}
\newcommand{\abar}{\bar{a}}
\newcommand{\sres}{s_\text{res}}
\newcommand{\aF}{a_\text{res}}

\newcommand{\rtext}[1]{\textcolor{black}{{{#1}}}}

\newlist{myitemize}{enumerate}{10}
\setlist[myitemize]{label*=\alph*.),align=left, leftmargin=0em,
labelindent=\parindent, listparindent=\parindent, labelwidth=0pt, itemindent=!}

\begin{document}

\title{Efimov states near a Feshbach resonance and the limits of van der Waals universality at finite background scattering length}

\author{Christian Langmack}
\affiliation{Technische\,Universit{\"a}t\,M{\"u}nchen,\,Physik\,Department, 85748 Garching, Germany}

\author{Richard Schmidt}
\affiliation{Department of Physics, Harvard University, Cambridge MA 02138, USA}
\affiliation{Institute for Quantum Electronics,  ETH Z\"urich, 8093 Z\"urich, Switzerland}
\affiliation{Max-Planck-Institut f\"ur Quantenoptik, 85748 Garching, Germany}

\author{Wilhelm Zwerger}
\affiliation{Technische\,Universit{\"a}t\,M{\"u}nchen,\,Physik\,Department, 85748 Garching, Germany}

\begin{abstract}

We calculate the spectrum of three-body Efimov bound states near a Feshbach resonance 
within a model which accounts both for the finite range of interactions and the presence
of background scattering. The latter may be due to direct interactions in an open channel 
or a second overlapping Feshbach resonance. It is found that background scattering gives 
rise to substantial changes in the trimer spectrum as a function of the detuning away
from a Feshbach resonance, in particular in the regime where the background channel
supports Efimov states on its own. Compared to the situation with negligible background 
scattering, the regime where van der Waals universality applies is shifted to larger values of
the resonance strength if the background scattering length is positive. For negative background 
scattering lengths, in turn, van der Waals universality extends to 
even small values of the resonance strength parameter, consistent with experimental results
on Efimov states in $^{39}$K. Within a simple model, we show that short-range
three-body forces do not affect van der Waals universality significantly. Repulsive three-body 
forces may, however, explain the observed variation between around $-8$ and $-10$
of the ratio  between the scattering length where the first Efimov trimer appears and the van der 
Waals length.

\end{abstract}

\color{black}
\date{\today}

\maketitle

\section{Introduction}
\label{chp:intro}

It is a surprising consequence of quantum mechanics that three particles may
be bound together even in a situation where no two-body bound states exist.
This effect was first discussed by Efimov in the context of nuclear physics~\cite{Efimov70}. 
As a possible realization of his prediction that near-resonant two-body interactions give rise to an infinite sequence of three-body bound states, Efimov suggested that the low-lying excited 
state of $^{12}$C --- the so called Hoyle state \cite{Hoyle54} which is of fundamental importance for
nucleosynthesis --- might be a weakly bound state of three alpha-particles. 
Their interaction, however, is quite different from those in ultracold atoms, where genuine Efimov states 
were first seen in a gas of $^{133}$Cs atoms near a Feshbach resonance~\cite{Grimm06}. 
Subsequently, Efimov states have been observed in radio-frequency spectroscopy and loss measurements in 
a variety of  Bose gases~\cite{Knoop2009,Zaccanti2009,Gross2009,Pollack2009,Gross2010}, in three-component 
Fermi gases~\cite{Ottenstein2008,Lompe2010,Huckans2009,Williams2009,Nakajima2011} 
and also in mass-imbalanced mixtures~\cite{Barontini2009,Pires2014,Pires2014b,Tung2014,Ulmanis2015,Johansen2016,Ulmanis2016b,Ulmanis2016c,Wacker2016}. 
More recently, it has been found that an analog of Efimov's original suggestion involving alpha-particles 
is realized in neutral $^{4}$He atoms, for which the lowest three-body bound state appears
at a binding energy which is about a hundred times larger than the extremely weakly bound
$^{4}$He dimer at $\epsilon_D\simeq k_B\cdot 1.3\,$ mK~\cite{Kunitski:2015}.

A striking feature of the Efimov effect is the appearance of universality near infinite two-body scattering length 
$a$, where the sequence of three-body bound states has an accumulation point. In 
particular, the scattering lengths $a^{(n)}_{-}<0$, where the $n$th Efimov state meets the atom threshold 
at energy $E=0$, form a geometric series with a ratio $a_{-}^{(n+1)}/a_{-}^{(n)}\!\to\! e^{\pi/s_0}$ of 
consecutive values for large $n$. Here, $s_0$ is a universal number which only depends on the mass 
ratios and the particle statistics. For identical mass bosons, it has the value $s_0\approx1.00624$ which 
gives rise to a large value $a_{-}^{(n+1)}/a_{-}^{(n)}\simeq 22.6944..$.
The origin of this universality can be understood from an effective field theory approach to the three-body problem~\cite{BHvK1999,Braaten:2004rn,Moroz2009,Floerchinger2009,Schmidt2010,Floerchinger2011,Schmidt,Horinouchi2015}. 
The underlying assumption of zero-range interactions --- with the scattering length as the single length scale --- implies
that universality only applies for the highly excited Efimov states with $n\gg 1$.
In practice, since atom losses on average increase with the fourth power of the scattering length, these states
are not accessible and it is only the lowest or maybe second Efimov state at $a_{-}\! =a_{-}^{(0)}$ or $a_{-}^{(1)}$ 
which can be observed.  In fact, the appearance of a second Efimov trimer at $a_{-}^{(1)}$ has been confirmed only recently in  $^{133}$Cs~\cite{Huang2014} and in a $^{6}$Li-$^{133}$Cs-mixture \cite{Pires2014,Tung2013,Tung2014,Johansen2016}.
Due to the finite range of interactions, deviations from universality for the lowest trimer states can be appreciable. 
As an example, the ratio $a_{-}^{(1)}/a_{-}=21 \pm 1.3$~\cite{Huang2014}
 observed for the Feshbach resonance in $^{133}$Cs near $B_0\simeq 787\,$G
differs substantially from the universal value $22.69\ldots$ reached as $n\gg 1$. 
On a more basic level, the absence of any scale beyond the scattering length itself 
necessarily implies that effective field theory can  account neither for the specific position $a_{-}<0$ where
the first Efimov state appears nor for its binding energy at infinite scattering length. 
To incorporate this, the field theory must be supplemented with a momentum cutoff $\Lambda^*$,
usually called the three-body parameter.  The three-body parameter was initially thought to exhibit no systematic dependence on low-energy observables, 
being sensitive to microscopic details of the  two-body potential at short distance
as well as to genuine three-body forces~\cite{DIncao2009}.\\

It was a surprising observation, therefore, that in ultracold atoms the measured values for $a_{-}$ for a number of different alkali atoms and for many 
Feshbach resonances all ranged in a regime between $-8\,l_{\text{vdw}}$  and $-10\,l_{\text{vdw}}$, with an average value $\langle a_{-}\rangle \approx -9.45\, l_{\text{vdw}}$
~\cite{Ottenstein2008,Huckans2009,Pollack2009,Zaccanti2009,Gross2009,Gross2010,Berninger2011,Wild2012}.
Here, $l_{\text{vdw}}=\left(mC_6/\hbar^2\right)^{1/4}/2$ is the van der Waals length, which is a measure of the strength 
of the attractive interaction $\sim -C_6/r^6$ at large interatomic distances
\footnote{throughout the paper we assume the particle masses to be identical. Our results can, however,
be easily extended to  the case of scattering of particles with different mass, where $m\to 2m_{\rm red}$ with the reduced mass $m_{\rm red}$.}. 
Since $a_{-}$ determines the overall scale of the complete Efimov spectrum,  this observation suggests a three-body parameter which 
is independent of the microscopic details~\cite{Berninger2011,Wild2012}. An explanation of this unexpected result, which has since been termed
the van der Waals universality in Efimov physics of ultracold gases, was given independently by Wang \textit{et al.}~\cite{Greene2012}
and by Schmidt \textit{et al.}~\cite{SRZ} (see also~\cite{chin2011,Sorensen:2012}). 

 The work by Wang \textit{et al.} is based on single channel potentials with a van der Waals tail $-C_6/r^6$ and 
 different forms of the short-range interaction, which may be adjusted to tune the scattering length to values much larger than the potential range. 
 For such models the origin of van der Waals universality  turns out to be connected to the 
 appearance of a barrier in the hyperspherical three-body  potential $V_{\rm eff}(R)$ 
for values of the hyperradius $R$ which are much smaller than $l_{\rm vdw}$. 
The detailed behavior of the interactions at small distances therefore does not play a role.
This result has been confirmed in a number of papers, in part also using different methods \cite{Wang2012,Greene2012,DIncao2013,Wang2014,Naidon2014,Blume2015,Mestrom2017}.

In practice, the two-body potentials which allow us to solve the three-body problem using hyperspherical coordinates are, 
unfortunately, quite different from those present in alkali atoms. In a recent  study of van der Waals universality~\cite{Mestrom2017} for instance, based on a Lennard-Jones potential $V(r)=-C_6(1-\bar\sigma^6/r^6)/r^6$,
the energy of the lowest three Efimov states is calculated as a function of the dimensionless inverse scattering length $l_{\text{vdw}}/a$  
for the three zero-energy resonances  with the largest values $\lambda_c=0.92, 0.57, 0.45$ of the ratio $\lambda=\bar\sigma/l_{\text{vdw}}$. The single-channel potentials are thus considered in a regime where the number of two-body bound states is of order $1$. 
This is the situation appropriate, e.g., for $^{4}$He, where just one such bound state exists
\footnote{It is instructive to note that the parameter $\lambda$ is simply related to the well known de Boer parameter 
$\Lambda^{\rm dB}=\hbar/(\bar\sigma\sqrt{m\epsilon})$ which quantifies the importance of quantum effects in the equation of 
state~\cite{deboer1948} by $\Lambda^{\rm dB}=\lambda^2/2$ (here, $\epsilon$ is the depth of the potential well). 
The maximum value of $\Lambda^{\rm dB}$ for which a single bound state exists is thus 
$\Lambda^{\rm dB}_c=(0.92)^2/2\simeq 0.42$, which is rather close to the de Boer parameter estimated for $^{4}$He. 
Remarkably, the fact that the de Boer parameter has a finite upper limit for all realistic two-body interaction potentials implies that the 
minimum of the shear viscosity for a purely classical fluid obeys the quantum inequality $\eta>\eta_{\rm min}=\alpha_{\eta}\cdot\hbar n$ 
with a numerical constant $\alpha_{\eta}\sim 1/\Lambda^{\rm dB}$ which is bounded below by a constant of order $1$, 
see~\cite{Enss2011}. }. By contrast,  the large polarizability of alkali atoms leads to ratios $\lambda=\bar\sigma/l_{\text{vdw}}$ which are much less than $1$.

A quite different and complementary approach to the problem has been developed by Schmidt \textit{et al.}~\cite{SRZ,Schmidt} who considered a standard two-channel model for a Feshbach resonance, 
including a finite range of the Feshbach coupling as a crucial new feature.  
Within this model, the spectrum of Efimov trimers can be calculated without any adjustable cutoff. It follows a universal set of binding energy curves which only 
 depends on the resonance strength parameter $s_{\rm res}=\bar{a}/r^{\star}$~\cite{Chin2010}, 
 defined as the ratio between the mean scattering length $\bar{a}=0.956\,l_{\text{vdw}}$ and 
 the intrinsic length $r^{\star}$ associated with a Feshbach resonance~\cite{Petrov2004,Gogolin2008}.
The position of the trimer states in the $(\abar/a,E)$ plane is therefore completely fixed 
by only two experimentally accessible parameters: the van der Waals length $l_{\rm vdw}$ and 
the intrinsic length $r^{\star}$. In the limit of closed-channel dominated resonances with $s_{\text{res}}\ll 1$, the spectrum gets pushed toward the unitarity point $E=1/a=0$ 
and the characteristic length and energy scale is determined by $r^{\star}$.
For open-channel dominated resonances with $s_{\text{res}}\gg 1$, in turn, 
the trimer spectrum reaches its maximal extent in the $(\bar{a}/a,E)$ plane.  
Specifically, the first trimer state detaches from the continuum at $a_{-}=-8.3\, l_\text{vdw}$,
which is about three times the effective interaction range $r_e\to 3\bar{a}$ in this limit. 

The model by Schmidt \etal thus provides a qualitative understanding of the observed 
van der Waals universality, which is generically observed for Efimov trimers near open 
channel dominated resonances~\cite{SRZ}.  Moreover, it also shows how ratios like 
$a_{-}^{(n+1)}/a_{-}^{(n)}$ approach their universal values only in the limit $n\gg 1$ or when the resonance strength takes on intermediate values $s_{\rm res}\simeq 1$ where the effective range is close to zero. The genuine Fesh\-bach resonance physics associated with the presence of two separate scattering channels has also been analyzed by Wang and Julienne~\cite{Wang2014} within a two-spin van der Waals model. In contrast to the earlier work \cite{SRZ},  however, only the open-channel dominated limit was studied.

A striking prediction of the  model by Schmidt  \etal  \cite{SRZ} is that the ratio $|a_{-}|/ l_{\rm vdw}\simeq 8.3$  increases if the Feshbach resonance is no longer in the open-channel dominated regime. In particular,
 in the limit  $s_{\rm res}\ll 1$, it approaches the value $a_-\to -10.3\, r^{\star}$, consistent with the exact solution of the 
 three-body problem for closed-channel dominated  Feshbach resonances by Petrov and Gogolin~\cite{Petrov2004,Gogolin2008}.  
 The scattering length at which the first Efimov trimer appears is thus again determined by the effective range parameter $r_e\simeq -2r^{\star}$, 
 which is now in magnitude much larger than $l_\text{vdw}$  relevant in the open-channel dominated case.
Experimentally, a trend to larger values of $a_-/l_\text{vdw}$ for decreasing values of $\sres$ has been observed for heteronuclear mixtures \cite{Johansen2016}.  By contrast, the observed ratio $|a_{-}|/ l_{\rm vdw}\simeq 7.75$ for the $^{7}$Li 
resonance at $737\,$G  still seems to follow the naive van der Waals universality despite a rather small value of $s_{\rm  res}\simeq 0.56$. 
Furthermore, measurements of the ratio  $|a_{-}|/ l_{\rm vdw}$ for a number of closed-channel dominated resonances in $^{39}$K~\cite{Roy2013} also do not observe the predicted strong increase with decreasing values of the resonance strength. Van der Waals universality thus appears to be much more robust than what a standard two-channel Feshbach resonance model suggests. \\

In our present work, we address the question of the range of validity and the limits of van der Waals universality for the
practically accessible first few Efimov states by extending our previous model to account for the effects of a nonvanishing background scattering length \rtext{$\abg$}. The latter may be due  to either direct interactions in the open channel or arising from  overlapping Feshbach resonances. We start with a model involving a sum of two-body interactions only. The three-body spectrum can then be calculated from data which are fully determined by two-body physics, with no adjustable parameters. As a result, the spectrum of Efimov states is completely 
fixed by the value of the van der Waals length $l_\text{vdw}$, the resonance strength $s_{\rm  res}$, and the   
standard dimensionless parameter $r_{\rm bg}=a_{\rm bg}/\bar{a}$ which measures the influence of a  
background scattering length $a_{\rm bg}$~\cite{Chin2010}. In a second step,
we provide a model calculation for Efimov states in the presence of short-range three-body forces which allows 
to see to which extent the Efimov spectrum is sensitive to genuine three-body forces.

\section{Microscopic Model}
\label{chp:model}

An observation and systematic study of the Efimov effect with ultracold atoms typically requires Feshbach resonances, where the
scattering length may be tuned externally, e.g., by a magnetic field.  Here one takes advantage of the coupling of atoms in an open channel to 
a closed-channel molecule which has different quantum numbers and can thus be shifted energetically with respect to the open-channel scattering continuum. 
In particular, when the molecular state becomes degenerate with the atomic scattering threshold at a specific value $B_0$ of
the magnetic field,  a resonant enhancement of $a$ is found.
A standard description of the resonant enhancement of the scattering length near a Feshbach resonance  is provided by a two-channel model
\begin{align}
\label{eq:ResModel}
\hat H_\text{res} & =  \int_{\mathbf{R}} \hspace{-0.15 cm} \left\{  \!
\hat{\psi}^{\dagger} \left(\! -\tfrac{\hbar^2 \nabla^2}{2m}\right)
\hat{\psi}(\mathbf{R}) 
\! + \hat{\phi}^{\dagger} \! \left(\! -\tfrac{\hbar^2\nabla^2}{4m}
 \! + \tilde\nu_{\phi}(B)\right) \! \hat{\phi}(\mathbf{R}) \right\}\nonumber\\
&+ \frac{g_{\phi}}{2} \!\!
\int_{\mathbf{R},\mathbf{r}} \hspace{-0.4 cm}
\chi_{\phi}(\mathbf{r}) \! \left[
\hat{\psi}^{\dagger}(\mathbf{R} \hspace{-0.05 cm} + \!\tfrac{\mathbf{r}}{2})
\hat{\psi}^{\dagger}(\mathbf{R} \hspace{-0.05 cm} - \!\tfrac{\mathbf{r}}{2})
\hat{\phi}(\mathbf{R}) 
 \hspace{-0.07 cm} + \hspace{-0.07 cm} {\rm h.c.} \right]\!\!,
\end{align}
where $\int_{\mathbf{R}}\equiv \int d^3R$. Here the first term represents the kinetic energy of the atoms in the open channel (represented by the operators $\hat \psi^\dagger, \hat \psi$), while the second term accounts for the kinetic energy of the Feshbach molecule ($\hat \phi^\dagger,\hat \phi$) in the closed channel. It has mass $2m$ and is energetically detuned from the open-channel scattering threshold by the bare detuning $\tilde \nu_\phi(B)=\delta\mu (B-B_{\rm c})$, where $\delta\mu$ is the  differential magnetic moment of the molecule, and $B_{\rm c}$ denotes the magnetic field at which the closed-channel molecule, when  not coupled to the open channel, crosses zero energy. 
The closed-channel (Feshbach) molecule is created  from two atoms in the open channel. This process is represented by the third term in Eq.~\eqref{eq:ResModel} where two atoms, separated by a distance $\vecr$, are converted into the molecule $\hat \phi$ at the center-of-mass coordinate 
$\mathbf{R}$.
The strength of the conversion is given by $g_\phi$ while the form factor $\chi_\phi(\vecr)$ accounts for the range of the atom-molecule  coupling. 
The model contains three adjustable parameters which eventually must be connected with experimentally measurable quantities. 
This has been done in~\cite{SRZ} using the following line of arguments:

\begin{enumerate}

\item The strength $g_{\phi}$ of the coupling is fixed by the intrinsic length parameter $r^{\star}$ of the Feshbach resonance defined by the relation
$a_{\rm res}(B)=-\hbar^2/(mr^{\star}{\nu}(B))$ between the resonant scattering length and the renormalized detuning 
${\nu}(B)=\delta\mu(B-B_0)$, where $B_0$ denotes the magnetic field where the Feshbach resonance appears. The known value of $r^{\star}$ fixes the coupling for identical bosons via $r^{\star}=8\pi\hbar^4/(m^2g_\phi^2)$ \cite{Gogolin2008,SRZ}.  For SU(2) fermions, this expression carries an additional factor of $1/2$, see e.g. Refs.~\cite{Zwerger2008,Zwerger2016,Cetina2016,Schmidt2017}.

\item  The range $\sigma_\phi$ of the Feshbach transfer function $\chi_\phi(\vecr)$, which is normalized such that 
$\int_{\vecr}\chi_\phi(\vecr)=1$, is obtained from the known value $\delta\mu(B_0-B_c)=\hbar^2/(2mr^{\star}\bar{a})$ of the resonance shift  
in potentials with a van der Waals tail in the absence of background scattering \cite{Julienne2004}. As shown in~\cite{SRZ}, 
for a transfer function $\chi(\vecr)\sim e^{-|\vecr|/\sigma_\phi}$, this  fixes the range to be $\sigma_\phi=\bar{a}$. 

\item The bare detuning $\tilde \nu_\phi(B)$ which involves the parameter $B_c$ is inferred as a function of $\sigma_\phi$, and thus of $\abar$, 
from the actual magnetic field position $B_0$  of the resonance where $1/a=0$. 

\end{enumerate}

In order to study the effect of a nonvanishing background scattering length on the spectrum of three-body bound states, 
an obvious strategy would be to introduce a direct two-body interaction 
\begin{equation}
\label{eq.directinteraction}
\hat{H}_{\rm dir}=\int_{\mathbf{r},\mathbf{r'}} \!\!\!\!\!
:\hat{\psi}^{\dagger}(\mathbf{r})\hat{\psi}(\mathbf{r})\, 
V(\mathbf{r}-\mathbf{r}')\,
\hat{\psi}^\dagger(\mathbf{r}')\hat{\psi}(\mathbf{r}'):
\end{equation}
which involves an interaction potential $V(\vecr)$ which has a van der Waals tail of range $l_\text{vdw}$. 
This normal-ordered expression describes a standard density-density interaction with $\hat n(\vecr)= \hat \psi^\dagger(\vecr) \hat\psi(\vecr)$. 
By contrast,  $\hat H_\text{res}$ in Eq.~\eqref{eq:ResModel} represents an interaction in the pairing channel that has a different singular transfer momentum. The scattering vertices for the combined model $\hat H=\hat H_\text{res}+\hat H_\text{dir}$ therefore develop a complicated momentum dependence which 
is  difficult to handle in analytical form (for a detailed study of the corresponding two-body problem see Ref.~\cite{Marcelis2004}). 

These problems may be avoided by replacing the interaction Eq.~\eqref{eq.directinteraction} by a  simpler model which leads to the same scattering length and a similar effective range. To this end one introduces an auxiliary molecular field 
$\hat{b}$  whose exchange accounts for the  background scattering. The corresponding interaction Hamiltonian  is given by
\begin{align}
\hat{H}_\text{bg} & = \int_{\mathbf{R}} 
\tilde \nu_{b} \;\hat{b}^{\dagger} \hat{b}(\mathbf{R})  +
\frac{g_{b}}{2} \!\!
\int_{\mathbf{R},\mathbf{r}} \hspace{-0.2 cm} 
\chi_{b}(\mathbf{r}) \nonumber\\
&\quad\quad\quad\quad\times\left[
\hat{\psi}^{\dagger}(\mathbf{R} + \!\tfrac{\mathbf{r}}{2})
\hat{\psi}^{\dagger}(\mathbf{R} - \!\tfrac{\mathbf{r}}{2})
\hat{b}(\mathbf{R})  + {\rm h.c.} \right] .
\label{eq:background}
\end{align}
We consider this model in the limit $g_b\to\infty$  where the field $\hat b$ can be regarded as a physical molecular state which mediates an
effective atom-atom interaction characteristic for an open-channel dominated Feshbach resonance~\cite{Marcelis2004}.  This becomes evident 
by using the Heisenberg equation of motion for the operator $\hat{b}$. In the limit  $g_b\to\infty$, it can be solved explicitly to give
$\hat{b}(\vecR)=-g_b/(2\tilde \nu_b)\int_\vecr \chi_b(\vecr)\hat{\psi}^{\dagger}(\vecR+\vecr/2) \hat{\psi}^{\dagger}(\vecR-\vecr/2)$. The auxiliary field 
$\hat b$ can thus be eliminated and leads to an open-channel interaction of strength $\sim g_b^2/\tilde \nu_b$. As a result, the Hamiltonian $\hat{H}_\text{bg}$ only depends on the ratio $g_b^2/\tilde \nu_b$ that is determined by the experimentally accessible value $a_\text{bg}$ of the background scattering length, see Eq.~\eqref{eq:ai} below. The reduction of $\hat{H}_\text{bg}$ to a single adjustable parameter relies on the fact that 
the range parameter  of the form factor $\chi_b(\mathbf{r})\sim e^{-r/\sigma_b}/r$ is again chosen to be $\sigma_b=\bar{a}$. This 
choice guarantees that the effective range of the induced atom-atom interaction in the case of pure background scattering is consistent with the generic result $r_e=2.92\bar{a}$ for direct two-body interaction potentials with a van der 
Waals tail in the relevant limit where $|a|\gg\bar{a}$ and the number of two-body bound states is large compared to one~\cite{Flambaum1999,SRZ}.

In the following we will thus consider the model
\begin{equation}
\label{eq.fullmodel}
\hat H = \hat H_\text{res}+ \hat H_\text{bg}.
\end{equation}
Apart from providing a simple model to account for a nonvanishing background scattering due to direct interactions in an open channel, this Hamiltonian also describes a second possible origin of a finite background scattering length, namely strongly overlapping Feshbach resonances. Specifically, the full model~\eqref{eq.fullmodel} provides an accurate microscopic description of strongly overlapping Feshbach resonances in cases where the background scattering length is due to an open-channel dominated Feshbach resonance whereas the primary resonance, described by Eq.~\eqref{eq:ResModel}, can be of arbitrary strength $\sres$.

\section{Two-body scattering}
\label{section.twobody}
\begin{figure}[t]
\includegraphics[width=\linewidth]{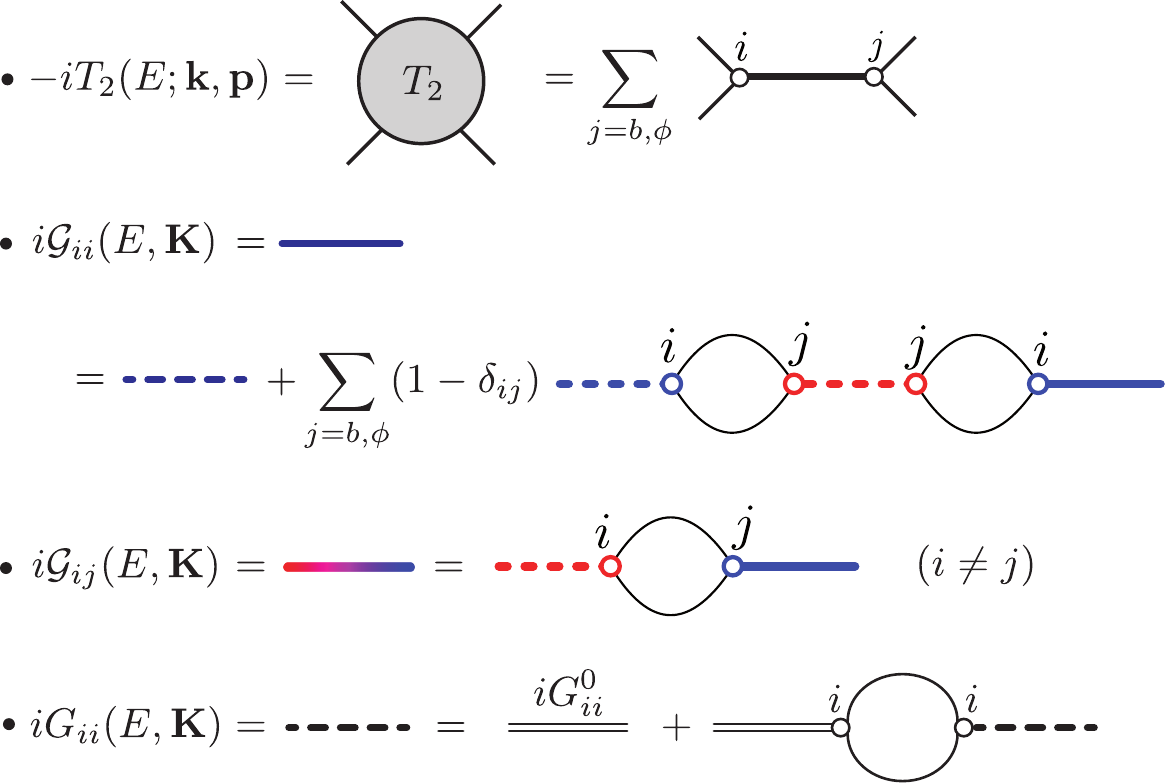}
\caption{
Diagrammatic solution of the two-body problem. The scattering T matrix is obtained from the matrix valued molecule Green's function (thick solid line). Blue color corresponds to the molecule in the $\phi$ channel, red color to the $b$ channel. Thin black lines denote atom propagators, and double lines  the bare molecule propagator. Small circles represent the atom-molecule conversion couplings $g_\phi$ and $g_b$.
}
\label{fig.twochannel}
\end{figure}

The solution of the two-body problem allows us to express all microscopic model parameters of the Hamiltonian~\eqref{eq.fullmodel} in terms of experimental observables. The two-body scattering T matrix is obtained from the exchange diagram shown in the first line in Fig.~\ref{fig.twochannel} leading to
\begin{equation}
T_{2}\left(E;\mathbf{k},\mathbf{p}\right)
= \sum_{i,j=b,\phi} g_i \chi_{i}(\mathbf{k}) \,
\mathcal{G}_{ij}(E) \,g_j \chi_{j}(\mathbf{p})  .
\label{eq:T_2_matrix}
\end{equation}

The coupling between the background and Feshbach channel leads to a hybridization of the matrix valued molecular propagator 
$\mathcal{G}(E)\equiv \mathcal{G}(E,\vecp=0) $  \footnote{By Galilean symmetry $\mathcal G(E,\vecp)=\mathcal G(E-\frac{\vecp^2}{4m})$  },
\begin{equation}
\label{eq:Gmat}
\begin{pmatrix}
\mathcal G_{bb}(E) & \mathcal G_{b\phi}(E) \\
\mathcal G_{\phi b}(E) & \mathcal G_{\phi \phi}(E)
\end{pmatrix}
=
\begin{pmatrix}
G^{-1}_{bb}(E) & -\Sigma_{b\phi}(E) \\
-\Sigma_{\phi b}(E) & G^{-1}_{\phi \phi}(E)
\end{pmatrix}^{-1}.
\end{equation}
A straightforward evaluation of the diagrams shown in Fig.~\ref{fig.twochannel} yields the Dyson equation for the diagonal elements
\begin{equation}
\label{eq:Dyson1}
G_{ii}^{-1}(E)=[G^{0}_{ii}(E)]^{-1} - \Sigma_{ii}(E)
\end{equation}
where
\begin{equation}
\label{eq:G0}
[G^{0}_{ii}(E)]^{-1} = (E + i 0^{+}) \delta_{i \phi} - \tilde\nu_i.
\end{equation}
The self-energies are given by
\begin{subequations}
\begin{align}
\Sigma_{ij}(E) & = \frac{g_i g_j}{2} \hspace{-0.1 cm} \int \hspace{-0.2 cm} \frac{d^3 q}{(2 \pi)^3}  \frac{\chi_i(\mathbf{q}) \chi_j(\mathbf{q})}{(E + i 0^+) - q^2/m}\\
& = -\frac{g_i g_j}{8 \pi} m\,Q_{ij}(E).
\end{align}
\end{subequations}
For a Lorentzian form factor $\chi_i(k)=1/(1+(k\sigma_i)^2)$ which mimics the wave function of a closed-channel molecule, the rescaled self-energy $Q_{ij}(E)$ approaches $Q_{ij}(0)=1/(\sigma_i+\sigma_j)$ at zero energy. From the T matrix, the $s$-wave scattering amplitude $f(k)$ is obtained by s-wave projection and on-shell evaluation,
\begin{equation}
\label{eq:f}
f(k)=-\frac{m}{8 \pi}\int \frac{d\cos\theta_{\angle(\vecp,\veck)}}{2} T_2(E=k^2/m,\veck,\vecp).
\end{equation}
Here $m/(8 \pi)$ is the phase space factor for a pair of incoming and outgoing identical bosons. 

Comparing this result to the low-energy expansion
\begin{equation}\label{scattampl}
f(k)\approx\frac{1}{-1/a+\frac{1}{2} r_e k^2-ik}
\end{equation}
 allows us to compute the scattering length
 \begin{equation}
a = \; a_b + \frac{(1 - a_b Q_{\phi b}(0))^2}{
a^{-1}_\phi - a_b Q^2_{\phi b}(0)}\;\;,
\label{eq:2-body-scattering-length}
\end{equation}
where $a_\phi$ and $a_b$ are given by ($i=\phi,b$)
\begin{equation}
\label{eq:ai}
a^{-1}_i  = -\frac{8 \pi}{m} \frac{\tilde \nu_i}{g^2_i} + \frac{1}{2\sigma_i}.
\end{equation}

\begin{figure}[t]
\includegraphics[width=0.85\linewidth]{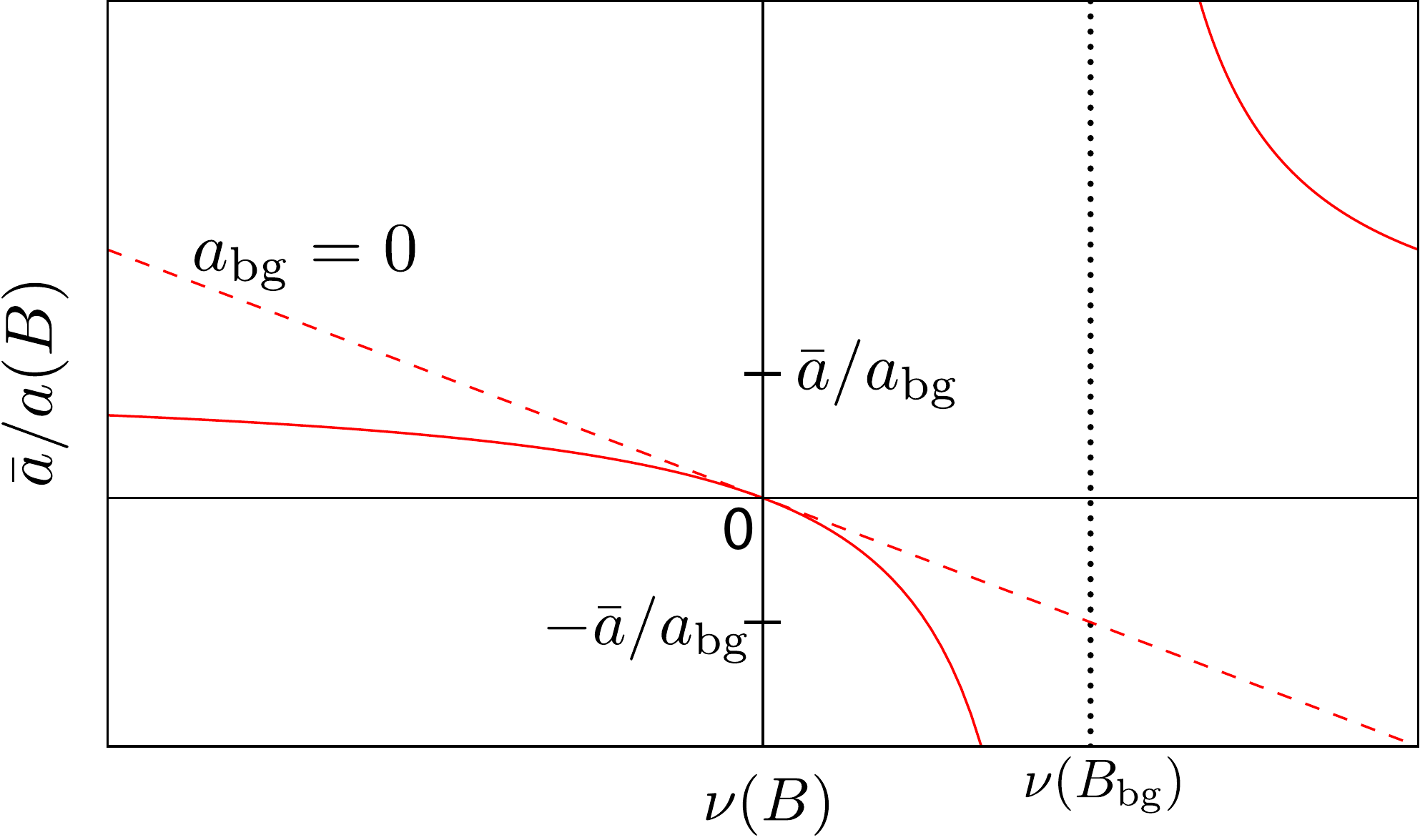}
\caption{Determination of the background scattering length. The inverse scattering length (solid) as a function of the detuning $\nu(B)$ from the position of the Feshbach resonance. The dotted  vertical line represents the detuning $\nu(B_{\rm bg})$ where the  scattering length crosses zero. The red dashed line shows the case of zero background scattering length. 
}
\label{fig.backgrounddet}
\end{figure}

The full scattering length~\eqref{eq:2-body-scattering-length} depends on the magnetic field $B$  only through $a_\phi$ 
which involves the bare detuning $\tilde \nu_\phi(B)=\delta\mu (B-B_{\rm c})$. In order to connect the model parameters with experimentally accessible quantities, we employ the fact that quite generally the magnetic field dependence of the scattering length close to a Feshbach resonance at $B=B_0$ 
can be parametrized in the form~\cite{Chin2010}
\begin{eqnarray}\label{aformula}
a(B)&=&a_\text{bg} +a_\text{res}(B)\nonumber\\
&=&a_\text{bg} -\frac{\hbar^2}{2\mu_\text{red} r^\star \delta \mu (B-B_0)}\, .
\end{eqnarray}
Here $\delta\mu$ is the differential magnetic moment of the closed-channel molecule and $\mu_\text{red}$ the reduced mass ($\mu_\text{red}=m/2$ for identical bosons). Moreover, the intrinsic length $r^\star>0$ may be determined from the dependence of the two-body bound state energy $\epsilon_D$ (we choose signs such that the dimer binding energy is positive) that exists for $a>0$ or the associated binding wave number $\kappa_{\rm D} = \sqrt{2 \mu_{\rm red} \epsilon_{\rm D}} / \hbar$ on the renormalized detuning 
$\nu(B)=\delta \mu (B-B_0)$ via
\begin{equation}
\label{eq:r*}
r^\star = - \left.\frac{\hbar^2}{2 \mu_{\rm red}} \frac{\partial \kappa_{\rm D}}{\partial \nu(B)}\right|_{B=B_0}\, .
\end{equation}
The background scattering length $\abg$, in turn, is determined by the value  of the detuning $\nu(B_\text{bg})$ where the full scattering length $a(B)$ crosses zero. Experimentally, its value can be inferred from a plot of the inverse scattering length as a function of the detuning, as  shown in Fig.~\ref{fig.backgrounddet}. It is important to emphasize that this definition of the background scattering length does not rely on any assumption about the magnetic field dependence of the total scattering length far away from the resonance.

The requirement that our result~\eqref{eq:2-body-scattering-length} is consistent with the generic behavior~\eqref{aformula} 
of the scattering length near the Feshbach resonance at $B=B_0$, immediately leads to the identification $a_b= a_{\rm bg}$. 
Choosing $\sigma_b=\bar{a}$ for the range, the single remaining parameter $g_b^2/\tilde \nu_b$ in Eq.~\eqref{eq:background}
is thus determined by the value of the background scattering length via Eq.~\eqref{eq:ai}. In addition, the condition
\begin{equation}
a^{-1}_\phi = a_{\rm bg} Q^2_{\phi b}(0) 
 + \left(1 - {a}_{\rm bg} Q_{\phi b}(0) \right)^2 a^{-1}_{\text{res}}(B)
\label{eq:scat_phi}
\end{equation}
allows one to express the bare detuning $\tilde \nu_\phi(B)=\delta\mu (B-B_{\rm c})$ which appears in the definition of $a_\phi$ in terms of 
the experimentally relevant renormalized value $\nu(B)=\delta \mu (B-B_0)$. It also determines the microscopic coupling
\begin{equation}
g^{2}_{\phi}  =  \frac{8 \pi}{m^2 r^\star} \frac{1}{\left(1 -  {a}_{\rm bg} Q_{\phi b}(0) \right)^2}
\end{equation}
in terms of $r^\star$ and a background scattering length dependent renormalization.

An important consistency check for our choice $\sigma_\phi=\sigma_b=\bar a$ for the two range parameters is provided by calculating the effective range of the full two-body scattering amplitude which is given by (see also \cite{Pricoupenko2011})
\begin{equation}
r_{\rm e} = 
-2 r^\star \left(1 - \frac{{a}_{\rm bg}}{a}\right)^2 
+ 3 \bar{a} \left(1 - \frac{4}{3} \frac{\bar{a}}{a} \right) .
\label{eq:two_re}
\end{equation}
In the limit of pure background scattering $a\to\abg$, the first term vanishes and the resulting effective range is rather close to the expression 
\begin{equation}\label{revdw}
r_e^\text{vdw}=2.9179 \bar a \left(1-\frac{2\bar a}{a}+\frac{2 \bar a^2}{a^2}\right)
\end{equation} 
obtained by Flambaum, Gribakin, and Harabati for low-energy scattering in deep potentials with a van der Waals tail~\cite{Flambaum1999}.
For vanishing background scattering, in turn, the latter result is recovered  in the limit of an open-channel dominated resonance where $s_{\rm res}\gg 1$. 
With decreasing values of the resonance strength, the effective range crosses zero near $s_{\rm res}=2/3$ and eventually approaches the well known result 
$r_e\to-2r^\star$ in the closed-channel dominated limit~\cite{Chin2010}.

\begin{figure}[t]
\includegraphics*[width=\columnwidth,clip=true]{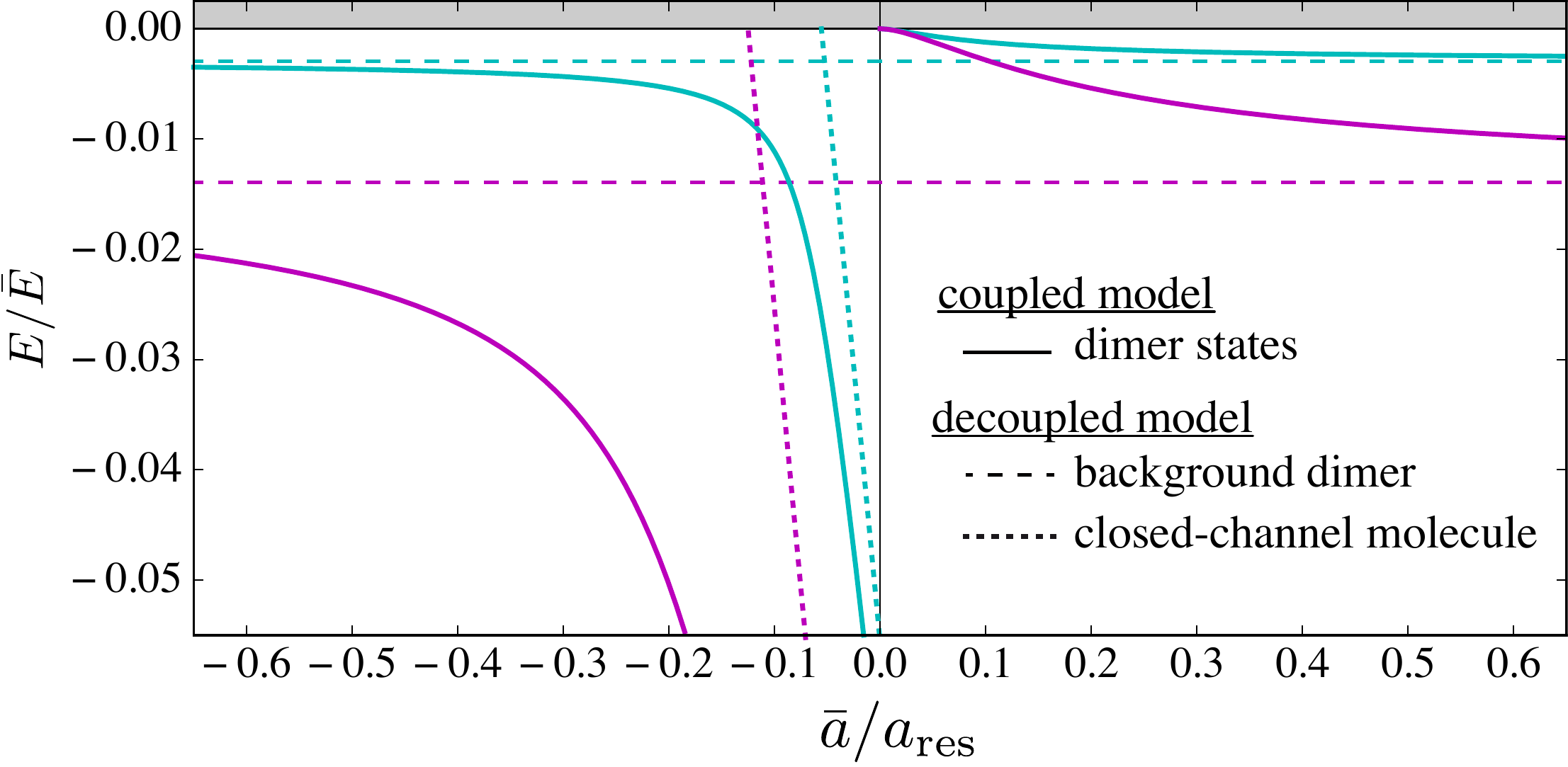}
\caption{
Two-body spectrum as a function of the resonant part of the scattering length $\bar a /a_\text{res}$. Energies are given in units of $\bar E = \hbar^2/m \bar a^2$. The spectrum (solid lines) is shown for an open-channel dominated resonance with $s_{\rm res}=10$ with positive background scattering length ${a}_{\rm bg}/\bar{a} = 10$ (magenta) and $20$ (cyan). Also shown are the energy of the bare Feshbach molecule  (dotted)  and the background dimer (dashed) in the absence of a coupling to the Feshbach channel. }
\label{fig.twobodyspectrum}
\end{figure}

The two-body bound state spectrum is determined by the poles of the T matrix in the complex energy plane. 
In Fig.~\ref{fig.twobodyspectrum}  the two-body spectrum is shown for large positive values of the background scattering length $\abg$  as a function of the resonant 
part of the scattering length $\aF(B)$  defined by Eq.~\eqref{aformula}, i.e., $1/\aF(B)=-mr^\star\nu(B)$.  In this case the background channel supports a bound 
state with binding energy $\epsilon_\text{bg}$ when decoupled from the Feshbach channel (dashed lines). The coupling to the Feshbach channel   leads to a 
level repulsion between the background dimer and the bare Feshbach molecular state (dotted). Note that at resonance, where the dimer state reaches the 
continuum in the coupled model, it is  accompanied by a second deeply bound state in the spectrum.

\section{Three-Body Scattering}
\label{chp:3-body}

\subsection{Coupled-channel STM equation}

\begin{figure}[t]
\includegraphics[width=\linewidth]{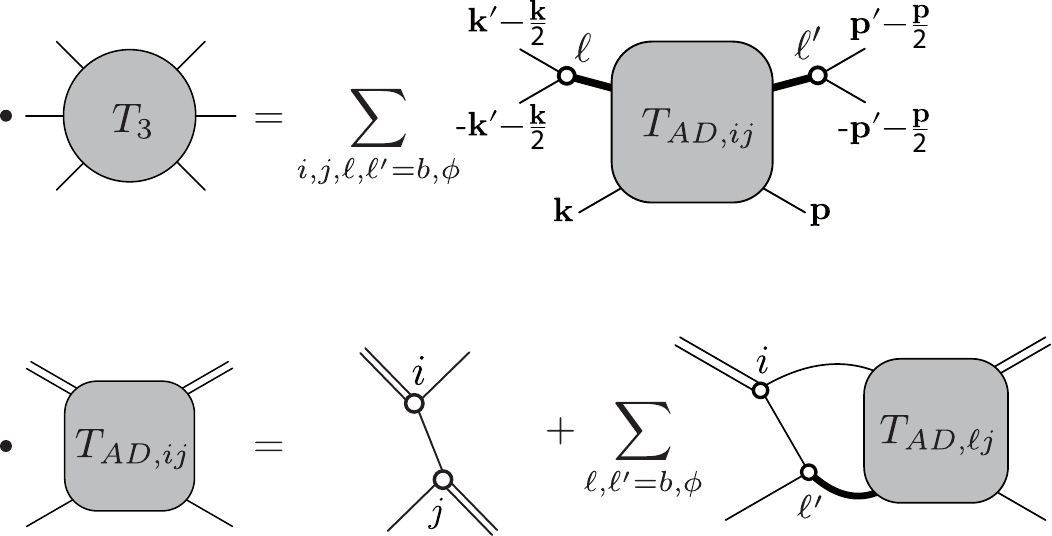}
\caption{Diagrammatic solution of the three-body problem in the absence of a three-body force. The three-body T matrix $T_3$ is determined by the atom-dimer scattering vertex $T_{AD}$ and the exchange of the molecule fields $\phi$ and $b$ (thick lines). Thin (double) lines correspond to atom (bare molecules) propagators. The vertex $T_{AD}$ is given by the matrix-valued STM equation represented diagrammatically in the last line.
}
\label{fig.T3Matrix}
\end{figure}

A standard approach to calculate the spectrum of Efimov states starts from the three-atom scattering matrix $T_3$,  which exhibits poles at energies
where three-body bound states appear.  In the absence of a genuine three-body force, $T_3$ is solely determined by two-body scattering processes. In our model, those are mediated by the exchange of the dimers $b$ and $\phi$ as shown in Fig.~\ref{fig.T3Matrix}. Evaluating these diagrammatic expressions gives
\begin{align}
&T_{3}(E;\mathbf{k}',\mathbf{k},\mathbf{p},\mathbf{p}') = \hspace{-0.4 cm}
\sum_{i,j,\ell,\ell'=b,\phi} \hspace{-0.4 cm}
g_\ell g_{\ell'}
\chi_{\ell}(\mathbf{k}') \mathcal{G}_{\ell i}(E-\tfrac{k^2}{2 m}) \nonumber\\
& \quad\quad\quad\times 
T_{{\rm AD},i j}(E;\mathbf{k},\mathbf{p}) \,
\mathcal{G}_{j \ell'}(E-\tfrac{p^2}{2 m}) \chi_{\ell'}(\mathbf{p}').
\label{eq:TAAA}
\end{align}
Here $E$ is the total energy  of the three incoming atoms, and the momenta are specified in Fig.~\ref{fig.T3Matrix}. The vertex $T_{{\rm AD},ij}$ denotes the atom-dimer scattering matrix with $(i,j)\in (b,\phi)$. This vertex  is determined by the matrix generalization of the Skornyakov-Ter-Martirosian (STM) equation as depicted in Fig.~\ref{fig.T3Matrix} which yields 
\begin{align}
& T_{{\rm AD},i j}(E;\mathbf{k},\mathbf{p})\!  = T^{\,0}_{{\rm AD},i j}(E;\mathbf{k},\mathbf{p})
\!+\! \sum_{\ell,\ell'} \! \int_\mathbf{q} \!
T^{\,0}_{{\rm AD},i \ell'}(E;\mathbf{k},\mathbf{q}) \nonumber \\
& \hspace{1.5 cm} \times \mathcal{G}_{\ell' \ell}(E-\tfrac{q^2}{2 m},-\mathbf{q})
T_{{\rm AD},\ell j}(E;\mathbf{q},\mathbf{p})
\label{eq:TAD}
\end{align}
where $\int_\mathbf{q}\equiv\int d^3 q/(2 \pi)^3$, and the sum extends over the fields $(b,\phi)$. The function $T^{\,0}_{{\rm AD},i j}$ refers to the tree-level diagram of the atom dimer scattering, evaluated in the center-of-mass frame. It is given by
\begin{align}
T^{\,0}_{{\rm AD},i j}(E;\mathbf{k},\mathbf{p})  = 
\frac{g_i \, g_j\chi_{i}(\mathbf{p} \! +\! \tfrac{\mathbf{k}}{2}) 
\chi_{j}(\mathbf{k} \! + \! \tfrac{\mathbf{p}}{2})}{ (E+i0^+) - (k^2 \!+\! p^2 
\!+ \!\mathbf{k} \cdot \mathbf{p})/m} .
\label{eq:T0AD}
\end{align}
For the following discussion it is convenient to define rescaled atom-dimer vertices 
\begin{equation}
t_{{\rm AD},i j}(E;\mathbf{k},\mathbf{p})=\frac{k \,p }{m \,g_i\, g_j}T_{{\rm AD},i j}(E;\mathbf{k},\mathbf{p})
\end{equation}
and  analogously  for $t_{{\rm AD},i j}^{(0)}$, where $k=|\mathbf{k}|$ and $p=|\mathbf{p}|$ \cite{Braaten:2004rn,Moroz2009,Floerchinger2011}. 
Since we consider only low-energy scattering, we may perform an s-wave projection of the atom-dimer vertex
\begin{equation}
t_{{\rm AD},i j}(E;k,p) 
= \frac{1}{2} \int_{-1}^{\,1} \!\! d\cos\theta_{\angle \mathbf{k},\mathbf{p}}
\;\;t_{{\rm AD},i j}(E;\mathbf{k},\mathbf{p}) ,
\label{eq:s-wave_projection}
\end{equation}
and  analogously  for $t_{{\rm AD},i j}^{(0)}$. This leads to the coupled-channel $s$-wave projected STM equation
\begin{align}
&\!\!\! t_{{\rm AD},i j}(E;{k},{p}) \! = \! t^{\,0}_{{\rm AD},i j}(E;{k},{p})
\! + \! \sum_{\ell,\ell'} \! \int \hspace{-0.2 cm} \frac{dq}{2\pi^2}
t^{\,0}_{{\rm AD},i \ell'}(E;{k},{q})\nonumber\\
& \hspace{1.5 cm} \times m \,g_{\ell} \,g_{\ell'} \,\mathcal{G}_{\ell' \ell}(E-\tfrac{3 q^2}{4 m})
t_{{\rm AD},\ell j}(E;{q},{p}).
\label{eq:s-wave_STM}
\end{align}

The trimer binding energies $E_{\rm T}$ are found from the poles of the retarded atom-dimer T matrix in the complex energy plane. Near those poles, $t_{\rm AD}$ can be expressed in terms of a pole expansion 
\begin{equation}
t_{{\rm{AD}},i j}(E;k,p) \underset{E \to E_{\rm T}}{\xrightarrow{\hspace*{0.8 cm}}} 
\frac{b_{ij}(k,p)}{E - E_{\rm T} + i 0^+}
\label{eq:trimer_pole}
\end{equation}
where $b_{ij}(k,p)$ is the residue function.
The term $t^{\,0}_{{\rm{AD}},i j}(E;k,p)$ in Eq.~\eqref{eq:s-wave_STM} 
is suppressed by the pole in $t_{{\rm{AD}},i j}(E;k,p)$. Hence, when inserting Eq.~\eqref{eq:trimer_pole} into Eq.~\eqref{eq:s-wave_STM}, the problem is reduced to the solution of the Fredholm integral equation
\begin{equation}
b_{ij}(k,p) \! = \!\!\!
\sum_{\ell=d,\phi} \! \int_{0}^{\infty}dq\, K_{i \ell}(E;k,q)\,
b_{\ell j}(q,p)
\label{eq:BE_STM}
\end{equation}
with kernel
\begin{align}
& \hspace{-0.3 cm} K_{i \ell}(E_{\rm T};k,q)  = \nonumber \\
& \hspace{ 0.3 cm}  \sum_{\ell'=b,\phi} \!\!\!
\frac{t^{\, 0}_{\rm{AD},i \ell'}(E_{\rm T};k,q)}{2 \pi^2} m \,g_\ell \, g_{\ell'}
\mathcal{G}_{\ell' \ell}(E_{\rm T} \! - \! \tfrac{3q^2}{4m}) \;,
\label{eq:kernel}
\end{align}
which can be solved via discretization~\cite{NR}.
Regarding discretized momenta as indices, Eq.~\eqref{eq:BE_STM} can be cast in matrix form
\begin{equation}
\underline{\mathbf{C}}(E) \cdot \underline{\mathbf{B}} \!
\equiv \!
\begin{pmatrix}
\mathbbm{1} - \mathbf{K}_{b b} & -\mathbf{K}_{b \phi}    \\
-\mathbf{K}_{\phi b}          & \mathbbm{1} - \mathbf{K}_{\phi \phi}
\end{pmatrix}
\begin{pmatrix} 
\mathbf{b}_{b b}  & \mathbf{b}_{b \phi}  \\
\mathbf{b}_{\phi b} & \mathbf{b}_{\phi \phi} \\
\end{pmatrix} \!
= \!
\mathbf{0}
\label{eq:residue_equation}
\end{equation}
which admits a non-trivial bound-state solution at the zeros of the determinant, $\det[\underline{\mathbf{C}}(E=E_{\rm T})]=0$.

\subsection{Bound states far from the Feshbach resonance}\label{subsec_faraway}

We first consider the simplified problem of pure background scattering which is relevant for the description of the fully coupled system far from the Feshbach resonance. Of particular interest in this context is the regime where 
$a_\text{bg}$ is so large  that the background channel can support Efimov trimers on its own. Moreover, for $a_\text{bg}>0$, it also supports a two-body bound state. 

\begin{figure}[t]
	\centering
	\includegraphics[width=0.9\linewidth]{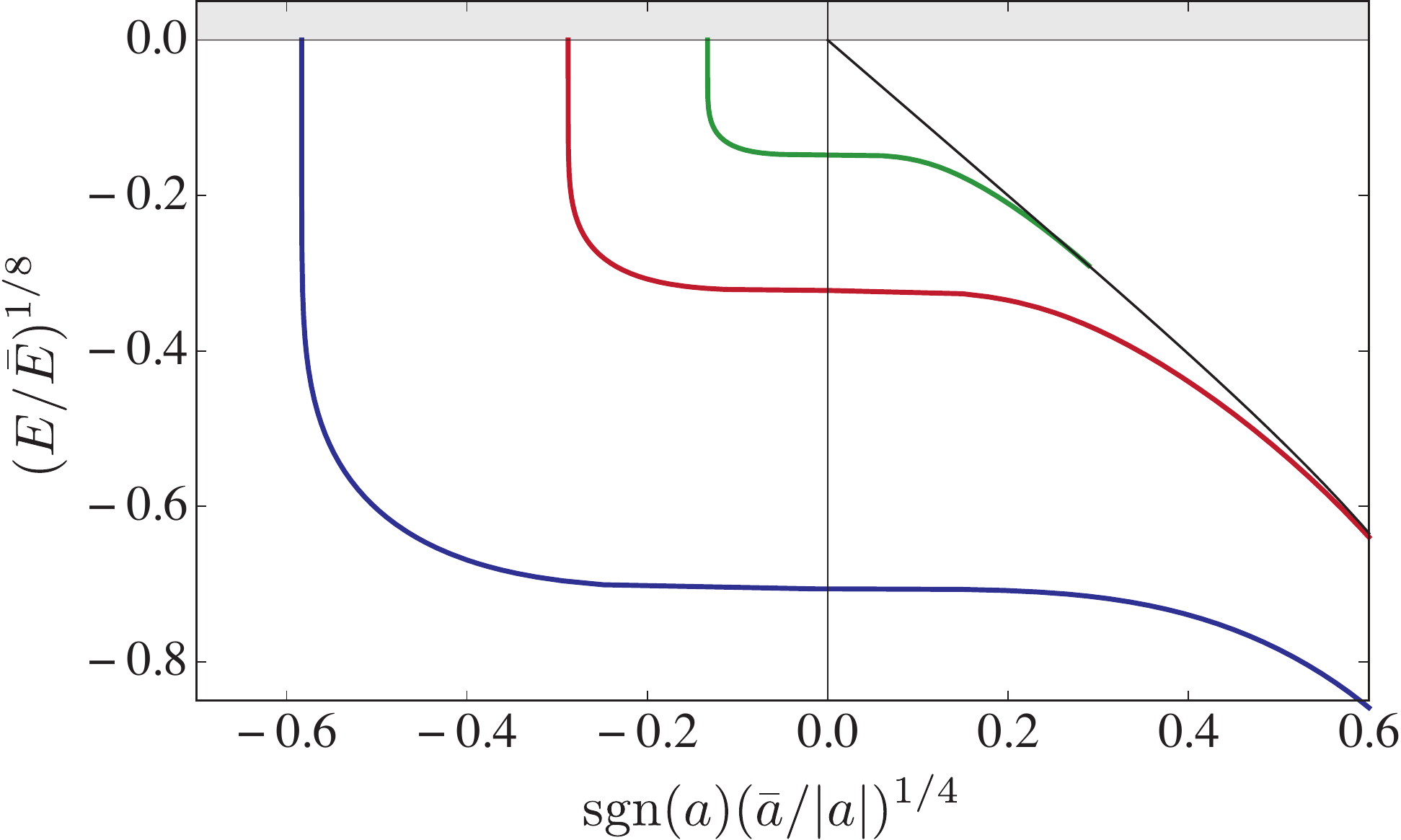}
\caption{ Efimov three-body bound state spectrum for an open-channel dominated Feshbach resonance as accurately described by our background scattering model where the two-body scattering is solely mediated by the field $b$. Colored lines are the binding energies of the three lowest Efimov 
trimers, $E_{{\rm T}}$. The black line is the energy of the two-body dimer state.}
\label{fig:d_channel}
\end{figure}

For pure background scattering, the determination of the three-body bound state spectrum reduces to the solution of $\det[1-\mathbf{K}_{bb}(E)]=0$. The resulting spectrum is shown in Fig.~\ref{fig:d_channel} as a function of the dimensionless inverse scattering length $\bar a/a$.  Due to the finite range of the two-body interaction $\sigma_\phi=\bar a$, the Thomas collapse is avoided~\cite{Thomas35} and a deepest trimer state exists which meets the atom threshold at a scattering length $a_-/\abar=-8.64$. At unitarity it has a binding wave number $\kappa_* \abar=0.249$. As expected, these results are identical to those obtained in our earlier calculation of the Efimov spectrum within a two-channel Feshbach resonance model in the limit $s_{\rm res}\gg 1$ and in the
absence of some further background scattering~\cite{SRZ}. Remarkably, they also agree well with those obtained for direct van der Waals interactions with a shallow potential well within the hyperspherical approach by Wang \textit{et al.}~\cite{Greene2012} where the deepest trimer state at unitarity  appears at a wave number $\kappa_* \bar a \approx 0.216$, while $a_-/\abar\approx-10.2$ is the scattering length where the first trimer detaches from the two-atom continuum. 

\begin{table}[b]
\scalebox{1}{
\begin{tabular}{c|rrr}
& \hspace{0.7 cm} $n=0$ & \hspace{0.7 cm} $n=1$ & \hspace{0.7 cm} $n=2$ 
\\[0.01 cm]
\hline \rule{0pt}{0.4 cm}
$a_{-}^{(n)} / \, \bar{a}$ \;\;\;\;        
& $-8.644$   &    $-145.97$      &    $-3181.9$        \\[0.1 cm]
$\kappa_{*}^{(n)} \, \bar{a}$ \;\;
& $0.2489$   & \; $0.0108$     & \; $0.000475$     \\
[0.1 cm]
\hline \rule{0pt}{0.4 cm}
$a_{-}^{(n+1)} /   a_{-}^{(n)}$   \;\;
& $16.886$    & $21.798$          & $22.630$            \\[0.1 cm]
$\kappa_{*}^{(n)} /\kappa_{*}^{(n+1)}$ \;\;
& $23.068$    & $22.698$          & $22.694$            \\[0.1 cm]
\hline \rule{0pt}{0.4 cm}
$\kappa_{*}^{(n)} \, a_{-}^{(n)}$ \;\;
& $-2.151$   & $-1.575$         & $-1.512$       
\end{tabular}
}
\caption{\label{tab:d_channel} Approach of universal scaling relations for a open-channel dominated resonance where $\sres\to\infty$ as obtained from our background scattering model.}
\end{table}

An independent check of the reliability of our model for describing open-channel scattering in potentials with a van der Waals tail
is obtained by studying the deviations from universality for the three lowest Efimov states due to the finite range of the interaction \footnote{For a recent study on the related effect of finite-range interactions on three-body recombination rates see \cite{Soerensen2013}}. This is shown in Table \ref{tab:d_channel}, where each trimer bound state is labeled by an index $n$, with $n=0$ for the lowest one. 
Characteristic features of the spectrum are the scattering lengths  $a^{(n)}_{-}$ where the $n$-th trimer meets the two-atom threshold and the binding wavenumber at unitarity $\kappa^{(n)}_{*}$. Obviously, the universal limit where ratios like $a_{-}^{(n+1)} /   a_{-}^{(n)}$ and $\kappa_{*}^{(n)} /\kappa_{*}^{(n+1)}$ are given by the discrete scaling factor  $e^{\pi/s_0}\approx22.6944$, is approached rather quickly as $n\geq 2$. 

These results may be compared with systematic studies of the deviations from universal scaling by Deltuva~\cite{Deltuva2012} and by Gattobigio \textit{et~al.}~\cite{Gattobigio2014}. In particular, Deltuva has solved three- and four-body integral equations with form factors chosen to accurately reproduce the behavior of Lennard-Jones potentials. Remarkably, our values for the ratios  $a^{(1)}_{-}/a_{-}$, $a^{(2)}_{-}/a^{(1)}_{-}$, and $a^{(3)}_{-}/a^{(2)}_{-}$ are within $4.9\%$, $0.7\%$, and $0.04\%$ of Deltuva's values of $17.752$, $21.935$, and $22.639$. A simple alternative to account for finite range effects was developed by Gattobigio \textit{et al.} \cite{Gattobigio2014}, introducing an additive correction to the three-body parameter. From a hyperspherical formalism  they find for a Gaussian potential $\kappa_{*} /\kappa_{*}^{(1)}\approx23.0$ which is within $0.2\%$ of our prediction. A similar level of agreement is obtained with the results obtained within effective field theory by Ji \textit{et~al.}~\cite{Ji2010,Ji2011,Ji2015}.

\section{Efimov Spectrum at Negative Background Scattering Length}\label{subsecaneg}

\begin{figure}[b]
	\includegraphics[width=0.96\linewidth]{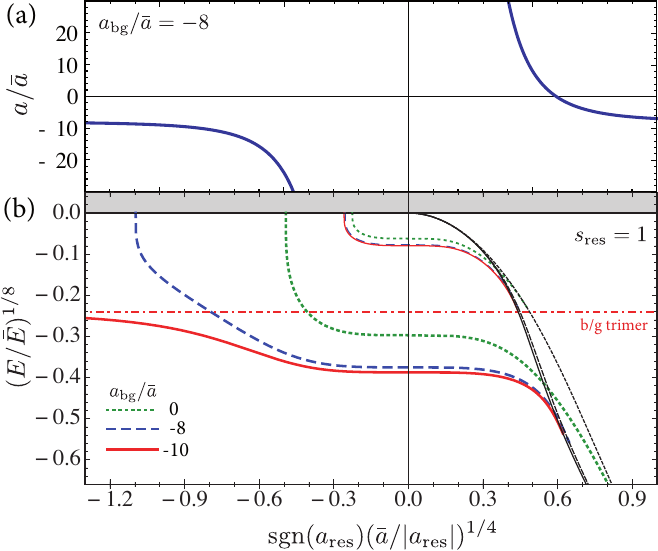}
\caption{ (a) Scattering length $a(B)/\abar$ as a function of the inverse, resonant contribution $\abar/a_\text{res}$ for $\abg/\bar a = -8$. (b) Efimov spectrum for the two deepest Efimov states for $s_{\rm res}=1$. Green, blue, and red 
curves correspond to ${a}_{\rm bg}/\bar{a} =0,-8,$ and $-10$, respectively.  Black lines give the respective
atom-dimer thresholds.  The red dot-dashed line shows the trimer state in the background channel for the case of $\abg/\bar{a} =-10$.
}\label{fig:negspec}
\end{figure}

We now turn to the fully coupled model, starting with the case of negative background scattering length. In Fig.~\ref{fig:negspec} we show the 
Efimov spectrum for fixed resonance strength $s_\text{res}=1$ and three different values of $a_\text{bg}<0$.  In order to better understand the underlying 
physics, the spectrum is displayed as a function of the inverse resonant part $1/a_\text{res}=-m r^\star \nu(B)$ of the full scattering length  which is proportional 
to the magnetic detuning from resonance $\nu(B)$.  As shown in Fig.~\ref{fig:negspec}, the relevant regime is $\bar a/|a_\text{res}|<1$ 
while for values $\bar a/|a_\text{res}|$ of order $1$ or larger, the physics is completely determined by the background scattering. 

The different values of $a_\text{bg}$ chosen in Fig.~\ref{fig:negspec} are representative for three different physical situations: For background scattering 
lengths whose magnitude is smaller or of order $\bar{a}$, the resonant and full scattering lengths are practically indistinguishable in the relevant regime. The 
Efimov spectrum is then identical with the one obtained in our earlier work~\cite{SRZ}, giving rise to a set of universal curves $E/\bar{E}$ as a function of 
$\bar{a}/a$ which only depend on the resonance strength parameter $s_\text{res}$. This is the case shown as dotted green lines in Fig.~\ref{fig:negspec}(b) 
which correspond to the choice where $a_\text{bg}=0$.  Considerably larger values 
$|{a}_{\rm bg}| >\bar{a}$ of the background scattering length, 
however, lead to a qualitatively different form of the Efimov spectrum. As will be discussed in more detail in Section~\ref{subsec_expcom} below, this is relevant 
e.g. for Feshbach resonances in $^{85}$Rb, where ${a}_{\rm bg}$  significantly exceeds $\bar a$~\cite{Wild2012,Blackley2013}.  In this regime two different 
cases have to be distinguished, depending on whether or not the background interaction is able to develop a three-body bound state on its own. 
In our model, this happens if the dimensionless background scattering length $a_{\rm bg}/\bar{a}$ is either below or above $-8.64$. 
Correspondingly, the two lowest Efimov states are shown in Fig.~\ref{fig:negspec} for $\abg/\bar{a}=-8$ and $\abg/\bar{a}=-10$ as blue and red lines, 
respectively.

Apparently, the finite background scattering length leads to a distorted spectrum and the Efimov states are shifted to lower energies. Even right at resonance, where the scattering is  dominated by the Feshbach channel, the trimer energy is substantially modified.  In particular, in a situation where the background channel supports a trimer bound state on its own, the lowest Efimov state of the fully coupled problem does not reach the scattering continuum as one moves far away from the Feshbach resonance. Instead, as shown for the specific choice $\abg/\bar{a}=-10$, the lowest trimer adiabatically  evolves into the background dominated Efimov state whose energy is shown as the dot-dashed horizontal line in Fig.~\ref{fig:negspec}(b). The position where the trimers meet the atom-dimer thresholds, $\bar a/a_*^{(n)}$, are also affected. The reason is two-fold: first, the trimer spectrum itself is modified by the presence of a finite background scattering, and, second, also the specific shape of the atom-dimer threshold depends on $\abg$.

\begin{figure}[t]
\centerline{ 
\includegraphics*[width=0.95\linewidth]{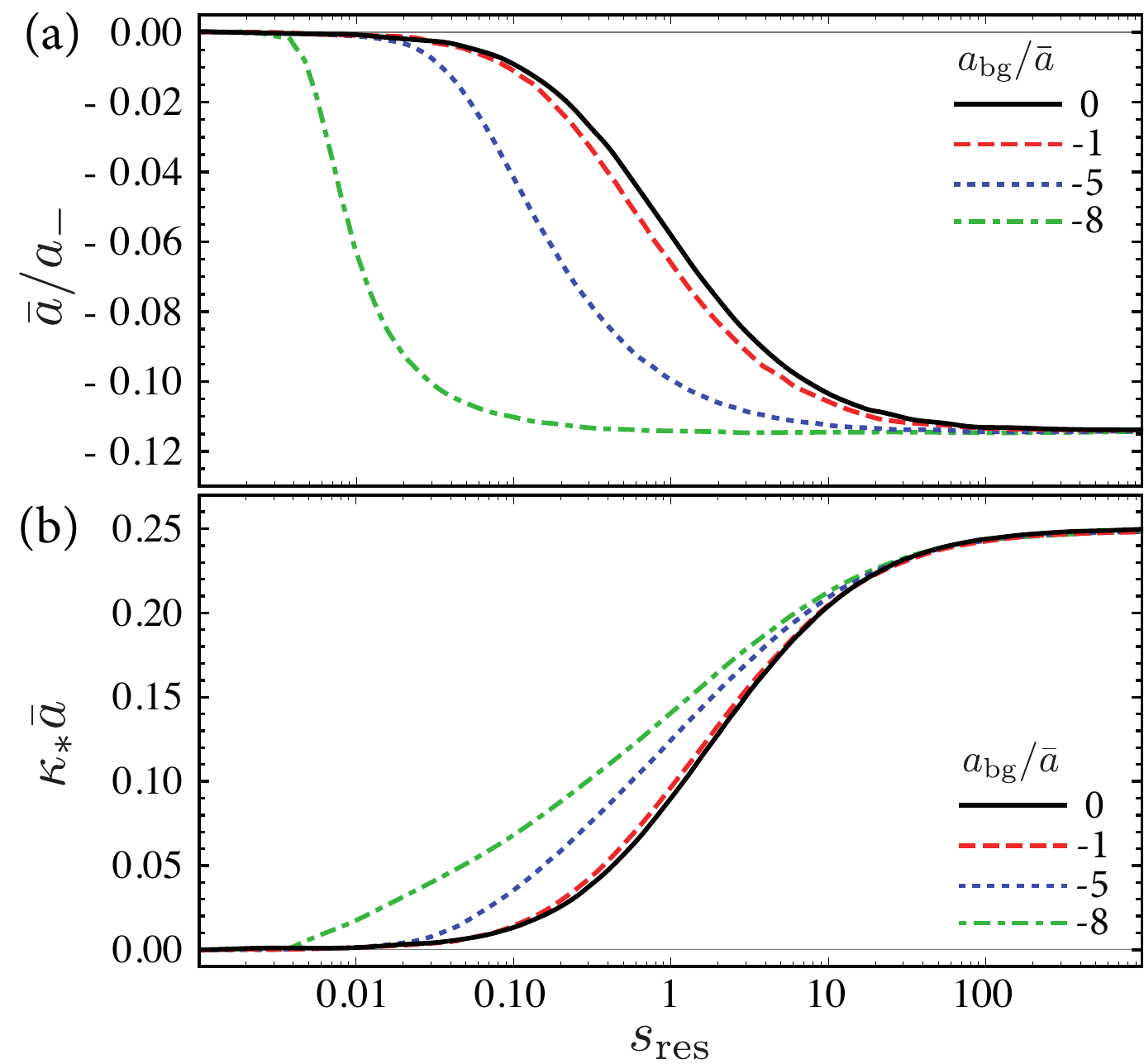}}
\vspace*{0.0cm}
\caption{
Dependence of (a) the ratio $\abar/a_-$ and (b) Efimov binding wave number at unitarity $\abar \kappa_*$ on $s_\text{res}$, for various values of a negative background scattering length. The black solid lines reproduce the results of Ref.~\cite{SRZ}.} 
\label{fig:neg_abg1}
\end{figure}

In Fig.~\ref{fig:neg_abg1}(a) we study the dependence of the ratio $a_-/\abar$ as a function of the resonance strength $s_\text{res}$ for a range of values of the background scattering lengths $\abg<0$. As expected and emphasized in previous work~\cite{SRZ}, the result $a_-/\abar \approx -8.64$ obtained for $\abg=0$ which gives rise to van der Waals universality is unaffected by even quite substantial values of the background scattering length in the open-channel dominated limit $s_\text{res}\gg1$.  By contrast, for intermediate or closed-channel dominated Feshbach resonances, a finite negative background scattering length has the effect to shift $a_-$ toward the open-channel dominated result $a_-/\abar \approx-8.64$. Only for $r^\star\gg |\abg|$  the background scattering becomes again irrelevant and the universal result $a_-\to -10.3\, r^{\star}$ for closed-channel dominated resonances is reached.

In Fig.~\ref{fig:neg_abg1}(b), the effect of background scattering on the dimensionless binding wave number $\kappa_* \abar$ of the lowest Efimov state at infinite scattering length is shown as a function of the resonance strength $s_\text{res}$. Apparently, background scattering now has a less dramatic impact 
compared to the change in $a_-$. In particular, unlike for the value $\abar/a_-$,  $\kappa_*\abar$ does not approach the open-channel dominated results even when $\abg\to-8.64\,\bar{a}$.

\begin{figure}[t]
\includegraphics*[width=8.5cm,clip=true]{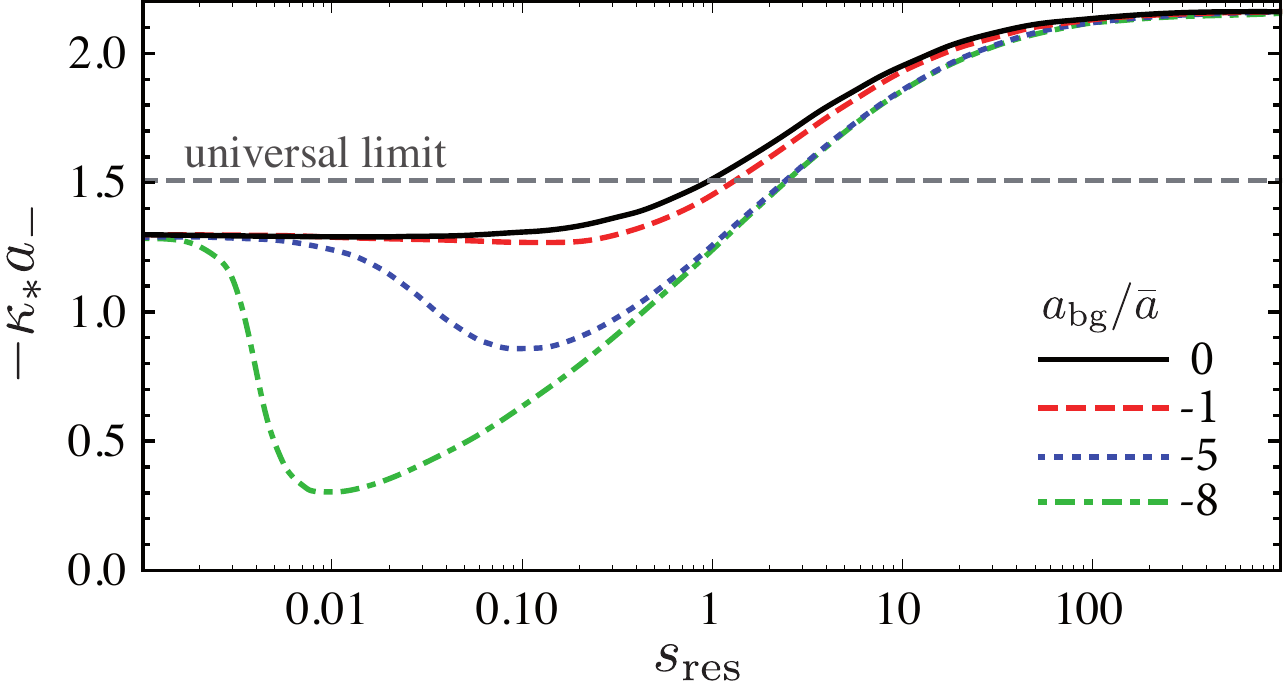}
\caption{
Dependence of  $\kappa_* a_-$ on the resonance strength $s_\text{res}$, for various values of a negative background scattering length.
The black solid line reproduces the result of Ref.~\cite{SRZ} for $\abg=0$. The horizontal dashed line shows the prediction from universal theory $\kappa_* a_-=1.50763\ldots$~\cite{Gogolin2008,Braaten:2004rn,Naidon2017}.
} 
\label{fig:neg_abg2}
\end{figure}

The finding that $\kappa_*$ is much less affected by background scattering than $a_-$ is further confirmed in Fig.~\ref{fig:neg_abg2} where we study the ratio $\kappa_* a_-$ as a function of $s_\text{res}$. Here the dashed line represents the universal result $\kappa_* a_-\approx-1.508$ valid for highly excited Efimov states in the universal regime $n\gg 1$. The qualitatively different impact of $\abg$ on $\kappa_*$ is apparent through the strong deformation of the observed ratio $a_-\kappa_*$. In particular, the pronounced minima are a consequence of the different dependence of $\abar / a_-$  and $\kappa_*\abar$ on the resonance strength $\sres$  which, e.g., for $\abg/\bar{a}=-8$, becomes maximal at $s_\text{res}\simeq 0.01$.

A qualitative understanding of the effects found above can be obtained by considering the result~\eqref{eq:two_re} for the effective range. 
In this expression, background scattering leads to a renormalization of the Feshbach strength parameter $r^\star$ of the form~\cite{Pricoupenko2011}
\begin{equation}\label{rstareff}
r^\star_\text{eff}=r^\star\left(1-\frac{a_\text{bg}}{a}\right)^2= r^\star\left(\frac{a_\text{res}}{a}\right)^2.
\end{equation}
Provided that both $a$ and the background scattering length are negative, the effective value $r^\star_\text{eff}$ of the characteristic length associated with 
the Feshbach resonance is thus reduced compared to its value in the case of pure resonant scattering. Recalling that the resonance strength is 
$s_\text{res}=\bar{a}/r^\star$, this shifts the physics toward the more open-channel dominated regime. This explains the observation in Fig.~\ref{fig:neg_abg1}(a) that for $\abg=-8\, \bar{a}$ the result $a_-=-8.64\,\bar{a}$ characteristic for an open-channel dominated resonance remains valid down to very small values of the resonance strength $s_\text{res}$. 
In fact, when $a\to \abg$ the parameter $r^\star_\text{eff}$ even approaches zero. If in this case the background scattering length is itself close to the `critical' value $-8.64\,\bar{a}$, one recovers the open-channel dominated result $a_-=-8.64\,\bar{a}$ independent of the resonance parameter  $r^\star$.

The observation that the wave number $\kappa_*$ is much less affected by background scattering can be explained by the fact that $\kappa_*$ is measured at $a\to\infty$ and thus the renormalization~\eqref{rstareff} of $r^\star$ is absent.  Changes in $\kappa_*$ therefore originate from beyond effective range effects which are probed only due to the large binding energies involved.

\section{Efimov Spectrum at Positive Background Scattering Length}\label{secthreepos}

Based on a simple statistical argument, van der Waals interactions in a single open channel give rise to a positive background scattering length in three out of four cases~\cite{Pethick2002}. In fact, the first observation of Efimov physics had been made close to a Feshbach resonance in $^{133}$Cs~\cite{Grimm06,Lange2009,Berninger2011} which is characterized by a rather 
 large positive value $\abg=18.6\,\bar{a}$ of the background scattering length. In such a case, the spectrum of the few lowest three-body bound states  exhibits a number of features which differ from the conventional picture with self-similar energy versus inverse scattering length curves applicable near the 
accumulation point at $E=1/a=0$. These features are caused by the presence of a mutual influence of both two- and three-body bound states in the background channel, a complication which is not present for negative values of $\abg$, because in this case the background channel does not support two-body bound states on its own. 

Quite generally, a weakly bound two-body state exists whenever the full scattering length $a(B)$ is positive. Near the resonance, 
its binding energy has the universal value $\epsilon_D=\hbar^2/ma^2$. Moving away from the resonance on the side where $a(B)$ is still positive, the two-body 
bound state becomes background dominated.  In the limit $\abg\gg \bar{a}$,  its energy approaches $\epsilon_{D}\to \hbar^2/m\abg^2$ which is independent of 
the detuning $\nu(B)\sim (B-B_0)$. A similar situation is present in the three-body sector, provided the background channel can support its own trimer bound 
states. 

\begin{figure}[t]
\includegraphics*[width=8cm,clip=true]{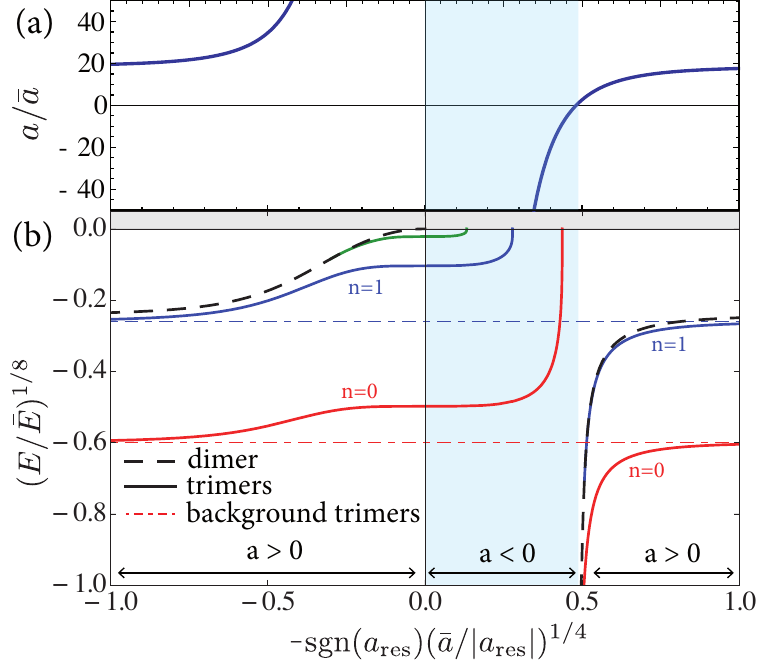}
\caption{
(a) Scattering length $a(B)/\abar$ as a function of the inverse, resonant contribution $\abar/a_\text{res}$ for $\abg/\bar a = 18.6$. (b) The resulting Efimov spectrum for the three deepest Efimov states (solid colored lines) for $s_{\rm res}=596$ as realized for the Feshbach resonance at $-12.38$G in $^{133}$Cs \cite{Berninger2013}.   Dashed black lines give the
atom-dimer thresholds. The  dot-dashed lines show the trimer states in the background channel.
}
\label{fig:E_spec_pos_abg}
\end{figure}

In order to exhibit the generic behavior of the Efimov spectrum found in such a case, we have chosen a strongly open-channel dominated Feshbach resonance with $\sres=596$ and a rather large value $\abg=18.6\abar$ of the background scattering length as realized for the Feshbach resonance at $-12.38$G in $^{133}$Cs~\cite{Berninger2013}.  For this value of $\abg$, the background channel 
supports two three-body bound states whose energy is unaffected by a change of the magnetic field near the resonance. They are shown as dot-dashed 
horizontal lines in Fig.~\ref{fig:E_spec_pos_abg}(b), where the spectrum of both two- and three-body bound states in the full model is plotted as a function 
of the rescaled detuning $-1/a_\text{res}=mr^\star\nu(B)$. The corresponding full scattering length $a(B)$ is shown in Fig.~\ref{fig:E_spec_pos_abg}(a).  
When plotting  the  spectra  as a function of $-1/a_\text{res}$, the regimes with $a>0$ appear on \textit{both} sides of the resonance.  For $a(B)>0$, a weakly bound dimer exists in the spectrum (dashed black line) with universal binding energy $\epsilon_D=\hbar^2/ma^2$ close to resonance at $\nu(B)=0$. As one moves away from the resonance to the left, $a(B)\to\abg$, the dimer state becomes background dominated. Its energy $\epsilon_D$ is then given by the dimer binding energy $\epsilon_{\text{bg}}\simeq\hbar^2/m\abg^2$ in the background channel. The dimer energy at negative detuning is adiabatically connected  to the right hand side  of Fig.~\ref{fig:E_spec_pos_abg}(b), where for large $\nu(B)>0$ again $a(B)\to\abg$.  As one moves from there inwards to the magnetic field where the full scattering length $a(B)$, shown in Fig.~\ref{fig:E_spec_pos_abg}(a), crosses zero, the dimer energy crosses over to very large values and will ultimately follow 
the energy of the closed-channel molecular state described by $\tilde \nu_\phi=\delta\mu (B-B_c)$; see Fig.~\ref{fig.twobodyspectrum}. 

To understand the dependence of the three-body binding energies as a function of the detuning away from resonance,  we first observe that the lowest Efimov state has to be adiabatically connected to the lowest Efimov state in the background channel as one moves from unitarity toward the far left side of Fig.~\ref{fig:E_spec_pos_abg}(b). This statement holds for every Efimov quantum number $n$ for which a trimer  state exists in the background channel. In Fig.~\ref{fig:E_spec_pos_abg}(b) these are two states. The corresponding states in the coupled problem remain necessarily \textit{below} the atom-dimer threshold as one moves away from unitarity to the left in Fig.~\ref{fig:E_spec_pos_abg}(b). As a consequence of the adiabatic connection of the left and right hand side of the spectrum those trimer states will merge with the atom-dimer threshold (dashed line) at \textit{positive} values of $\nu(B)$. In contrast,  trimer states with a quantum number $n$ that is larger than the number of trimers supported by the background channel will always meet the atom-dimer threshold at negative detuning $\nu(B)$; see for instance the green line in Fig.~\ref{fig:E_spec_pos_abg}(b).

\begin{figure}[t]
\includegraphics*[width=0.9\linewidth]{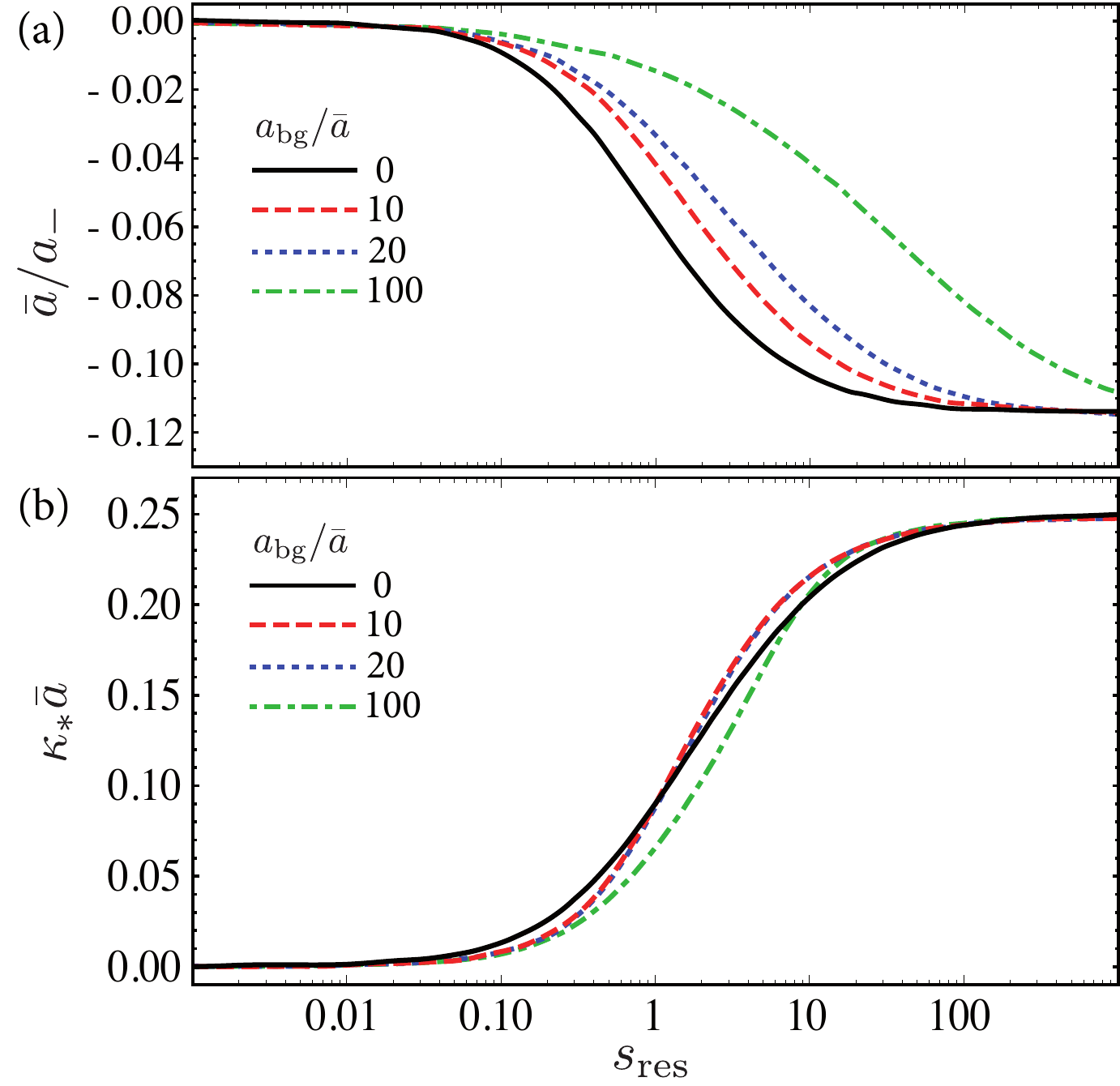}
\caption{
Dependence of (a) the ratio $\abar/a_-$ and (b) Efimov binding wave number at unitarity $\abar \kappa_*$ on the resonance strength $s_\text{res}$, for various values of a positive background scattering length. The black solid lines reproduce the results of Schmidt \textit{et al.}~\cite{SRZ}.
}
\label{fig:pos_abg1}
\end{figure}

Let us now consider an increase of the background scattering length $\abg$. This has a two-fold effect on the background energy spectrum: 
First, the background dimer becomes more weakly bound, and second, the background trimer states move to lower energies. As can be seen from 
Fig.~\ref{fig:d_channel},  the energy of the background trimer states depends only weakly on the background scattering length.  A substantial change only 
appears when $\abg$ is increased by approximately the Efimov factor $e^{\pi/s_0}\approx 22.6$, which lifts a new background trimer state into the 
spectrum. Together with the weak dependence of the background trimer energies on $\abg$ this has the consequence  that the background trimer energies  
set a lower bound on the energy of trimers in the fully coupled problem with the same quantum number $n$. Again, by continuity it follows that on the side of 
negative detuning from resonance, they are always more weakly bound than the background trimer while at positive detuning they must 
exceed the background trimer energy.  

Whether this effect  shifts energies of the trimers at unitarity upwards or downwards depends on their unperturbed energies for $\abg=0$. For instance, as the 
resonance becomes  closed-channel dominated, the wave number of the unperturbed trimer at unitarity decreases, see Fig.~\ref{fig:pos_abg1}(b), and 
ultimatively will scale as $\kappa_*\sim 1/r^\star$ \cite{Petrov2004,SRZ}. At the same time the trimer energies in the background channel are not affected 
by the change in $r^\star$. 
Thus, $r^\star$ effectively tunes the position of the trimer state in between the dashed horizontal lines in Fig.~\ref{fig:E_spec_pos_abg}(b). The binding wave 
number $\kappa_*$ can hence either grow or decrease with respect to its unperturbed value; cf.~Fig.~\ref{fig:pos_abg1}(b). For very large values of $\abg$, one finds in fact that the coupled-channel result oscillates several times around its value for $\abg=0$ (black solid line) as $\sres$ is varied.

Finally, we consider the values of $a_-^{(n)}$ on the right side of the resonance in Fig.~\ref{fig:E_spec_pos_abg}(b) which are also substantially affected by the 
presence of background scattering.  As is shown in Fig.~\ref{fig:pos_abg1}(a), the change in the scattering length $a_-$ depends strongly on the resonance strength parameter $s_\text{res}$. In contrast to the case of 
negative background scattering lengths, the ratio $a_-/\abar$ now evolves toward \textit{larger} values as  $\abg$ increases. 
This observation can again be understood in terms of the renormalized effective parameter $r_\text{eff}^\star=r^\star(1-\abg/a)^2$ which appears in Eq.~\eqref{eq:two_re}.  Obviously, $r_\text{eff}^\star$ is shifted to larger values for $a<0$ and $\abg>0$. A positive background scattering length therefore   shifts the condition for the threshold scattering length  $a_-$ to that of a more closed-channel dominated resonance.  For closed-channel dominated resonances the unperturbed~$\abg=0$ value of $a_-\to -10.3\, r^\star$ is very large [cf.~black line in  Fig.~\ref{fig:pos_abg1}(a)]. The effect of background scattering on the value of $a_-$ therefore becomes increasingly weaker, as is evident from Fig.~\ref{fig:pos_abg1}(a).  Since with increasing values of the 
total scattering length the effect of $a_\text{bg}>0$ is  weakened, also the value of $a_-^{(1)}$  for the first excited Efimov state is much less affected at given $\abg$.

\section{Comparison to Experiments}\label{subsec_expcom}

In the following, we will apply our model in quantitative terms to a selected number of Feshbach resonances where Efimov states have been observed. 
From the outset, it is important to realize that the scattering length $a$ is typically not a direct observable. Instead, a magnetic field is tuned across a given 
position $B=B_0$ of the Feshbach resonance and the associated scattering length $a(B)$ is inferred using a parametrization which either follows 
from a specific model calculation or a phenomenological parametrization as in Eq.~\eqref{aformula}. Obviously, this conversion is not unique, and 
it is important to keep this ambiguity in the translation of the magnetic field to a given scattering length in mind when the comparison to experiments is made.

The specific resonances considered are listed in Table~\ref{tab:3A_thres}, where we restrict ourselves to $s$-wave Feshbach resonances which are well 
isolated. Specifically, we apply the criterion proposed by D'Errico \textit{et al.}~\cite{DErrico2007} who define an isolated Feshbach 
resonance  as one where the distance in magnetic field between two resonances at fields $B_{0,1}$ and $B_{0,2}$ is much larger than the distance $\Delta B$ 
of either resonance to its respective zero crossing, i.e.,~$\beta\equiv|B_{0,1} - B_{0,2}|/{\rm max}\{|\Delta B_{1}|,|\Delta B_{2}|\} \gg 1$. For all the resonances 
studied in Table~\ref{tab:3A_thres} this criterion is well satisfied, with $\beta>5.9$.

In Table~\ref{tab:3A_thres}, we provide quantitative results for the magnetic field $B_-$ where the lowest Efimov state meets the atomic threshold with
respect to the position of $B_0$. Within our model, this quantity is given by $\delta B=B_{-} - B_0 = -\hbar^2 [m r^\star\delta\mu(a_{-} - a_\text{bg})]^{-1}$.  Here,
we assume the differential magnetic moment at the Feshbach resonance $\delta\mu(B_0)$ to be unchanged compared to its value at the three-atom threshold 
$\delta\mu(B_{-})$.  If this is not the case, a corrected value can be obtained by multiplication of $\delta B_-$ by the factor $\delta\mu(B_0)/\delta\mu(B_{-})$.

\begin{table*}[t]
\centering
\begin{tabular}{l|cccc|cc|ccc|cc|cc}
& $F,m_F$ & $B_{0}\,[G]$
& $s_{\rm res}$
& ${a}_{\rm bg}/\bar{a}$
& \multicolumn{2}{c|}{{$\delta B_-=B_{-}  -  B_0 \,[G]$}}
& \multicolumn{3}{c|}{$-a_{-}/\;\bar{a}$}
& \multicolumn{2}{c|}{$a_{-}^{(n+1)}/\;a_{-}^{(n)}$} 
& \multicolumn{2}{c}{$B_{-}^{(n)}  - B_0\,[G]$}   \\
\hline
& & & & & {\footnotesize (exp.)} & {\footnotesize (model)} & 
{\footnotesize (exp.)} & {\footnotesize (model)} & {\footnotesize ($\abg=0$)} 
&{\footnotesize ($n=0$)} & {\footnotesize ($n=1$)} 
&{\footnotesize ($n=1$)} & {\footnotesize ($n=2$)}\\
$^{7}$Li~\cite{Dyke:2013} \rule{0pt}{0.3 cm} 
& $1,1$ & $737.69$ & $0.56$ & $-0.644$ 
& $15.01$ & $5.48$ & $8.11$ & $21.08$ & $23.5$
& $24.94$ & $22.81$ & $0.213$ & $9.34 \! \times \!10^{-3}$ \\
$^{39}$K~\cite{Roy2013} \rule{0pt}{0.5 cm}
& $1,1$ & $402.4$ & $2.8$ & $-0.47$
& $2.28$ & $2.25$ & $11.19$ & $11.39$ & $11.8$
& $20.74$ & $22.39$ & $0.104$ & $4.64\! \times \! 10^{-3}$  \\
\phantom{$^{39}$K}~\cite{Roy2013}
& $1,-1$ & $560.72$  & $2.5$ & $-0.47$
& $2.66$ & $2.35$ & $10.38$ & $11.69$ & $12.2$
& $21.00$ & $22.42$ & $0.107$ & $4.79\! \times \! 10^{-3}$ \\
$^{85}$Rb~\cite{Wild2012}    \rule{0pt}{0.5 cm}
& $2,-2$ & $155.04$ & $28$   & ${-5.65}$  
& {-15.9} & $-20.03$ & ${9.68}$ & $8.68$ & $9.1$
& $17.90$ & $22.00$ & $-0.409$ & $-0.0180$  \\
\phantom{$^{85}$Rb}~\cite{Blackley2013}     
& $2,-2$ & $532.3$ & $4.39$   & $-6.05$  
& --- & $-5.01$ & --- & $8.82$ & $10.8$
& $21.32$ & $22.44$ & $-0.0764$ & $-0.00330$ \\
\phantom{$^{85}$Rb}~\cite{Blackley2013}   
& $2,2$ & $852.3$ & $2.41$ & $-5.01$ 
& --- & $1.42$ & --- & $9.24$ & $12.3$
& $23.38$ & $22.60$ & $0.0285$ & $0.00123$ \\
{$^{133}$Cs}~\cite{Berninger2013} \rule{0pt}{0.5 cm}
& $3,3$ & $-12.38$ & $596$ & $18.6$ 
& $19.94$ & $20.01$ & $9.11$ & $8.82$  & $8.7$
& $16.64$ & $21.78$ & $3.319$ & $0.171$ \\
\phantom{$^{133}$Cs}~\cite{Berninger2013}
& $3,3$ & $548.8$ & $157.7$ & $25.1$ 
& $4.5,5.91$ & $5.38$ & $10.75,10.00$ & $9.70$  & $8.7$
& $15.42$ & $21.65$ & $1.070$ & $0.0572$ \\
\phantom{$^{133}$Cs}~\cite{Huang2014} 
& $3,3$ & $786.8$ & $1692$ & $21.8$ 
& $66.1$ & $67.20$ & $10.06$ & $8.72$  & $8.6$
& $16.77$ & $21.79$ & $12.20$ & $0.639$ \\
\end{tabular}
\caption{Predictions related to the van der Waals universality of the ratio $a_-/\bar a$ for specific Feshbach resonances. The first columns give the  hyperfine states ($F$, $m_F$), the position of the Feshbach resonance $B_0$, its strength $s_\text{res}$, and background scattering length $\abg/\bar a$. Note that  $\sres$ is subject to systematic corrections since it has not necessarily been determined by the measurement of the detuning dependence of the dimer binding energy as given by Eq.~\eqref{eq:r*}. Predictions are compared to experimental observations where available.}
\label{tab:3A_thres}
\end{table*}

Remarkably, our results for $\delta B_-\equiv B_{-} - B_0$ are in good agreement with experimental data on three resonances in $^{133}$Cs which all feature
a large positive background scattering length $\abg$. We note that a narrow $d$-wave resonance cuts through the broad s-wave resonance  at $548.8$G which 
splits the profile of $a(B)$ into two branches just where the Efimov state would appear in an unperturbed situation. In consequence, the scattering length $a_-$ 
is realized experimentally for two closeby values of the magnetic field. Since in our model we do not account for the $d$-wave resonance, our prediction lies in 
between the two observed magnetic field values. 

A detailed study of Efimov loss features at Feshbach resonances in $^{39}$K was performed by Roy \textit{et al.}~\cite{Roy2013}. In particular,
 they studied  a number of resonances of intermediate strength $\sres\lesssim 1$. As shown in Table~\ref{tab:3A_thres}, we find good agreement between 
 theory and experiment for the two resonances which fulfill $\beta\gg 1$ and which also reach into the regime of intermediate strength. Quantitative predictions
for  the other resonances examined in Ref.~\cite{Roy2013} where $\beta\lesssim 1$ probably require a detailed coupled-channel calculation which accounts for multiple closed-channel dominated Feshbach 
resonances.

Concerning the case of $^{85}$Rb, the Efimov effect near an isolated resonance at $155.04$G has been measured by Wild \textit{et al.}~\cite{Wild2012}. 
The resonance is open-channel dominated and features a large negative large of $\abg$ \cite{Claussen03}. In the work of Wild \textit{et al.}, only the 
value of $a_-$ is given but not the specific magnetic field where the Efimov loss feature has been observed. Inferring $\delta B_-$ from the scattering length parametrization $a(B)=a_\text{bg}\left(1-\frac{\Delta B}{B-B_0}\right)$ with $\abg$ taken from Ref.~\cite{Wild2012}, we find reasonable agreement between theory and experiment, while a direct comparison with the given value of $a_-/\abar=-9.68$ is even better. 
In addition, in Table~\ref{tab:3A_thres} we provide predictions for two other resonances in $^{85}$Rb which so far have not been explored experimentally. 

Efimov states in $^7$Li have been studied by Dyke \textit{et al.}~\cite{Dyke:2013} and Gross \textit{et al.} \cite{Gross2009,Gross2010}. 
For the isolated Feshbach resonance at $B_0=737.7$G the background scattering length $\abg/\bar a=-0.644$ is rather small and one might expect a 
negligible effect.  It comes as a surprise, therefore, that the experimental result $a_-/\abar=-8.11$ still follows the open-channel dominated result
despite the rather small value $s_{\rm res}=0.56$ of the resonance strength. Since $\abg$ is so small, our theory predicts only a minor effect of background scattering on $a_-$.  The underlying reason for this discrepancy remains unclear. A possible route to resolve this puzzle may be found in the 
observation that the effective range calculated within our model at the position of $a_-$ differs significantly from the value stated in~\cite{Dyke:2013} which 
underlies their parametrization of the Feshbach resonance.

As discussed above, the different parametrizations of the scattering length profiles $a(B)$ as a function of the magnetic field can be a source of 
discrepancies when experimental results are compared to specific theoretical models. This is highlighted by a comparison of our predictions for $a_-$ 
to  the observed values as shown in the fourth column of Table~\ref{tab:3A_thres}; see in particular the case of $^{133}$Cs. For instance, 
while our findings for $B_{-} - B_0$ are in excellent agreement with the results found by Berninger \textit{et al.} \cite{Berninger2011} and Huang \textit{et al.} 
\cite{Huang2014}, the  prediction for $a_-$ differs substantially from the observed value. 
 A possible reason for this discrepancy may be found in the fact that in~\cite{Berninger2013,Huang2014} coupled-channel models (M2012) were used to 
 identify the magnetic field dependence of the scattering length which may not follow the simple analytical form of $a(B)$ in Eq.~\eqref{aformula}.  Also the 
 discrepancy in the magnetic field $\delta B_-$ for $^{85}$Rb is likely rooted in a similar parametrization dependence.  In the last two columns in Table~\ref{tab:3A_thres}, we show  predictions for $a_-^{(n)}$ and the corresponding values in terms of magnetic field for excited 
 Efimov states. It is evident that a verification of these predictions requires a high magnetic field sensitivity.

Finally, we turn to a more detailed discussion of the Efimov spectrum near Feshbach resonances in $^{133}$Cs which exhibit large values 
of the background scattering length. Specifically, observables which are affected quite strongly by a large value of ${a}_{\rm bg}$ are the scattering lengths 
$a_{*}^{(n)}$ where the trimers meet the atom-dimer threshold. In Fig.~\ref{fig:E_spec_pos_abg}(b) we have already studied the spectrum for parameters as 
realized for the Feshbach resonance at $-12.38$G \cite{Grimm06,Berninger2013}. For this resonance the lowest two trimers meet the atom-dimer 
threshold at positive values of $\nu(B)$. This behavior was theoretically described first by Massignan \textit{et al.} \cite{Massignan2008} and by 
Lasinio \textit{et al.}  \cite{Pricoupenko2010,Pricoupenko2011}. While our results are consistent with these model specific earlier ones --- in particular regarding 
the presence of two Efimov states merging into the atom-dimer threshold at positive values of $\nu(B)$ --- we find no evidence for an additional trimer state that 
never crosses the atom threshold as predicted in~\cite{Massignan2008}. 
Specifically, a quantitative comparison of our predictions for the merging of the lowest trimer state into the atom-dimer threshold can be made with the 
experimental results based on atom-dimer relaxation~\cite{Knoop2009}. The measured values $a_{*}^{(1)}=3.83\,\abar$ and $a_{*}^{(1)}/a_{-}=-0.43$ are significantly smaller than those obtained within universal theory, where $a_{*}^{(1)}/a_{-}=-1.07$. Our model, in turn, gives $a_{*}^{(1)}=2.71\,\bar{a}$ and a 
ratio $a_{*}^{(1)}/a_{-}=-0.31$, in quite reasonable agreement with the experimental results~\cite{Knoop2009}.

\section{Influence of a three-body force}

In this  section, we address the question of whether genuine three-body forces may change the spectrum of three-body bound states substantially compared 
to the results obtained under the assumption of a sum of two-body interactions. Three-body forces arise naturally when evaluating the induced dipole-dipole 
interactions between neutral atoms within third-order perturbation theory. This leads to a microscopic, long-range three-body interaction potential of the form
\begin{equation}\label{ef3_ATPot}
W_\text{AT}=\gamma \frac{1+3\cos\theta_{12}\cos\theta_{23}\cos\theta_{31}}{r_{12}^3r_{23}^3r_{31}^3}
\end{equation}
where $\theta_{ij}$ and $r_{ij}$ are the angles and side lengths of the triangle spanned by the three atoms. The interaction~\eqref{ef3_ATPot} is known as the 
Axilrod-Teller potential~\cite{Axilrod1943} and it is valid at atomic separations $r_{ij}$ much larger than a short distance scale $r_0$ where the effects of
exchange and details of the short-range interaction come into play~\cite{Tang2009,Babb2010,Tang2012}. 
The coefficient $\gamma$ depends on the specific atoms chosen and it has been calculated to high precision for various mixtures of atom species by Babb and 
coworkers~\cite{Tang2012}.  Long-range Axilrod-Teller  forces have only a small effect on the three-body spectrum provided  their typical energy scale $E_\text{AT,vdW}\equiv W_\text{AT}(r_{ij}\simeq l_{\rm vdW})$  is much smaller than the two-body van der Waals energy $E_{\rm vdW}$.  As shown in~\cite{Tang2009,Tang2012} this is indeed the case, with typical values 
$ E_\text{AT,vdW}/E_{\rm vdW}\simeq 0.005$, e.g., for the case of Lithium.

\begin{figure}[t]
  \includegraphics[width=0.55\linewidth]{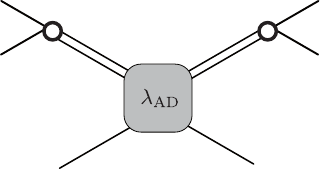} 
  \caption{\label{ef3_fig_3bforcetree} 
  Tree-level diagram which gives the effective, three-body force $\sim(\psi^*\psi)^3$ upon integrating out the  field $\phi$. }
\end{figure}

By contrast, the short-distance behavior of the three-body force, where exchange effects become relevant,  is largely unknown.  In  particular, the short-distance behavior is not expected to exhibit universal features like those present in the finite-range corrections of the two-body scattering amplitude, which are fixed by the van der Waals length.
Moreover, there is no \textit{a priori} reason why they should be negligible compared to those resulting from the sum of two-body
contributions. 
In the following, therefore, we study the consequences for the lowest Efimov trimers resulting from a three-body force which is short range in nature.  In order to simplify matters, we neglect background scattering; i.e., the system is described in the two-body sector by Eq.~\eqref{eq:ResModel}. For the description of a short-range three-body force, we introduce a phenomenological atom-dimer contact interaction
\begin{equation}\label{3BForce}
H_\text{3B}=\lambda_{\rm AD} \int d^3r \phi^*(\vecr)\phi(\vecr)\psi^*(\vecr)\psi(\vecr),
\end{equation}
where the dimensionless coupling strength $\lambda_{\rm AD}\Lambda$  is defined at a UV cutoff scale $\Lambda$, that is related to the  range of the microscopic three-body force. Integrating out the dimer field $\phi$ yields a momentum-dependent three-atom interaction $\sim (\psi^*\psi)^3$ which is determined by the evaluation of the tree-level diagram shown in Fig.~\ref{ef3_fig_3bforcetree}.

The introduction of the atom-dimer force $\lambda_{\rm AD}$ yields an additional contribution to the tree-level exchange diagram shown in the lower panel of Fig.~\ref{fig.T3Matrix}. This leads to a modification of the 
STM equation which reads (in this section we neglect background scattering, set $2m=1$, and define $t_{{\rm AD}}(E;{k},{p})=t_{{\rm AD},\phi\phi}(E;{k},{p})$)
\begin{align}
& t_{{\rm AD}}(E;{k},{p})  =  t^{\,0}_{{\rm AD}}(E;{k},{p})+2 k p \lambda_{ \rm AD}
\nonumber\\
&+ \int_0^\Lambda \frac{dq}{4\pi^2}
\left[t^{\,0}_{{\rm AD}}(E;{k},{q})+2 k p \lambda_{ \rm AD}\right]\nonumber\\
&  \hspace{1cm}\times  g_{\phi}^2 G_{\phi\phi}(E-\tfrac{3 q^2}{2})
t_{{\rm AD}}(E;{q},{p}).
\label{3BF_STM}
\end{align}
In the absence of $\lambda_{ \rm AD}$ the integral of the right-hand-side is UV convergent for any value of $\sres$ due to the functional form of $t^{\,0}_{{\rm AD}}(E;{k},{q})$. When $\lambda_{ \rm AD}\neq0$, explicit UV regularization with cutoff $\Lambda$ is, however, again required for open-channel dominated resonances. In contrast, for closed-channel dominated resonances the function $G_{\phi\phi}(E-\tfrac{3 q^2}{2})$ provides sufficient regularization due to the presence of the large value of $r^\star$. The bound state equation~\eqref{eq:BE_STM} is modified as well and becomes (we define $b(k,p)\equiv b_{\phi\phi}(k,p)$)
\begin{align}
b(k,p) \! =  \! \int_{0}^\Lambda dq\, & \frac{t^{\, 0}_{\rm{AD}}(E_{\rm T};k,q)+2 k p \lambda_{ \rm AD}}{4\pi^2} \nonumber\\
&\times g_\phi^2 G_{\phi\phi}(E_{\rm T} \! \,-\, \! \tfrac{3q^2}{4m})b(q,p)
\label{3BF_BE_STM}
\end{align}
which is  solved by discretization.

\begin{figure}[t]
  \includegraphics[width=0.95\linewidth]{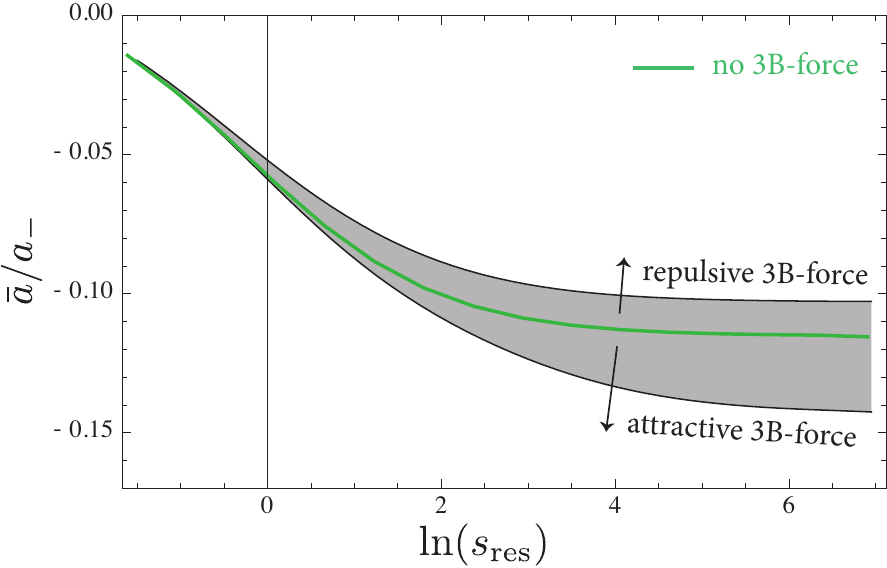} 
  \caption{\label{ef3_fig_3bforcecross} 
 The ratio $\bar a/a_-$ as  a function of $s_\text{res}$. The shaded region describes the effects of a short-range three-body force with values $\lambda_{\rm AD}\Lambda=0.1$ (repulsion) up to $\lambda_{\rm AD}\Lambda=-0.01$ (attraction) defined at an UV cutoff scale $\Lambda\abar=3$. }
\end{figure}

\begin{figure}[t]
  \includegraphics[width=0.9\linewidth]{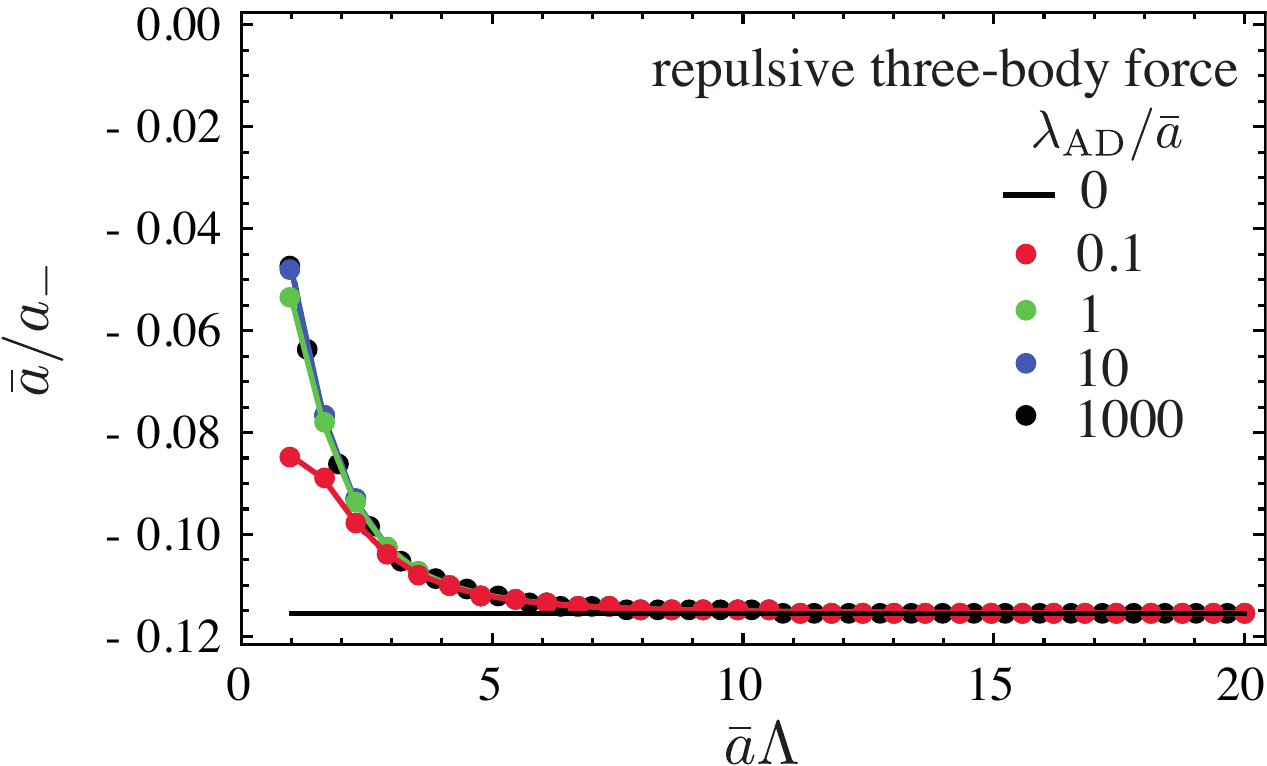} 
  \caption{\label{fig_3B_LambdaDep} UV cutoff dependence of $\abar/a_-$ for repulsive three-body forces given by the dimensionless coupling $\lambda_{\rm AD}/\abar$.}
\end{figure}

In Fig.~\ref{ef3_fig_3bforcecross} we show the predicted ratio $\bar a/a_-$ as a function of $s_\text{res}$ for the specific value $\Lambda \abar=3$ of the UV 
cutoff (the dependence on $\Lambda \abar$ is shown in Fig.~\ref{fig_3B_LambdaDep}). While the solid line displays the result for vanishing three-body forces already shown in Figs.~\ref{fig:neg_abg1}(a) and~\ref{fig:pos_abg1}(a) as solid 
black lines, the shaded region corresponds to finite values of $\lambda_{\rm AD}\Lambda$ in the interval $(0.1\ldots-0.01)$, covering the range from strongly 
repulsive to weakly attractive three-body interactions.  Within this range, three-body forces give rise to a change in the ratio $a_-/\bar a$ which 
varies between $-7$ and $-9.75$ in the open-channel dominated limit $s_\text{res}\gg 1$. 

For closed-channel dominated resonances our model 
predicts that  three-body forces have no effect at all on $\bar a/a_-$,  independently of the sign of $\lambda_{\rm AD}$. This  can be understood from the presence 
of  $G_{\phi\phi}$ in Eq.~\eqref{3BF_STM} which acts as a regulator of the integral and hence suppresses the influence of $\lambda_{\rm AD}$. As expected, 
repulsive interactions shift  $\abar/a_-$ toward larger values while attractive ones lead to a decrease of 
$\abar/a_-$. At fixed cutoff scale $\Lambda\abar$, the ratio $a_-/\abar$ converges to a finite value for an arbitrarily large, repulsive three-body 
force $\lambda_{\rm AD}\Lambda \gg 1$, with a shift of $a_-/\bar a$ not exceeding  $15\%$ for reasonable values of $\Lambda \abar>3$. This result reflects the fact that for repulsive $\lambda_{\rm AD}>0$ the three-body term in Eq.~\eqref{3BForce} becomes irrelevant in the limit $\Lambda\to\infty$ as can be seen in 
Fig.~\ref{fig_3B_LambdaDep} where the cutoff dependence of $\abar/a_-$ is shown.

By contrast,  attractive three-body forces have a  rather dramatic effect on the observed three-body physics. In particular, we find that for  sufficiently strong attraction  three-body forces can  give rise to a non-Efimov three-body bound state that is energetically well separated from the Efimov trimers. 
The fact that such an additional three-body bound state has never been observed might be seen as an indication that three-body forces, if purely attractive, have to be weak.

To conclude, our qualitative model  \eqref{3BForce} for short range three-body forces provides a possible explanation for the so far unexplained observation that the ratio $a_-/\abar$ takes values which depend on the specific alkali atom or Feshbach resonance but always remains confined within the range 
between $-8$ and $-10$. This explanation relies on the assumption that the underlying short-range three-body forces, which are clearly not universal between 
different Alkali atoms, are repulsive in nature. Note that a trivial source of the variation of the ratio $a_-/\abar$, namely the specific shape of the transfer function 
$\chi(\vecr)$ which reflects the detailed form of the closed-channel bound state wave function appears not to be sufficient to account for the observed range found for the 
ratio $a_-/\abar$. Indeed,  choosing, e.g., a Gaussian cutoff $\chi(q)=\exp{(-q^2\sigma_\phi^2/2})$, one finds  $a_-=-7.4\, l_\text{vdw}$ for open-channel dominated resonances instead of $a_-=-8.3\, l_\text{vdw}$ for the exponential cutoff used here and in the previous work~\cite{SRZ}. We have found no reasonable choice of $\chi$ that is able to yield values $|a_-|/\abar$ as large as $10$~\cite{Schmidt}.

\section{Conclusion}

The unexpected observation of van der Waals universality in the Efimov effect of ultracold atoms has triggered a large number of theoretical investigations trying to understand the physics behind this quite surprising effect.  In simple  terms, the origin of van der Waals universality is connected with the fact that the finite range, which appears in the low-energy scattering of ultracold atoms, is 
fixed by the van der Waals or mean scattering length.  At the two-body level, this was realized some time ago by Flambaum, Gribakin, and Harabati~
\cite{Flambaum1999}. The fact that the van der Waals length sets the scale not only for the two-body 
scattering amplitude at low energies but also for the three-body amplitude, whose poles near zero energy determine the scattering lengths $a_{-}^{(n)}$,
can be seen explicitly within the 
hyperspherical approach; see, e.g., Refs.~\cite{Thogersen2009,Wang2012,Greene2012,DIncao2013,Wang2014}.

A complementary approach to understand the origin of van der Waals universality, which covers Feshbach resonances of arbitrary strength $s_{\rm res}$, was developed by Schmidt \etal \cite{SRZ}. Here, $\bar{a}\simeq l_{\rm vdw}$ appears as the natural unit of Efimov physics as a consequence of the finite range $\sigma_\phi$ of the Feshbach coupling, which is  connected with the mean scattering length by $\sigma_\phi=\bar{a}$. In addition, the approach shows that van der Waals universality is restricted to open-channel dominated 
resonances, while in the opposite regime  $s_{\rm res}\ll 1$, it is $r^\star$ rather than the van der Waals which sets the relevant length scale. 

In our present work, we have studied how these results are changed by finite background scattering and  short-range genuine three-body forces. As already pointed out in~\cite{SRZ},  it turns out that van der Waals universality is unaffected by background scattering if the Feshbach resonance is open-channel dominated. In practice, for finite values of the Feshbach resonance strength $\sres$, ratios like $a_-/\bar{a}$ change according to the value of background scattering length, with a restricted or an extended regime of validity of van der Waals universality depending on whether 
$\abg$ is either positive or negative.

Concerning repulsive, short-range three-body forces we find  the intuitively expected trend that $-a_-/\abar$ is shifted toward larger values which is, however, bounded from above by a value close to $10$ for an arbitrary large short-range repulsion.  Despite a possibly wide atom-specific variation of short-range repulsive three-body forces, the ratio $-a_-/\abar$ thus remains within a narrow range, consistent with the different values found experimentally.

As is shown in Table \ref{tab:3A_thres}, quantitative comparison of our model with experiment gives rise to reasonable agreement for a number of Feshbach resonances. In addition, our predictions for the merging of the lowest trimer state 
into the atom-dimer threshold for the Feshbach resonance in $^{133}$Cs  at $B_0=-12.38$ G is encouraging; see the discussion at the end of Sec.~\ref{secthreepos}. Yet, a number of unexplained features remain.  For instance, the observed ratio $a_-/\bar{a}\simeq -8.1$ for the $^7$Li resonance at  $B_0\simeq 738\,$G is still in the regime expected for an open-channel dominated Feshbach resonance, despite the apparently rather small value $s_{\rm res}=0.56$, for which our model predicts  $a_-/\bar{a}\simeq -21$. As discussed above, this may be related to the problem of inferring a proper value of the $r^\star$ parameter and thus the resonance strength. A second major 
discrepancy  appears in the case of the  $^{133}$Cs  resonance at $B_0\simeq 787\,$G, where the observed ratio 
$a_{-}^{(1)}/a_{-}=21\pm 1.3$~\cite{Huang2014} differs  substantially from our prediction  $a_{-}^{(1)}/a_{-}= 16.77$. In fact, as expected for a large value $s_\text{res}\simeq 1692$, 
our result is not far from the value $a_{-}^{(1)}/a_{-}= 17.75$ obtained by Deltuva for a single channel Lennard-Jones potential~\cite{Deltuva2012}. By contrast, the two-spin generalization of this model by Wang and Julienne~\cite{Wang2014} 
gives $a_{-}^{(1)}/a_{-}= 20.7$ for the resonance at $B_0=-12.38\,G$, which has a similar value of the background scattering length and is also in the open-channel dominated limit with $s_\text{res}\simeq 596$.  The origin of this considerable discrepancy between our model and the experimentally observed value remains an open question.

Finally, we emphasize a basic point about the distinction between van der Waals universality and the one 
found in ratios like $a_{-}^{(n+1)}/a_{-}^{(n)}\!\to\! e^{\pi/s_0}$ for large $n$. The latter kind of universality is exact and it results from a limit cycle flow in the three-body scattering amplitude
at low energies. Van der Waals universality, in turn, is not an exact statement. It just tells us that --- for open-channel dominated Feshbach resonances --- the van der Waals length 
sets the characteristic scale both for the scattering lengths $a_{-}^{(n)}$ where Efimov 
trimers first appear and for the corresponding inverse wave numbers $1/\kappa_{*}^{(n)}$
of their binding energies at infinite scattering length. The precise value of ratios like 
$-a_{-}/\bar{a}$ or $\kappa_{*}\bar{a}$ is, however, \emph{not} universal. It is only
within a single channel model for zero-energy resonances in deep potentials with a van der Waals tail
where the ratio $a_{-}/l_{\rm vdw}=-9.45$ becomes a universal number~\cite{Wang2012}. In the realistic case of Efimov states accessible via Feshbach resonances, however, the numerical value of dimensionless ratios like $a_{-}/l_{\rm vdw}$ depends on the specific choice of the transfer
function $\chi(r)$. In practice, fortunately, the ratios are not very sensitive to this choice, leading to a narrow range of possible values,  e.g., for  $a_{-}/l_{\rm vdw}$, and thus an apparent universality which, moreover, is only weakly broken by genuine three-body forces.

Another point which should be mentioned in this context is related to the question of whether van der  Waals universality,
which so far has been tested only in observables like $a_{-}$ or $a_{-}^{(1)}$ (i.e., low-energy properties of the three-body scattering amplitude), extends to larger wave numbers, in particular to the binding wave number $\kappa_{*}$ of the deepest Efimov trimer. An experimental test requires a measurement of binding energies near 
infinite scattering length, which is rather difficult due to the short lifetimes involved. Recent progress in this 
direction has been achieved in an experiment at JILA by ramping a unitary Bose gas of $^{85}$Rb to the weakly interacting regime and measuring the number of Efimov trimers from the density independent contribution to the lifetime \cite{Klauss2017}.  A different way of inferring the spectrum of three-body bound states in a moderately degenerate gas 
at unitarity has been discussed by Barth and Hofmann \cite{Barth2015}. Based on experimental results on the 
behavior of the momentum distribution $n(q)$ at large wave vectors $q$ by Makotyn \etal \cite{Makotyn2014}, they have shown that there are oscillations on top of the dominant dependence $q^4\, n(q)\to\mathcal{C}_2$ determined by the two-body contact $\mathcal{C}_2$ which are sensitive to the position of the lowest Efimov bound state at wave vector $\kappa_{*}$.  A test of the prediction that $\kappa_* \abar\simeq 0.25$ for this state is again determined by a quasi-universal value in terms of $\bar{a}$ would provide a much stronger test of van der Waals universality than has been possible so far. 
In particular, this would also shed light on the question of whether the lowest state can indeed be interpreted in terms of an Efimov picture. This is not the case in helium 4, where the wave function of the lowest three-body bound state differs substantially from what is predicted in Efimov's scenario; see \cite{Kunitski:2015,Blume2015}. For ultracold atoms, in turn, the Efimov description might hold even for the lowest state because --- in contrast to helium 4 ---  there is a wide separation between the short-range scale $\bar\sigma$ and the van der Waals length.

\section*{Acknowledgements}
We thank Marcus Barth for his insightful suggestions and inspiring discussions, and acknowledge James Babb, Doerte Blume, Cheng Chin, Matthew Frye, Bo Gao, Rudolf Grimm, Jeremy Hutson, Randy Hulet, Servaas Kokkelmans, Ronen Kroeze,  Sanjukta Roy, and Nikolaj Zinner for valuable input.


\begin{thebibliography}{90}%
\makeatletter
\providecommand \@ifxundefined [1]{%
 \@ifx{#1\undefined}
}%
\providecommand \@ifnum [1]{%
 \ifnum #1\expandafter \@firstoftwo
 \else \expandafter \@secondoftwo
 \fi
}%
\providecommand \@ifx [1]{%
 \ifx #1\expandafter \@firstoftwo
 \else \expandafter \@secondoftwo
 \fi
}%
\providecommand \natexlab [1]{#1}%
\providecommand \enquote  [1]{``#1''}%
\providecommand \bibnamefont  [1]{#1}%
\providecommand \bibfnamefont [1]{#1}%
\providecommand \citenamefont [1]{#1}%
\providecommand \href@noop [0]{\@secondoftwo}%
\providecommand \href [0]{\begingroup \@sanitize@url \@href}%
\providecommand \@href[1]{\@@startlink{#1}\@@href}%
\providecommand \@@href[1]{\endgroup#1\@@endlink}%
\providecommand \@sanitize@url [0]{\catcode `\\12\catcode `\$12\catcode
  `\&12\catcode `\#12\catcode `\^12\catcode `\_12\catcode `\%12\relax}%
\providecommand \@@startlink[1]{}%
\providecommand \@@endlink[0]{}%
\providecommand \url  [0]{\begingroup\@sanitize@url \@url }%
\providecommand \@url [1]{\endgroup\@href {#1}{\urlprefix }}%
\providecommand \urlprefix  [0]{URL }%
\providecommand \Eprint [0]{\href }%
\providecommand \doibase [0]{http://dx.doi.org/}%
\providecommand \selectlanguage [0]{\@gobble}%
\providecommand \bibinfo  [0]{\@secondoftwo}%
\providecommand \bibfield  [0]{\@secondoftwo}%
\providecommand \translation [1]{[#1]}%
\providecommand \BibitemOpen [0]{}%
\providecommand \bibitemStop [0]{}%
\providecommand \bibitemNoStop [0]{.\EOS\space}%
\providecommand \EOS [0]{\spacefactor3000\relax}%
\providecommand \BibitemShut  [1]{\csname bibitem#1\endcsname}%
\let\auto@bib@innerbib\@empty
\bibitem [{\citenamefont {Efimov}(1970)}]{Efimov70}%
  \BibitemOpen
  \bibfield  {author} {\bibinfo {author} {\bibfnamefont {V.}~\bibnamefont
  {Efimov}},\ }\href@noop {} {\bibfield  {journal} {\bibinfo  {journal} {Phys.
  Lett. B}\ }\textbf {\bibinfo {volume} {33}},\ \bibinfo {pages} {563}
  (\bibinfo {year} {1970})}\BibitemShut {NoStop}%
\bibitem [{\citenamefont {{Hoyle}}(1954)}]{Hoyle54}%
  \BibitemOpen
  \bibfield  {author} {\bibinfo {author} {\bibfnamefont {F.}~\bibnamefont
  {{Hoyle}}},\ }\href {\doibase 10.1086/190005} {\bibfield  {journal} {\bibinfo
   {journal} {Astrophysical Journal Supplement}\ }\textbf {\bibinfo {volume}
  {1}},\ \bibinfo {pages} {121} (\bibinfo {year} {1954})}\BibitemShut {NoStop}%
\bibitem [{\citenamefont {Kraemer}\ \emph {et~al.}(2006)\citenamefont
  {Kraemer}, \citenamefont {Mark}, \citenamefont {Waldburger}, \citenamefont
  {Danzl}, \citenamefont {Chin}, \citenamefont {Engeser}, \citenamefont
  {Lange}, \citenamefont {Pilch}, \citenamefont {Jaakkola}, \citenamefont
  {N\"{a}gerl},\ and\ \citenamefont {Grimm}}]{Grimm06}%
  \BibitemOpen
  \bibfield  {author} {\bibinfo {author} {\bibfnamefont {T.}~\bibnamefont
  {Kraemer}}, \bibinfo {author} {\bibfnamefont {M.}~\bibnamefont {Mark}},
  \bibinfo {author} {\bibfnamefont {P.}~\bibnamefont {Waldburger}}, \bibinfo
  {author} {\bibfnamefont {J.}~\bibnamefont {Danzl}}, \bibinfo {author}
  {\bibfnamefont {C.}~\bibnamefont {Chin}}, \bibinfo {author} {\bibfnamefont
  {B.}~\bibnamefont {Engeser}}, \bibinfo {author} {\bibfnamefont
  {A.}~\bibnamefont {Lange}}, \bibinfo {author} {\bibfnamefont
  {K.}~\bibnamefont {Pilch}}, \bibinfo {author} {\bibfnamefont
  {A.}~\bibnamefont {Jaakkola}}, \bibinfo {author} {\bibfnamefont {H.-C.}\
  \bibnamefont {N\"{a}gerl}}, \ and\ \bibinfo {author} {\bibfnamefont
  {R.}~\bibnamefont {Grimm}},\ }\href@noop {} {\bibfield  {journal} {\bibinfo
  {journal} {Nature}\ }\textbf {\bibinfo {volume} {440}},\ \bibinfo {pages}
  {315} (\bibinfo {year} {2006})}\BibitemShut {NoStop}%
\bibitem [{\citenamefont {Knoop}\ \emph {et~al.}(2009)\citenamefont {Knoop},
  \citenamefont {Ferlaino}, \citenamefont {Mark}, \citenamefont {Berninger},
  \citenamefont {Schobel}, \citenamefont {Nagerl},\ and\ \citenamefont
  {Grimm}}]{Knoop2009}%
  \BibitemOpen
  \bibfield  {author} {\bibinfo {author} {\bibfnamefont {S.}~\bibnamefont
  {Knoop}}, \bibinfo {author} {\bibfnamefont {F.}~\bibnamefont {Ferlaino}},
  \bibinfo {author} {\bibfnamefont {M.}~\bibnamefont {Mark}}, \bibinfo {author}
  {\bibfnamefont {M.}~\bibnamefont {Berninger}}, \bibinfo {author}
  {\bibfnamefont {H.}~\bibnamefont {Schobel}}, \bibinfo {author} {\bibfnamefont
  {H.~C.}\ \bibnamefont {Nagerl}}, \ and\ \bibinfo {author} {\bibfnamefont
  {R.}~\bibnamefont {Grimm}},\ }\href {http://dx.doi.org/10.1038/nphys1203}
  {\bibfield  {journal} {\bibinfo  {journal} {Nat Phys}\ }\textbf {\bibinfo
  {volume} {5}},\ \bibinfo {pages} {227} (\bibinfo {year} {2009})}\BibitemShut
  {NoStop}%
\bibitem [{\citenamefont {Zaccanti}\ \emph {et~al.}(2009)\citenamefont
  {Zaccanti}, \citenamefont {Deissler}, \citenamefont {Derrico}, \citenamefont
  {Fattori}, \citenamefont {Jona-Lasinio}, \citenamefont {M{\"u}ller},
  \citenamefont {Roati}, \citenamefont {Inguscio},\ and\ \citenamefont
  {Modugno}}]{Zaccanti2009}%
  \BibitemOpen
  \bibfield  {author} {\bibinfo {author} {\bibfnamefont {M.}~\bibnamefont
  {Zaccanti}}, \bibinfo {author} {\bibfnamefont {B.}~\bibnamefont {Deissler}},
  \bibinfo {author} {\bibfnamefont {C.}~\bibnamefont {Derrico}}, \bibinfo
  {author} {\bibfnamefont {M.}~\bibnamefont {Fattori}}, \bibinfo {author}
  {\bibfnamefont {M.}~\bibnamefont {Jona-Lasinio}}, \bibinfo {author}
  {\bibfnamefont {S.}~\bibnamefont {M{\"u}ller}}, \bibinfo {author}
  {\bibfnamefont {G.}~\bibnamefont {Roati}}, \bibinfo {author} {\bibfnamefont
  {M.}~\bibnamefont {Inguscio}}, \ and\ \bibinfo {author} {\bibfnamefont
  {G.}~\bibnamefont {Modugno}},\ }\href@noop {} {\bibfield  {journal} {\bibinfo
   {journal} {Nature Physics}\ }\textbf {\bibinfo {volume} {5}},\ \bibinfo
  {pages} {586} (\bibinfo {year} {2009})}\BibitemShut {NoStop}%
\bibitem [{\citenamefont {Gross}\ \emph {et~al.}(2009)\citenamefont {Gross},
  \citenamefont {Shotan}, \citenamefont {Kokkelmans},\ and\ \citenamefont
  {Khaykovich}}]{Gross2009}%
  \BibitemOpen
  \bibfield  {author} {\bibinfo {author} {\bibfnamefont {N.}~\bibnamefont
  {Gross}}, \bibinfo {author} {\bibfnamefont {Z.}~\bibnamefont {Shotan}},
  \bibinfo {author} {\bibfnamefont {S.}~\bibnamefont {Kokkelmans}}, \ and\
  \bibinfo {author} {\bibfnamefont {L.}~\bibnamefont {Khaykovich}},\ }\href
  {\doibase 10.1103/PhysRevLett.103.163202} {\bibfield  {journal} {\bibinfo
  {journal} {Phys. Rev. Lett.}\ }\textbf {\bibinfo {volume} {103}},\ \bibinfo
  {pages} {163202} (\bibinfo {year} {2009})}\BibitemShut {NoStop}%
\bibitem [{\citenamefont {Pollack}\ \emph {et~al.}(2009)\citenamefont
  {Pollack}, \citenamefont {Dries},\ and\ \citenamefont {Hulet}}]{Pollack2009}%
  \BibitemOpen
  \bibfield  {author} {\bibinfo {author} {\bibfnamefont {S.~E.}\ \bibnamefont
  {Pollack}}, \bibinfo {author} {\bibfnamefont {D.}~\bibnamefont {Dries}}, \
  and\ \bibinfo {author} {\bibfnamefont {R.~G.}\ \bibnamefont {Hulet}},\
  }\href@noop {} {\bibfield  {journal} {\bibinfo  {journal} {Science}\ }\textbf
  {\bibinfo {volume} {326}},\ \bibinfo {pages} {1683} (\bibinfo {year}
  {2009})}\BibitemShut {NoStop}%
\bibitem [{\citenamefont {Gross}\ \emph {et~al.}(2010)\citenamefont {Gross},
  \citenamefont {Shotan}, \citenamefont {Kokkelmans},\ and\ \citenamefont
  {Khaykovich}}]{Gross2010}%
  \BibitemOpen
  \bibfield  {author} {\bibinfo {author} {\bibfnamefont {N.}~\bibnamefont
  {Gross}}, \bibinfo {author} {\bibfnamefont {Z.}~\bibnamefont {Shotan}},
  \bibinfo {author} {\bibfnamefont {S.}~\bibnamefont {Kokkelmans}}, \ and\
  \bibinfo {author} {\bibfnamefont {L.}~\bibnamefont {Khaykovich}},\ }\href
  {\doibase 10.1103/PhysRevLett.105.103203} {\bibfield  {journal} {\bibinfo
  {journal} {Phys. Rev. Lett.}\ }\textbf {\bibinfo {volume} {105}},\ \bibinfo
  {pages} {103203} (\bibinfo {year} {2010})}\BibitemShut {NoStop}%
\bibitem [{\citenamefont {Ottenstein}\ \emph {et~al.}(2008)\citenamefont
  {Ottenstein}, \citenamefont {Lompe}, \citenamefont {Kohnen}, \citenamefont
  {Wenz},\ and\ \citenamefont {Jochim}}]{Ottenstein2008}%
  \BibitemOpen
  \bibfield  {author} {\bibinfo {author} {\bibfnamefont {T.~B.}\ \bibnamefont
  {Ottenstein}}, \bibinfo {author} {\bibfnamefont {T.}~\bibnamefont {Lompe}},
  \bibinfo {author} {\bibfnamefont {M.}~\bibnamefont {Kohnen}}, \bibinfo
  {author} {\bibfnamefont {A.~N.}\ \bibnamefont {Wenz}}, \ and\ \bibinfo
  {author} {\bibfnamefont {S.}~\bibnamefont {Jochim}},\ }\href {\doibase
  10.1103/PhysRevLett.101.203202} {\bibfield  {journal} {\bibinfo  {journal}
  {Phys. Rev. Lett.}\ }\textbf {\bibinfo {volume} {101}},\ \bibinfo {pages}
  {203202} (\bibinfo {year} {2008})}\BibitemShut {NoStop}%
\bibitem [{\citenamefont {Lompe}\ \emph {et~al.}(2010)\citenamefont {Lompe},
  \citenamefont {Ottenstein}, \citenamefont {Serwane}, \citenamefont {Wenz},
  \citenamefont {Z\"urn},\ and\ \citenamefont {Jochim}}]{Lompe2010}%
  \BibitemOpen
  \bibfield  {author} {\bibinfo {author} {\bibfnamefont {T.}~\bibnamefont
  {Lompe}}, \bibinfo {author} {\bibfnamefont {T.~B.}\ \bibnamefont
  {Ottenstein}}, \bibinfo {author} {\bibfnamefont {F.}~\bibnamefont {Serwane}},
  \bibinfo {author} {\bibfnamefont {A.~N.}\ \bibnamefont {Wenz}}, \bibinfo
  {author} {\bibfnamefont {G.}~\bibnamefont {Z\"urn}}, \ and\ \bibinfo {author}
  {\bibfnamefont {S.}~\bibnamefont {Jochim}},\ }\href {\doibase
  10.1126/science.1193148} {\bibfield  {journal} {\bibinfo  {journal}
  {Science}\ }\textbf {\bibinfo {volume} {330}},\ \bibinfo {pages} {940}
  (\bibinfo {year} {2010})}\BibitemShut {NoStop}%
\bibitem [{\citenamefont {Huckans}\ \emph {et~al.}(2009)\citenamefont
  {Huckans}, \citenamefont {Spielman}, \citenamefont {Tolra}, \citenamefont
  {Phillips},\ and\ \citenamefont {Porto}}]{Huckans2009}%
  \BibitemOpen
  \bibfield  {author} {\bibinfo {author} {\bibfnamefont {J.~H.}\ \bibnamefont
  {Huckans}}, \bibinfo {author} {\bibfnamefont {I.~B.}\ \bibnamefont
  {Spielman}}, \bibinfo {author} {\bibfnamefont {B.~L.}\ \bibnamefont {Tolra}},
  \bibinfo {author} {\bibfnamefont {W.~D.}\ \bibnamefont {Phillips}}, \ and\
  \bibinfo {author} {\bibfnamefont {J.~V.}\ \bibnamefont {Porto}},\ }\href
  {\doibase 10.1103/PhysRevA.80.043609} {\bibfield  {journal} {\bibinfo
  {journal} {Phys. Rev. A}\ }\textbf {\bibinfo {volume} {80}},\ \bibinfo
  {pages} {043609} (\bibinfo {year} {2009})}\BibitemShut {NoStop}%
\bibitem [{\citenamefont {Williams}\ \emph {et~al.}(2009)\citenamefont
  {Williams}, \citenamefont {Hazlett}, \citenamefont {Huckans}, \citenamefont
  {Stites}, \citenamefont {Zhang},\ and\ \citenamefont
  {O'Hara}}]{Williams2009}%
  \BibitemOpen
  \bibfield  {author} {\bibinfo {author} {\bibfnamefont {J.~R.}\ \bibnamefont
  {Williams}}, \bibinfo {author} {\bibfnamefont {E.~L.}\ \bibnamefont
  {Hazlett}}, \bibinfo {author} {\bibfnamefont {J.~H.}\ \bibnamefont
  {Huckans}}, \bibinfo {author} {\bibfnamefont {R.~W.}\ \bibnamefont {Stites}},
  \bibinfo {author} {\bibfnamefont {Y.}~\bibnamefont {Zhang}}, \ and\ \bibinfo
  {author} {\bibfnamefont {K.~M.}\ \bibnamefont {O'Hara}},\ }\href {\doibase
  10.1103/PhysRevLett.103.130404} {\bibfield  {journal} {\bibinfo  {journal}
  {Phys. Rev. Lett.}\ }\textbf {\bibinfo {volume} {103}},\ \bibinfo {pages}
  {130404} (\bibinfo {year} {2009})}\BibitemShut {NoStop}%
\bibitem [{\citenamefont {Nakajima}\ \emph {et~al.}(2011)\citenamefont
  {Nakajima}, \citenamefont {Horikoshi}, \citenamefont {Mukaiyama},
  \citenamefont {Naidon},\ and\ \citenamefont {Ueda}}]{Nakajima2011}%
  \BibitemOpen
  \bibfield  {author} {\bibinfo {author} {\bibfnamefont {S.}~\bibnamefont
  {Nakajima}}, \bibinfo {author} {\bibfnamefont {M.}~\bibnamefont {Horikoshi}},
  \bibinfo {author} {\bibfnamefont {T.}~\bibnamefont {Mukaiyama}}, \bibinfo
  {author} {\bibfnamefont {P.}~\bibnamefont {Naidon}}, \ and\ \bibinfo {author}
  {\bibfnamefont {M.}~\bibnamefont {Ueda}},\ }\href {\doibase
  10.1103/PhysRevLett.106.143201} {\bibfield  {journal} {\bibinfo  {journal}
  {Phys. Rev. Lett.}\ }\textbf {\bibinfo {volume} {106}},\ \bibinfo {pages}
  {143201} (\bibinfo {year} {2011})}\BibitemShut {NoStop}%
\bibitem [{\citenamefont {Barontini}\ \emph {et~al.}(2009)\citenamefont
  {Barontini}, \citenamefont {Weber}, \citenamefont {Rabatti}, \citenamefont
  {Catani}, \citenamefont {Thalhammer}, \citenamefont {Inguscio},\ and\
  \citenamefont {Minardi}}]{Barontini2009}%
  \BibitemOpen
  \bibfield  {author} {\bibinfo {author} {\bibfnamefont {G.}~\bibnamefont
  {Barontini}}, \bibinfo {author} {\bibfnamefont {C.}~\bibnamefont {Weber}},
  \bibinfo {author} {\bibfnamefont {F.}~\bibnamefont {Rabatti}}, \bibinfo
  {author} {\bibfnamefont {J.}~\bibnamefont {Catani}}, \bibinfo {author}
  {\bibfnamefont {G.}~\bibnamefont {Thalhammer}}, \bibinfo {author}
  {\bibfnamefont {M.}~\bibnamefont {Inguscio}}, \ and\ \bibinfo {author}
  {\bibfnamefont {F.}~\bibnamefont {Minardi}},\ }\href {\doibase
  10.1103/PhysRevLett.103.043201} {\bibfield  {journal} {\bibinfo  {journal}
  {Phys. Rev. Lett.}\ }\textbf {\bibinfo {volume} {103}},\ \bibinfo {pages}
  {043201} (\bibinfo {year} {2009})}\BibitemShut {NoStop}%
\bibitem [{\citenamefont {Pires}\ \emph
  {et~al.}(2014{\natexlab{a}})\citenamefont {Pires}, \citenamefont {Repp},
  \citenamefont {Ulmanis}, \citenamefont {Kuhnle}, \citenamefont
  {Weidem\"uller}, \citenamefont {Tiecke}, \citenamefont {Greene},
  \citenamefont {Ruzic}, \citenamefont {Bohn},\ and\ \citenamefont
  {Tiemann}}]{Pires2014}%
  \BibitemOpen
  \bibfield  {author} {\bibinfo {author} {\bibfnamefont {R.}~\bibnamefont
  {Pires}}, \bibinfo {author} {\bibfnamefont {M.}~\bibnamefont {Repp}},
  \bibinfo {author} {\bibfnamefont {J.}~\bibnamefont {Ulmanis}}, \bibinfo
  {author} {\bibfnamefont {E.~D.}\ \bibnamefont {Kuhnle}}, \bibinfo {author}
  {\bibfnamefont {M.}~\bibnamefont {Weidem\"uller}}, \bibinfo {author}
  {\bibfnamefont {T.~G.}\ \bibnamefont {Tiecke}}, \bibinfo {author}
  {\bibfnamefont {C.~H.}\ \bibnamefont {Greene}}, \bibinfo {author}
  {\bibfnamefont {B.~P.}\ \bibnamefont {Ruzic}}, \bibinfo {author}
  {\bibfnamefont {J.~L.}\ \bibnamefont {Bohn}}, \ and\ \bibinfo {author}
  {\bibfnamefont {E.}~\bibnamefont {Tiemann}},\ }\href {\doibase
  10.1103/PhysRevA.90.012710} {\bibfield  {journal} {\bibinfo  {journal} {Phys.
  Rev. A}\ }\textbf {\bibinfo {volume} {90}},\ \bibinfo {pages} {012710}
  (\bibinfo {year} {2014}{\natexlab{a}})}\BibitemShut {NoStop}%
\bibitem [{\citenamefont {Pires}\ \emph
  {et~al.}(2014{\natexlab{b}})\citenamefont {Pires}, \citenamefont {Ulmanis},
  \citenamefont {H\"afner}, \citenamefont {Repp}, \citenamefont {Arias},
  \citenamefont {Kuhnle},\ and\ \citenamefont {Weidem\"uller}}]{Pires2014b}%
  \BibitemOpen
  \bibfield  {author} {\bibinfo {author} {\bibfnamefont {R.}~\bibnamefont
  {Pires}}, \bibinfo {author} {\bibfnamefont {J.}~\bibnamefont {Ulmanis}},
  \bibinfo {author} {\bibfnamefont {S.}~\bibnamefont {H\"afner}}, \bibinfo
  {author} {\bibfnamefont {M.}~\bibnamefont {Repp}}, \bibinfo {author}
  {\bibfnamefont {A.}~\bibnamefont {Arias}}, \bibinfo {author} {\bibfnamefont
  {E.~D.}\ \bibnamefont {Kuhnle}}, \ and\ \bibinfo {author} {\bibfnamefont
  {M.}~\bibnamefont {Weidem\"uller}},\ }\href {\doibase
  10.1103/PhysRevLett.112.250404} {\bibfield  {journal} {\bibinfo  {journal}
  {Phys. Rev. Lett.}\ }\textbf {\bibinfo {volume} {112}},\ \bibinfo {pages}
  {250404} (\bibinfo {year} {2014}{\natexlab{b}})}\BibitemShut {NoStop}%
\bibitem [{\citenamefont {Tung}\ \emph {et~al.}(2014)\citenamefont {Tung},
  \citenamefont {Jim\'enez-Garc\'{\i}a}, \citenamefont {Johansen},
  \citenamefont {Parker},\ and\ \citenamefont {Chin}}]{Tung2014}%
  \BibitemOpen
  \bibfield  {author} {\bibinfo {author} {\bibfnamefont {S.-K.}\ \bibnamefont
  {Tung}}, \bibinfo {author} {\bibfnamefont {K.}~\bibnamefont
  {Jim\'enez-Garc\'{\i}a}}, \bibinfo {author} {\bibfnamefont {J.}~\bibnamefont
  {Johansen}}, \bibinfo {author} {\bibfnamefont {C.~V.}\ \bibnamefont
  {Parker}}, \ and\ \bibinfo {author} {\bibfnamefont {C.}~\bibnamefont
  {Chin}},\ }\href {\doibase 10.1103/PhysRevLett.113.240402} {\bibfield
  {journal} {\bibinfo  {journal} {Phys. Rev. Lett.}\ }\textbf {\bibinfo
  {volume} {113}},\ \bibinfo {pages} {240402} (\bibinfo {year}
  {2014})}\BibitemShut {NoStop}%
\bibitem [{\citenamefont {Ulmanis}\ \emph {et~al.}(2015)\citenamefont
  {Ulmanis}, \citenamefont {H{\"a}fner}, \citenamefont {Pires}, \citenamefont
  {Kuhnle}, \citenamefont {Weidem{\"u}ller},\ and\ \citenamefont
  {Tiemann}}]{Ulmanis2015}%
  \BibitemOpen
  \bibfield  {author} {\bibinfo {author} {\bibfnamefont {J.}~\bibnamefont
  {Ulmanis}}, \bibinfo {author} {\bibfnamefont {S.}~\bibnamefont {H{\"a}fner}},
  \bibinfo {author} {\bibfnamefont {R.}~\bibnamefont {Pires}}, \bibinfo
  {author} {\bibfnamefont {E.~D.}\ \bibnamefont {Kuhnle}}, \bibinfo {author}
  {\bibfnamefont {M.}~\bibnamefont {Weidem{\"u}ller}}, \ and\ \bibinfo {author}
  {\bibfnamefont {E.}~\bibnamefont {Tiemann}},\ }\href
  {http://stacks.iop.org/1367-2630/17/i=5/a=055009} {\bibfield  {journal}
  {\bibinfo  {journal} {New Journal of Physics}\ }\textbf {\bibinfo {volume}
  {17}},\ \bibinfo {pages} {055009} (\bibinfo {year} {2015})}\BibitemShut
  {NoStop}%
\bibitem [{\citenamefont {Johansen}\ \emph {et~al.}(2017)\citenamefont
  {Johansen}, \citenamefont {DeSalvo}, \citenamefont {Patel},\ and\
  \citenamefont {Chin}}]{Johansen2016}%
  \BibitemOpen
  \bibfield  {author} {\bibinfo {author} {\bibfnamefont {J.}~\bibnamefont
  {Johansen}}, \bibinfo {author} {\bibfnamefont {B.~J.}\ \bibnamefont
  {DeSalvo}}, \bibinfo {author} {\bibfnamefont {K.}~\bibnamefont {Patel}}, \
  and\ \bibinfo {author} {\bibfnamefont {C.}~\bibnamefont {Chin}},\ }\href
  {http://dx.doi.org/10.1038/nphys4130} {\bibfield  {journal} {\bibinfo
  {journal} {Nature Physics}\ }\textbf {\bibinfo {volume} {13}},\ \bibinfo
  {pages} {731} (\bibinfo {year} {2017})}\BibitemShut {NoStop}%
\bibitem [{\citenamefont {Ulmanis}\ \emph
  {et~al.}(2016{\natexlab{a}})\citenamefont {Ulmanis}, \citenamefont
  {H\"afner}, \citenamefont {Pires}, \citenamefont {Kuhnle}, \citenamefont
  {Wang}, \citenamefont {Greene},\ and\ \citenamefont
  {Weidem\"uller}}]{Ulmanis2016b}%
  \BibitemOpen
  \bibfield  {author} {\bibinfo {author} {\bibfnamefont {J.}~\bibnamefont
  {Ulmanis}}, \bibinfo {author} {\bibfnamefont {S.}~\bibnamefont {H\"afner}},
  \bibinfo {author} {\bibfnamefont {R.}~\bibnamefont {Pires}}, \bibinfo
  {author} {\bibfnamefont {E.~D.}\ \bibnamefont {Kuhnle}}, \bibinfo {author}
  {\bibfnamefont {Y.}~\bibnamefont {Wang}}, \bibinfo {author} {\bibfnamefont
  {C.~H.}\ \bibnamefont {Greene}}, \ and\ \bibinfo {author} {\bibfnamefont
  {M.}~\bibnamefont {Weidem\"uller}},\ }\href {\doibase
  10.1103/PhysRevLett.117.153201} {\bibfield  {journal} {\bibinfo  {journal}
  {Phys. Rev. Lett.}\ }\textbf {\bibinfo {volume} {117}},\ \bibinfo {pages}
  {153201} (\bibinfo {year} {2016}{\natexlab{a}})}\BibitemShut {NoStop}%
\bibitem [{\citenamefont {Ulmanis}\ \emph
  {et~al.}(2016{\natexlab{b}})\citenamefont {Ulmanis}, \citenamefont
  {H\"afner}, \citenamefont {Pires}, \citenamefont {Werner}, \citenamefont
  {Petrov}, \citenamefont {Kuhnle},\ and\ \citenamefont
  {Weidem\"uller}}]{Ulmanis2016c}%
  \BibitemOpen
  \bibfield  {author} {\bibinfo {author} {\bibfnamefont {J.}~\bibnamefont
  {Ulmanis}}, \bibinfo {author} {\bibfnamefont {S.}~\bibnamefont {H\"afner}},
  \bibinfo {author} {\bibfnamefont {R.}~\bibnamefont {Pires}}, \bibinfo
  {author} {\bibfnamefont {F.}~\bibnamefont {Werner}}, \bibinfo {author}
  {\bibfnamefont {D.~S.}\ \bibnamefont {Petrov}}, \bibinfo {author}
  {\bibfnamefont {E.~D.}\ \bibnamefont {Kuhnle}}, \ and\ \bibinfo {author}
  {\bibfnamefont {M.}~\bibnamefont {Weidem\"uller}},\ }\href {\doibase
  10.1103/PhysRevA.93.022707} {\bibfield  {journal} {\bibinfo  {journal} {Phys.
  Rev. A}\ }\textbf {\bibinfo {volume} {93}},\ \bibinfo {pages} {022707}
  (\bibinfo {year} {2016}{\natexlab{b}})}\BibitemShut {NoStop}%
\bibitem [{\citenamefont {Wacker}\ \emph {et~al.}(2016)\citenamefont {Wacker},
  \citenamefont {J\o{}rgensen}, \citenamefont {Birkmose}, \citenamefont
  {Winter}, \citenamefont {Mikkelsen}, \citenamefont {Sherson}, \citenamefont
  {Zinner},\ and\ \citenamefont {Arlt}}]{Wacker2016}%
  \BibitemOpen
  \bibfield  {author} {\bibinfo {author} {\bibfnamefont {L.~J.}\ \bibnamefont
  {Wacker}}, \bibinfo {author} {\bibfnamefont {N.~B.}\ \bibnamefont
  {J\o{}rgensen}}, \bibinfo {author} {\bibfnamefont {D.}~\bibnamefont
  {Birkmose}}, \bibinfo {author} {\bibfnamefont {N.}~\bibnamefont {Winter}},
  \bibinfo {author} {\bibfnamefont {M.}~\bibnamefont {Mikkelsen}}, \bibinfo
  {author} {\bibfnamefont {J.}~\bibnamefont {Sherson}}, \bibinfo {author}
  {\bibfnamefont {N.}~\bibnamefont {Zinner}}, \ and\ \bibinfo {author}
  {\bibfnamefont {J.~J.}\ \bibnamefont {Arlt}},\ }\href {\doibase
  10.1103/PhysRevLett.117.163201} {\bibfield  {journal} {\bibinfo  {journal}
  {Phys. Rev. Lett.}\ }\textbf {\bibinfo {volume} {117}},\ \bibinfo {pages}
  {163201} (\bibinfo {year} {2016})}\BibitemShut {NoStop}%
\bibitem [{\citenamefont {Kunitski}\ \emph {et~al.}(2015)\citenamefont
  {Kunitski}, \citenamefont {Zeller}, \citenamefont {Voigtsberger},
  \citenamefont {Kalinin}, \citenamefont {Schmidt}, \citenamefont
  {Sch{\"o}ffler}, \citenamefont {Czasch}, \citenamefont {Sch{\"o}llkopf},
  \citenamefont {Grisenti}, \citenamefont {Jahnke}, \citenamefont {Blume},\
  and\ \citenamefont {D{\"o}rner}}]{Kunitski:2015}%
  \BibitemOpen
  \bibfield  {author} {\bibinfo {author} {\bibfnamefont {M.}~\bibnamefont
  {Kunitski}}, \bibinfo {author} {\bibfnamefont {S.}~\bibnamefont {Zeller}},
  \bibinfo {author} {\bibfnamefont {J.}~\bibnamefont {Voigtsberger}}, \bibinfo
  {author} {\bibfnamefont {A.}~\bibnamefont {Kalinin}}, \bibinfo {author}
  {\bibfnamefont {L.~P.~H.}\ \bibnamefont {Schmidt}}, \bibinfo {author}
  {\bibfnamefont {M.}~\bibnamefont {Sch{\"o}ffler}}, \bibinfo {author}
  {\bibfnamefont {A.}~\bibnamefont {Czasch}}, \bibinfo {author} {\bibfnamefont
  {W.}~\bibnamefont {Sch{\"o}llkopf}}, \bibinfo {author} {\bibfnamefont
  {R.~E.}\ \bibnamefont {Grisenti}}, \bibinfo {author} {\bibfnamefont
  {T.}~\bibnamefont {Jahnke}}, \bibinfo {author} {\bibfnamefont
  {D.}~\bibnamefont {Blume}}, \ and\ \bibinfo {author} {\bibfnamefont
  {R.}~\bibnamefont {D{\"o}rner}},\ }\href {\doibase 10.1126/science.aaa5601}
  {\bibfield  {journal} {\bibinfo  {journal} {Science}\ }\textbf {\bibinfo
  {volume} {348}},\ \bibinfo {pages} {551} (\bibinfo {year}
  {2015})}\BibitemShut {NoStop}%
\bibitem [{\citenamefont {Bedaque}\ \emph {et~al.}(1999)\citenamefont
  {Bedaque}, \citenamefont {Hammer},\ and\ \citenamefont {van
  Kolck}}]{BHvK1999}%
  \BibitemOpen
  \bibfield  {author} {\bibinfo {author} {\bibfnamefont {P.~F.}\ \bibnamefont
  {Bedaque}}, \bibinfo {author} {\bibfnamefont {H.-W.}\ \bibnamefont {Hammer}},
  \ and\ \bibinfo {author} {\bibfnamefont {U.}~\bibnamefont {van Kolck}},\
  }\href {\doibase 10.1103/PhysRevLett.82.463} {\bibfield  {journal} {\bibinfo
  {journal} {Phys. Rev. Lett.}\ }\textbf {\bibinfo {volume} {82}},\ \bibinfo
  {pages} {463} (\bibinfo {year} {1999})}\BibitemShut {NoStop}%
\bibitem [{\citenamefont {Braaten}\ and\ \citenamefont
  {Hammer}(2006)}]{Braaten:2004rn}%
  \BibitemOpen
  \bibfield  {author} {\bibinfo {author} {\bibfnamefont {E.}~\bibnamefont
  {Braaten}}\ and\ \bibinfo {author} {\bibfnamefont {H.-W.}\ \bibnamefont
  {Hammer}},\ }\href@noop {} {\bibfield  {journal} {\bibinfo  {journal} {Phys.
  Rep.}\ }\textbf {\bibinfo {volume} {428}},\ \bibinfo {pages} {259} (\bibinfo
  {year} {2006})}\BibitemShut {NoStop}%
\bibitem [{\citenamefont {Moroz}\ \emph {et~al.}(2009)\citenamefont {Moroz},
  \citenamefont {Floerchinger}, \citenamefont {Schmidt},\ and\ \citenamefont
  {Wetterich}}]{Moroz2009}%
  \BibitemOpen
  \bibfield  {author} {\bibinfo {author} {\bibfnamefont {S.}~\bibnamefont
  {Moroz}}, \bibinfo {author} {\bibfnamefont {S.}~\bibnamefont {Floerchinger}},
  \bibinfo {author} {\bibfnamefont {R.}~\bibnamefont {Schmidt}}, \ and\
  \bibinfo {author} {\bibfnamefont {C.}~\bibnamefont {Wetterich}},\ }\href
  {\doibase 10.1103/PhysRevA.79.042705} {\bibfield  {journal} {\bibinfo
  {journal} {Phys. Rev. A}\ }\textbf {\bibinfo {volume} {79}},\ \bibinfo
  {pages} {042705} (\bibinfo {year} {2009})}\BibitemShut {NoStop}%
\bibitem [{\citenamefont {Floerchinger}\ \emph {et~al.}(2009)\citenamefont
  {Floerchinger}, \citenamefont {Schmidt}, \citenamefont {Moroz},\ and\
  \citenamefont {Wetterich}}]{Floerchinger2009}%
  \BibitemOpen
  \bibfield  {author} {\bibinfo {author} {\bibfnamefont {S.}~\bibnamefont
  {Floerchinger}}, \bibinfo {author} {\bibfnamefont {R.}~\bibnamefont
  {Schmidt}}, \bibinfo {author} {\bibfnamefont {S.}~\bibnamefont {Moroz}}, \
  and\ \bibinfo {author} {\bibfnamefont {C.}~\bibnamefont {Wetterich}},\ }\href
  {\doibase 10.1103/PhysRevA.79.013603} {\bibfield  {journal} {\bibinfo
  {journal} {Phys. Rev. A}\ }\textbf {\bibinfo {volume} {79}},\ \bibinfo
  {pages} {013603} (\bibinfo {year} {2009})}\BibitemShut {NoStop}%
\bibitem [{\citenamefont {Schmidt}\ and\ \citenamefont
  {Moroz}(2010)}]{Schmidt2010}%
  \BibitemOpen
  \bibfield  {author} {\bibinfo {author} {\bibfnamefont {R.}~\bibnamefont
  {Schmidt}}\ and\ \bibinfo {author} {\bibfnamefont {S.}~\bibnamefont
  {Moroz}},\ }\href {\doibase 10.1103/PhysRevA.81.052709} {\bibfield  {journal}
  {\bibinfo  {journal} {Phys. Rev. A}\ }\textbf {\bibinfo {volume} {81}},\
  \bibinfo {pages} {052709} (\bibinfo {year} {2010})}\BibitemShut {NoStop}%
\bibitem [{\citenamefont {Floerchinger}\ \emph {et~al.}(2011)\citenamefont
  {Floerchinger}, \citenamefont {Moroz},\ and\ \citenamefont
  {Schmidt}}]{Floerchinger2011}%
  \BibitemOpen
  \bibfield  {author} {\bibinfo {author} {\bibfnamefont {S.}~\bibnamefont
  {Floerchinger}}, \bibinfo {author} {\bibfnamefont {S.}~\bibnamefont {Moroz}},
  \ and\ \bibinfo {author} {\bibfnamefont {R.}~\bibnamefont {Schmidt}},\ }\href
  {\doibase 10.1007/s00601-011-0231-z} {\bibfield  {journal} {\bibinfo
  {journal} {Few-Body Systems}\ }\textbf {\bibinfo {volume} {51}},\ \bibinfo
  {pages} {153} (\bibinfo {year} {2011})}\BibitemShut {NoStop}%
\bibitem [{\citenamefont {Schmidt}(2013)}]{Schmidt}%
  \BibitemOpen
  \bibfield  {author} {\bibinfo {author} {\bibfnamefont {R.}~\bibnamefont
  {Schmidt}},\ }\emph {\bibinfo {title} {From few- to many-body physics with
  ultracold atoms}},\ \href@noop {} {Ph.D. thesis},\ \bibinfo  {school}
  {Technical University Munich} (\bibinfo {year} {2013})\BibitemShut {NoStop}%
\bibitem [{\citenamefont {Horinouchi}\ and\ \citenamefont
  {Ueda}(2015)}]{Horinouchi2015}%
  \BibitemOpen
  \bibfield  {author} {\bibinfo {author} {\bibfnamefont {Y.}~\bibnamefont
  {Horinouchi}}\ and\ \bibinfo {author} {\bibfnamefont {M.}~\bibnamefont
  {Ueda}},\ }\href {\doibase 10.1103/PhysRevLett.114.025301} {\bibfield
  {journal} {\bibinfo  {journal} {Phys. Rev. Lett.}\ }\textbf {\bibinfo
  {volume} {114}},\ \bibinfo {pages} {025301} (\bibinfo {year}
  {2015})}\BibitemShut {NoStop}%
\bibitem [{\citenamefont {Huang}\ \emph {et~al.}(2014)\citenamefont {Huang},
  \citenamefont {Sidorenkov}, \citenamefont {Grimm},\ and\ \citenamefont
  {Hutson}}]{Huang2014}%
  \BibitemOpen
  \bibfield  {author} {\bibinfo {author} {\bibfnamefont {B.}~\bibnamefont
  {Huang}}, \bibinfo {author} {\bibfnamefont {L.~A.}\ \bibnamefont
  {Sidorenkov}}, \bibinfo {author} {\bibfnamefont {R.}~\bibnamefont {Grimm}}, \
  and\ \bibinfo {author} {\bibfnamefont {J.~M.}\ \bibnamefont {Hutson}},\
  }\href {\doibase 10.1103/PhysRevLett.112.190401} {\bibfield  {journal}
  {\bibinfo  {journal} {Phys. Rev. Lett.}\ }\textbf {\bibinfo {volume} {112}},\
  \bibinfo {pages} {190401} (\bibinfo {year} {2014})}\BibitemShut {NoStop}%
\bibitem [{\citenamefont {Tung}\ \emph {et~al.}(2013)\citenamefont {Tung},
  \citenamefont {Parker}, \citenamefont {Johansen}, \citenamefont {Chin},
  \citenamefont {Wang},\ and\ \citenamefont {Julienne}}]{Tung2013}%
  \BibitemOpen
  \bibfield  {author} {\bibinfo {author} {\bibfnamefont {S.-K.}\ \bibnamefont
  {Tung}}, \bibinfo {author} {\bibfnamefont {C.}~\bibnamefont {Parker}},
  \bibinfo {author} {\bibfnamefont {J.}~\bibnamefont {Johansen}}, \bibinfo
  {author} {\bibfnamefont {C.}~\bibnamefont {Chin}}, \bibinfo {author}
  {\bibfnamefont {Y.}~\bibnamefont {Wang}}, \ and\ \bibinfo {author}
  {\bibfnamefont {P.~S.}\ \bibnamefont {Julienne}},\ }\href {\doibase
  10.1103/PhysRevA.87.010702} {\bibfield  {journal} {\bibinfo  {journal} {Phys.
  Rev. A}\ }\textbf {\bibinfo {volume} {87}},\ \bibinfo {pages} {010702}
  (\bibinfo {year} {2013})}\BibitemShut {NoStop}%
\bibitem [{\citenamefont {D'Incao}\ \emph {et~al.}(2009)\citenamefont
  {D'Incao}, \citenamefont {Greene},\ and\ \citenamefont {Esry}}]{DIncao2009}%
  \BibitemOpen
  \bibfield  {author} {\bibinfo {author} {\bibfnamefont {J.~P.}\ \bibnamefont
  {D'Incao}}, \bibinfo {author} {\bibfnamefont {C.~H.}\ \bibnamefont {Greene}},
  \ and\ \bibinfo {author} {\bibfnamefont {B.~D.}\ \bibnamefont {Esry}},\
  }\href {http://stacks.iop.org/0953-4075/42/i=4/a=044016} {\bibfield
  {journal} {\bibinfo  {journal} {Journal of Physics B: Atomic, Molecular and
  Optical Physics}\ }\textbf {\bibinfo {volume} {42}},\ \bibinfo {pages}
  {044016} (\bibinfo {year} {2009})}\BibitemShut {NoStop}%
\bibitem [{\citenamefont {Berninger}\ \emph {et~al.}(2011)\citenamefont
  {Berninger}, \citenamefont {Zenesini}, \citenamefont {Huang}, \citenamefont
  {Harm}, \citenamefont {N\"agerl}, \citenamefont {Ferlaino}, \citenamefont
  {Grimm}, \citenamefont {Julienne},\ and\ \citenamefont
  {Hutson}}]{Berninger2011}%
  \BibitemOpen
  \bibfield  {author} {\bibinfo {author} {\bibfnamefont {M.}~\bibnamefont
  {Berninger}}, \bibinfo {author} {\bibfnamefont {A.}~\bibnamefont {Zenesini}},
  \bibinfo {author} {\bibfnamefont {B.}~\bibnamefont {Huang}}, \bibinfo
  {author} {\bibfnamefont {W.}~\bibnamefont {Harm}}, \bibinfo {author}
  {\bibfnamefont {H.-C.}\ \bibnamefont {N\"agerl}}, \bibinfo {author}
  {\bibfnamefont {F.}~\bibnamefont {Ferlaino}}, \bibinfo {author}
  {\bibfnamefont {R.}~\bibnamefont {Grimm}}, \bibinfo {author} {\bibfnamefont
  {P.~S.}\ \bibnamefont {Julienne}}, \ and\ \bibinfo {author} {\bibfnamefont
  {J.~M.}\ \bibnamefont {Hutson}},\ }\href {\doibase
  10.1103/PhysRevLett.107.120401} {\bibfield  {journal} {\bibinfo  {journal}
  {Phys. Rev. Lett.}\ }\textbf {\bibinfo {volume} {107}},\ \bibinfo {pages}
  {120401} (\bibinfo {year} {2011})}\BibitemShut {NoStop}%
\bibitem [{\citenamefont {Wild}\ \emph {et~al.}(2012)\citenamefont {Wild},
  \citenamefont {Makotyn}, \citenamefont {Pino}, \citenamefont {Cornell},\ and\
  \citenamefont {Jin}}]{Wild2012}%
  \BibitemOpen
  \bibfield  {author} {\bibinfo {author} {\bibfnamefont {R.~J.}\ \bibnamefont
  {Wild}}, \bibinfo {author} {\bibfnamefont {P.}~\bibnamefont {Makotyn}},
  \bibinfo {author} {\bibfnamefont {J.~M.}\ \bibnamefont {Pino}}, \bibinfo
  {author} {\bibfnamefont {E.~A.}\ \bibnamefont {Cornell}}, \ and\ \bibinfo
  {author} {\bibfnamefont {D.~S.}\ \bibnamefont {Jin}},\ }\href {\doibase
  10.1103/PhysRevLett.108.145305} {\bibfield  {journal} {\bibinfo  {journal}
  {Phys. Rev. Lett.}\ }\textbf {\bibinfo {volume} {108}},\ \bibinfo {pages}
  {145305} (\bibinfo {year} {2012})}\BibitemShut {NoStop}%
\bibitem [{Note1()}]{Note1}%
  \BibitemOpen
  \bibinfo {note} {Throughout the paper we assume the particle masses to be
  identical. Our results can, however, be easily extended to the case of
  scattering of particles with different mass, where $m\to 2m_{\protect \rm
  red}$ with the reduced mass $m_{\protect \rm red}$.}\BibitemShut {Stop}%
\bibitem [{\citenamefont {Wang}\ \emph
  {et~al.}(2012{\natexlab{a}})\citenamefont {Wang}, \citenamefont {D'Incao},
  \citenamefont {Esry},\ and\ \citenamefont {Greene}}]{Greene2012}%
  \BibitemOpen
  \bibfield  {author} {\bibinfo {author} {\bibfnamefont {J.}~\bibnamefont
  {Wang}}, \bibinfo {author} {\bibfnamefont {J.~P.}\ \bibnamefont {D'Incao}},
  \bibinfo {author} {\bibfnamefont {B.~D.}\ \bibnamefont {Esry}}, \ and\
  \bibinfo {author} {\bibfnamefont {C.~H.}\ \bibnamefont {Greene}},\ }\href
  {\doibase 10.1103/PhysRevLett.108.263001} {\bibfield  {journal} {\bibinfo
  {journal} {Phys. Rev. Lett.}\ }\textbf {\bibinfo {volume} {108}},\ \bibinfo
  {pages} {263001} (\bibinfo {year} {2012}{\natexlab{a}})}\BibitemShut
  {NoStop}%
\bibitem [{\citenamefont {Schmidt}\ \emph {et~al.}(2012)\citenamefont
  {Schmidt}, \citenamefont {Rath},\ and\ \citenamefont {Zwerger}}]{SRZ}%
  \BibitemOpen
  \bibfield  {author} {\bibinfo {author} {\bibfnamefont {R.}~\bibnamefont
  {Schmidt}}, \bibinfo {author} {\bibfnamefont {S.}~\bibnamefont {Rath}}, \
  and\ \bibinfo {author} {\bibfnamefont {W.}~\bibnamefont {Zwerger}},\ }\href
  {http://dx.doi.org/10.1140/epjb/e2012-30841-3} {\bibfield  {journal}
  {\bibinfo  {journal} {The European Physical Journal B}\ }\textbf {\bibinfo
  {volume} {85}},\ \bibinfo {eid} {386} (\bibinfo {year} {2012})}\BibitemShut
  {NoStop}%
\bibitem [{\citenamefont {Chin}(2011)}]{chin2011}%
  \BibitemOpen
  \bibfield  {author} {\bibinfo {author} {\bibfnamefont {C.}~\bibnamefont
  {Chin}},\ }\href@noop {} {\bibfield  {journal} {\bibinfo  {journal}
  {arXiv:1111.1484}\ } (\bibinfo {year} {2011})}\BibitemShut {NoStop}%
\bibitem [{\citenamefont {S\o{}rensen}\ \emph {et~al.}(2012)\citenamefont
  {S\o{}rensen}, \citenamefont {Fedorov}, \citenamefont {Jensen},\ and\
  \citenamefont {Zinner}}]{Sorensen:2012}%
  \BibitemOpen
  \bibfield  {author} {\bibinfo {author} {\bibfnamefont {P.~K.}\ \bibnamefont
  {S\o{}rensen}}, \bibinfo {author} {\bibfnamefont {D.~V.}\ \bibnamefont
  {Fedorov}}, \bibinfo {author} {\bibfnamefont {A.~S.}\ \bibnamefont {Jensen}},
  \ and\ \bibinfo {author} {\bibfnamefont {N.~T.}\ \bibnamefont {Zinner}},\
  }\href {\doibase 10.1103/PhysRevA.86.052516} {\bibfield  {journal} {\bibinfo
  {journal} {Phys. Rev. A}\ }\textbf {\bibinfo {volume} {86}},\ \bibinfo
  {pages} {052516} (\bibinfo {year} {2012})}\BibitemShut {NoStop}%
\bibitem [{\citenamefont {Wang}\ \emph
  {et~al.}(2012{\natexlab{b}})\citenamefont {Wang}, \citenamefont {Wang},
  \citenamefont {D'Incao},\ and\ \citenamefont {Greene}}]{Wang2012}%
  \BibitemOpen
  \bibfield  {author} {\bibinfo {author} {\bibfnamefont {Y.}~\bibnamefont
  {Wang}}, \bibinfo {author} {\bibfnamefont {J.}~\bibnamefont {Wang}}, \bibinfo
  {author} {\bibfnamefont {J.~P.}\ \bibnamefont {D'Incao}}, \ and\ \bibinfo
  {author} {\bibfnamefont {C.~H.}\ \bibnamefont {Greene}},\ }\href {\doibase
  10.1103/PhysRevLett.109.243201} {\bibfield  {journal} {\bibinfo  {journal}
  {Phys. Rev. Lett.}\ }\textbf {\bibinfo {volume} {109}},\ \bibinfo {pages}
  {243201} (\bibinfo {year} {2012}{\natexlab{b}})}\BibitemShut {NoStop}%
\bibitem [{\citenamefont {D'Incao}\ \emph {et~al.}(2013)\citenamefont
  {D'Incao}, \citenamefont {Wang}, \citenamefont {Esry},\ and\ \citenamefont
  {Greene}}]{DIncao2013}%
  \BibitemOpen
  \bibfield  {author} {\bibinfo {author} {\bibfnamefont {J.~P.}\ \bibnamefont
  {D'Incao}}, \bibinfo {author} {\bibfnamefont {J.}~\bibnamefont {Wang}},
  \bibinfo {author} {\bibfnamefont {B.~D.}\ \bibnamefont {Esry}}, \ and\
  \bibinfo {author} {\bibfnamefont {C.~H.}\ \bibnamefont {Greene}},\ }\href
  {\doibase 10.1007/s00601-013-0616-2} {\bibfield  {journal} {\bibinfo
  {journal} {Few-Body Systems}\ }\textbf {\bibinfo {volume} {54}},\ \bibinfo
  {pages} {1523} (\bibinfo {year} {2013})}\BibitemShut {NoStop}%
\bibitem [{\citenamefont {Wang}\ and\ \citenamefont
  {Julienne}(2014)}]{Wang2014}%
  \BibitemOpen
  \bibfield  {author} {\bibinfo {author} {\bibfnamefont {Y.}~\bibnamefont
  {Wang}}\ and\ \bibinfo {author} {\bibfnamefont {P.~S.}\ \bibnamefont
  {Julienne}},\ }\href {http://dx.doi.org/10.1038/nphys3071} {\bibfield
  {journal} {\bibinfo  {journal} {Nat Phys}\ }\textbf {\bibinfo {volume}
  {10}},\ \bibinfo {pages} {768} (\bibinfo {year} {2014})}\BibitemShut
  {NoStop}%
\bibitem [{\citenamefont {Naidon}\ \emph {et~al.}(2014)\citenamefont {Naidon},
  \citenamefont {Endo},\ and\ \citenamefont {Ueda}}]{Naidon2014}%
  \BibitemOpen
  \bibfield  {author} {\bibinfo {author} {\bibfnamefont {P.}~\bibnamefont
  {Naidon}}, \bibinfo {author} {\bibfnamefont {S.}~\bibnamefont {Endo}}, \ and\
  \bibinfo {author} {\bibfnamefont {M.}~\bibnamefont {Ueda}},\ }\href {\doibase
  10.1103/PhysRevLett.112.105301} {\bibfield  {journal} {\bibinfo  {journal}
  {Phys. Rev. Lett.}\ }\textbf {\bibinfo {volume} {112}},\ \bibinfo {pages}
  {105301} (\bibinfo {year} {2014})}\BibitemShut {NoStop}%
\bibitem [{\citenamefont {Blume}(2015)}]{Blume2015}%
  \BibitemOpen
  \bibfield  {author} {\bibinfo {author} {\bibfnamefont {D.}~\bibnamefont
  {Blume}},\ }\href {\doibase 10.1007/s00601-015-0996-6} {\bibfield  {journal}
  {\bibinfo  {journal} {Few-Body Systems}\ }\textbf {\bibinfo {volume} {56}},\
  \bibinfo {pages} {859} (\bibinfo {year} {2015})}\BibitemShut {NoStop}%
\bibitem [{\citenamefont {Mestrom}\ \emph {et~al.}(2017)\citenamefont
  {Mestrom}, \citenamefont {Wang}, \citenamefont {Greene},\ and\ \citenamefont
  {D'Incao}}]{Mestrom2017}%
  \BibitemOpen
  \bibfield  {author} {\bibinfo {author} {\bibfnamefont {P.~M.~A.}\
  \bibnamefont {Mestrom}}, \bibinfo {author} {\bibfnamefont {J.}~\bibnamefont
  {Wang}}, \bibinfo {author} {\bibfnamefont {C.~H.}\ \bibnamefont {Greene}}, \
  and\ \bibinfo {author} {\bibfnamefont {J.~P.}\ \bibnamefont {D'Incao}},\
  }\href {\doibase 10.1103/PhysRevA.95.032707} {\bibfield  {journal} {\bibinfo
  {journal} {Phys. Rev. A}\ }\textbf {\bibinfo {volume} {95}},\ \bibinfo
  {pages} {032707} (\bibinfo {year} {2017})}\BibitemShut {NoStop}%
\bibitem [{Note2()}]{Note2}%
  \BibitemOpen
  \bibinfo {note} {It is instructive to note that the parameter $\lambda $ is
  simply related to the well known de Boer parameter $\Lambda ^{\protect \rm
  dB}=\hbar /(\protect \mathaccentV {bar}016\sigma \protect \sqrt {m\epsilon
  })$ which quantifies the importance of quantum effects in the equation of
  state~\cite {deboer1948} by $\Lambda ^{\protect \rm dB}=\lambda ^2/2$ (here,
  $\epsilon $ is the depth of the potential well). The maximum value of
  $\Lambda ^{\protect \rm dB}$ for which a single bound state exists is thus
  $\Lambda ^{\protect \rm dB}_c=(0.92)^2/2\simeq 0.42$, which is rather close
  to the de Boer parameter estimated for $^{4}$He. Remarkably, the fact that
  the de Boer parameter has a finite upper limit for all realistic two-body
  interaction potentials implies that the minimum of the shear viscosity for a
  purely classical fluid obeys the quantum inequality $\eta >\eta _{\protect
  \rm min}=\alpha _{\eta }\cdot \hbar n$ with a numerical constant $\alpha
  _{\eta }\sim 1/\Lambda ^{\protect \rm dB}$ which is bounded below by a
  constant of order $1$, see~\cite {Enss2011}.}\BibitemShut {Stop}%
\bibitem [{\citenamefont {Chin}\ \emph {et~al.}(2010)\citenamefont {Chin},
  \citenamefont {Grimm}, \citenamefont {Julienne},\ and\ \citenamefont
  {Tiesinga}}]{Chin2010}%
  \BibitemOpen
  \bibfield  {author} {\bibinfo {author} {\bibfnamefont {C.}~\bibnamefont
  {Chin}}, \bibinfo {author} {\bibfnamefont {R.}~\bibnamefont {Grimm}},
  \bibinfo {author} {\bibfnamefont {P.}~\bibnamefont {Julienne}}, \ and\
  \bibinfo {author} {\bibfnamefont {E.}~\bibnamefont {Tiesinga}},\ }\href
  {\doibase 10.1103/RevModPhys.82.1225} {\bibfield  {journal} {\bibinfo
  {journal} {Rev. Mod. Phys.}\ }\textbf {\bibinfo {volume} {82}},\ \bibinfo
  {pages} {1225} (\bibinfo {year} {2010})}\BibitemShut {NoStop}%
\bibitem [{\citenamefont {Petrov}(2004)}]{Petrov2004}%
  \BibitemOpen
  \bibfield  {author} {\bibinfo {author} {\bibfnamefont {D.~S.}\ \bibnamefont
  {Petrov}},\ }\href {\doibase 10.1103/PhysRevLett.93.143201} {\bibfield
  {journal} {\bibinfo  {journal} {Phys. Rev. Lett.}\ }\textbf {\bibinfo
  {volume} {93}},\ \bibinfo {pages} {143201} (\bibinfo {year}
  {2004})}\BibitemShut {NoStop}%
\bibitem [{\citenamefont {Gogolin}\ \emph {et~al.}(2008)\citenamefont
  {Gogolin}, \citenamefont {Mora},\ and\ \citenamefont {Egger}}]{Gogolin2008}%
  \BibitemOpen
  \bibfield  {author} {\bibinfo {author} {\bibfnamefont {A.~O.}\ \bibnamefont
  {Gogolin}}, \bibinfo {author} {\bibfnamefont {C.}~\bibnamefont {Mora}}, \
  and\ \bibinfo {author} {\bibfnamefont {R.}~\bibnamefont {Egger}},\ }\href
  {\doibase 10.1103/PhysRevLett.100.140404} {\bibfield  {journal} {\bibinfo
  {journal} {Phys. Rev. Lett.}\ }\textbf {\bibinfo {volume} {100}},\ \bibinfo
  {pages} {140404} (\bibinfo {year} {2008})}\BibitemShut {NoStop}%
\bibitem [{\citenamefont {Roy}\ \emph {et~al.}(2013)\citenamefont {Roy},
  \citenamefont {Landini}, \citenamefont {Trenkwalder}, \citenamefont
  {Semeghini}, \citenamefont {Spagnolli}, \citenamefont {Simoni}, \citenamefont
  {Fattori}, \citenamefont {Inguscio},\ and\ \citenamefont
  {Modugno}}]{Roy2013}%
  \BibitemOpen
  \bibfield  {author} {\bibinfo {author} {\bibfnamefont {S.}~\bibnamefont
  {Roy}}, \bibinfo {author} {\bibfnamefont {M.}~\bibnamefont {Landini}},
  \bibinfo {author} {\bibfnamefont {A.}~\bibnamefont {Trenkwalder}}, \bibinfo
  {author} {\bibfnamefont {G.}~\bibnamefont {Semeghini}}, \bibinfo {author}
  {\bibfnamefont {G.}~\bibnamefont {Spagnolli}}, \bibinfo {author}
  {\bibfnamefont {A.}~\bibnamefont {Simoni}}, \bibinfo {author} {\bibfnamefont
  {M.}~\bibnamefont {Fattori}}, \bibinfo {author} {\bibfnamefont
  {M.}~\bibnamefont {Inguscio}}, \ and\ \bibinfo {author} {\bibfnamefont
  {G.}~\bibnamefont {Modugno}},\ }\href {\doibase
  10.1103/PhysRevLett.111.053202} {\bibfield  {journal} {\bibinfo  {journal}
  {Phys. Rev. Lett.}\ }\textbf {\bibinfo {volume} {111}},\ \bibinfo {pages}
  {053202} (\bibinfo {year} {2013})}\BibitemShut {NoStop}%
\bibitem [{\citenamefont {Bloch}\ \emph {et~al.}(2008)\citenamefont {Bloch},
  \citenamefont {Dalibard},\ and\ \citenamefont {Zwerger}}]{Zwerger2008}%
  \BibitemOpen
  \bibfield  {author} {\bibinfo {author} {\bibfnamefont {I.}~\bibnamefont
  {Bloch}}, \bibinfo {author} {\bibfnamefont {J.}~\bibnamefont {Dalibard}}, \
  and\ \bibinfo {author} {\bibfnamefont {W.}~\bibnamefont {Zwerger}},\ }\href
  {\doibase 10.1103/RevModPhys.80.885} {\bibfield  {journal} {\bibinfo
  {journal} {Rev. Mod. Phys.}\ }\textbf {\bibinfo {volume} {80}},\ \bibinfo
  {pages} {885} (\bibinfo {year} {2008})}\BibitemShut {NoStop}%
\bibitem [{\citenamefont {{Zwerger}}(2016)}]{Zwerger2016}%
  \BibitemOpen
  \bibfield  {author} {\bibinfo {author} {\bibfnamefont {W.}~\bibnamefont
  {{Zwerger}}},\ }\href@noop {} {\bibfield  {journal} {\bibinfo  {journal}
  {arXiv:1608.00457}\ } (\bibinfo {year} {2016})},\ \Eprint
  {http://arxiv.org/abs/1608.00457} {1608.00457} \BibitemShut {NoStop}%
\bibitem [{\citenamefont {Cetina}\ \emph {et~al.}(2016)\citenamefont {Cetina},
  \citenamefont {Jag}, \citenamefont {Lous}, \citenamefont {Fritsche},
  \citenamefont {Walraven}, \citenamefont {Grimm}, \citenamefont {Levinsen},
  \citenamefont {Parish}, \citenamefont {Schmidt}, \citenamefont {Knap},\ and\
  \citenamefont {Demler}}]{Cetina2016}%
  \BibitemOpen
  \bibfield  {author} {\bibinfo {author} {\bibfnamefont {M.}~\bibnamefont
  {Cetina}}, \bibinfo {author} {\bibfnamefont {M.}~\bibnamefont {Jag}},
  \bibinfo {author} {\bibfnamefont {R.~S.}\ \bibnamefont {Lous}}, \bibinfo
  {author} {\bibfnamefont {I.}~\bibnamefont {Fritsche}}, \bibinfo {author}
  {\bibfnamefont {J.~T.~M.}\ \bibnamefont {Walraven}}, \bibinfo {author}
  {\bibfnamefont {R.}~\bibnamefont {Grimm}}, \bibinfo {author} {\bibfnamefont
  {J.}~\bibnamefont {Levinsen}}, \bibinfo {author} {\bibfnamefont {M.~M.}\
  \bibnamefont {Parish}}, \bibinfo {author} {\bibfnamefont {R.}~\bibnamefont
  {Schmidt}}, \bibinfo {author} {\bibfnamefont {M.}~\bibnamefont {Knap}}, \
  and\ \bibinfo {author} {\bibfnamefont {E.}~\bibnamefont {Demler}},\ }\href
  {\doibase 10.1126/science.aaf5134} {\bibfield  {journal} {\bibinfo  {journal}
  {Science}\ }\textbf {\bibinfo {volume} {354}},\ \bibinfo {pages} {96}
  (\bibinfo {year} {2016})}\BibitemShut {NoStop}%
\bibitem [{\citenamefont {Schmidt}\ \emph {et~al.}(2018)\citenamefont
  {Schmidt}, \citenamefont {Knap}, \citenamefont {Ivanov}, \citenamefont {You},
  \citenamefont {Cetina},\ and\ \citenamefont {Demler}}]{Schmidt2017}%
  \BibitemOpen
  \bibfield  {author} {\bibinfo {author} {\bibfnamefont {R.}~\bibnamefont
  {Schmidt}}, \bibinfo {author} {\bibfnamefont {M.}~\bibnamefont {Knap}},
  \bibinfo {author} {\bibfnamefont {D.~A.}\ \bibnamefont {Ivanov}}, \bibinfo
  {author} {\bibfnamefont {J.-S.}\ \bibnamefont {You}}, \bibinfo {author}
  {\bibfnamefont {M.}~\bibnamefont {Cetina}}, \ and\ \bibinfo {author}
  {\bibfnamefont {E.}~\bibnamefont {Demler}},\ }\href
  {http://stacks.iop.org/0034-4885/81/i=2/a=024401} {\bibfield  {journal}
  {\bibinfo  {journal} {Reports on Progress in Physics}\ }\textbf {\bibinfo
  {volume} {81}},\ \bibinfo {pages} {024401} (\bibinfo {year}
  {2018})}\BibitemShut {NoStop}%
\bibitem [{\citenamefont {G\"oral}\ \emph {et~al.}(2004)\citenamefont
  {G\"oral}, \citenamefont {K\"ohler}, \citenamefont {Gardiner}, \citenamefont
  {Tiesinga},\ and\ \citenamefont {Julienne}}]{Julienne2004}%
  \BibitemOpen
  \bibfield  {author} {\bibinfo {author} {\bibfnamefont {K.}~\bibnamefont
  {G\"oral}}, \bibinfo {author} {\bibfnamefont {T.}~\bibnamefont {K\"ohler}},
  \bibinfo {author} {\bibfnamefont {S.~A.}\ \bibnamefont {Gardiner}}, \bibinfo
  {author} {\bibfnamefont {E.}~\bibnamefont {Tiesinga}}, \ and\ \bibinfo
  {author} {\bibfnamefont {P.~S.}\ \bibnamefont {Julienne}},\ }\href
  {http://stacks.iop.org/0953-4075/37/i=17/a=006} {\bibfield  {journal}
  {\bibinfo  {journal} {Journal of Physics B: Atomic, Molecular and Optical
  Physics}\ }\textbf {\bibinfo {volume} {37}},\ \bibinfo {pages} {3457}
  (\bibinfo {year} {2004})}\BibitemShut {NoStop}%
\bibitem [{\citenamefont {Marcelis}\ \emph {et~al.}(2004)\citenamefont
  {Marcelis}, \citenamefont {van Kempen}, \citenamefont {Verhaar},\ and\
  \citenamefont {Kokkelmans}}]{Marcelis2004}%
  \BibitemOpen
  \bibfield  {author} {\bibinfo {author} {\bibfnamefont {B.}~\bibnamefont
  {Marcelis}}, \bibinfo {author} {\bibfnamefont {E.~G.~M.}\ \bibnamefont {van
  Kempen}}, \bibinfo {author} {\bibfnamefont {B.~J.}\ \bibnamefont {Verhaar}},
  \ and\ \bibinfo {author} {\bibfnamefont {S.~J. J. M.~F.}\ \bibnamefont
  {Kokkelmans}},\ }\href {\doibase 10.1103/PhysRevA.70.012701} {\bibfield
  {journal} {\bibinfo  {journal} {Phys. Rev. A}\ }\textbf {\bibinfo {volume}
  {70}},\ \bibinfo {pages} {012701} (\bibinfo {year} {2004})}\BibitemShut
  {NoStop}%
\bibitem [{\citenamefont {Flambaum}\ \emph {et~al.}(1999)\citenamefont
  {Flambaum}, \citenamefont {Gribakin},\ and\ \citenamefont
  {Harabati}}]{Flambaum1999}%
  \BibitemOpen
  \bibfield  {author} {\bibinfo {author} {\bibfnamefont {V.~V.}\ \bibnamefont
  {Flambaum}}, \bibinfo {author} {\bibfnamefont {G.~F.}\ \bibnamefont
  {Gribakin}}, \ and\ \bibinfo {author} {\bibfnamefont {C.}~\bibnamefont
  {Harabati}},\ }\href {\doibase 10.1103/PhysRevA.59.1998} {\bibfield
  {journal} {\bibinfo  {journal} {Phys. Rev. A}\ }\textbf {\bibinfo {volume}
  {59}},\ \bibinfo {pages} {1998} (\bibinfo {year} {1999})}\BibitemShut
  {NoStop}%
\bibitem [{Note3()}]{Note3}%
  \BibitemOpen
  \bibinfo {note} {By Galilean symmetry $\protect \mathcal G(E,\protect \mathbf
  p)=\protect \mathcal G(E-\protect \frac {\protect \mathbf
  p^2}{4m})$}\BibitemShut {NoStop}%
\bibitem [{\citenamefont {Pricoupenko}\ and\ \citenamefont
  {Jona-Lasinio}(2011)}]{Pricoupenko2011}%
  \BibitemOpen
  \bibfield  {author} {\bibinfo {author} {\bibfnamefont {L.}~\bibnamefont
  {Pricoupenko}}\ and\ \bibinfo {author} {\bibfnamefont {M.}~\bibnamefont
  {Jona-Lasinio}},\ }\href {\doibase 10.1103/PhysRevA.84.062712} {\bibfield
  {journal} {\bibinfo  {journal} {Phys. Rev. A}\ }\textbf {\bibinfo {volume}
  {84}},\ \bibinfo {pages} {062712} (\bibinfo {year} {2011})}\BibitemShut
  {NoStop}%
\bibitem [{\citenamefont {Press}\ \emph {et~al.}(2007)\citenamefont {Press},
  \citenamefont {Teukolsky}, \citenamefont {Vetterling},\ and\ \citenamefont
  {Flannery}}]{NR}%
  \BibitemOpen
  \bibfield  {author} {\bibinfo {author} {\bibfnamefont {W.~H.}\ \bibnamefont
  {Press}}, \bibinfo {author} {\bibfnamefont {S.~A.}\ \bibnamefont
  {Teukolsky}}, \bibinfo {author} {\bibfnamefont {W.~T.}\ \bibnamefont
  {Vetterling}}, \ and\ \bibinfo {author} {\bibfnamefont {B.~P.}\ \bibnamefont
  {Flannery}},\ }\href@noop {} {\emph {\bibinfo {title} {Numerical Recipes 3rd
  Edition: The Art of Scientific Computing}}},\ \bibinfo {edition} {3rd}\ ed.\
  (\bibinfo  {publisher} {Cambridge University Press},\ \bibinfo {address} {New
  York, NY, USA},\ \bibinfo {year} {2007})\BibitemShut {NoStop}%
\bibitem [{\citenamefont {Thomas}(1935)}]{Thomas35}%
  \BibitemOpen
  \bibfield  {author} {\bibinfo {author} {\bibfnamefont {L.~H.}\ \bibnamefont
  {Thomas}},\ }\href {\doibase 10.1103/PhysRev.47.903} {\bibfield  {journal}
  {\bibinfo  {journal} {Phys. Rev.}\ }\textbf {\bibinfo {volume} {47}},\
  \bibinfo {pages} {903} (\bibinfo {year} {1935})}\BibitemShut {NoStop}%
\bibitem [{Note4()}]{Note4}%
  \BibitemOpen
  \bibinfo {note} {For a recent study on the related effect of finite-range
  interactions on three-body recombination rates see \cite
  {Soerensen2013}}\BibitemShut {NoStop}%
\bibitem [{\citenamefont {Deltuva}(2012)}]{Deltuva2012}%
  \BibitemOpen
  \bibfield  {author} {\bibinfo {author} {\bibfnamefont {A.}~\bibnamefont
  {Deltuva}},\ }\href {\doibase 10.1103/PhysRevA.85.012708} {\bibfield
  {journal} {\bibinfo  {journal} {Phys. Rev. A}\ }\textbf {\bibinfo {volume}
  {85}},\ \bibinfo {pages} {012708} (\bibinfo {year} {2012})}\BibitemShut
  {NoStop}%
\bibitem [{\citenamefont {Gattobigio}\ and\ \citenamefont
  {Kievsky}(2014)}]{Gattobigio2014}%
  \BibitemOpen
  \bibfield  {author} {\bibinfo {author} {\bibfnamefont {M.}~\bibnamefont
  {Gattobigio}}\ and\ \bibinfo {author} {\bibfnamefont {A.}~\bibnamefont
  {Kievsky}},\ }\href {\doibase 10.1103/PhysRevA.90.012502} {\bibfield
  {journal} {\bibinfo  {journal} {Phys. Rev. A}\ }\textbf {\bibinfo {volume}
  {90}},\ \bibinfo {pages} {012502} (\bibinfo {year} {2014})}\BibitemShut
  {NoStop}%
\bibitem [{\citenamefont {Ji}\ \emph {et~al.}(2010)\citenamefont {Ji},
  \citenamefont {Phillips},\ and\ \citenamefont {Platter}}]{Ji2010}%
  \BibitemOpen
  \bibfield  {author} {\bibinfo {author} {\bibfnamefont {C.}~\bibnamefont
  {Ji}}, \bibinfo {author} {\bibfnamefont {D.~R.}\ \bibnamefont {Phillips}}, \
  and\ \bibinfo {author} {\bibfnamefont {L.}~\bibnamefont {Platter}},\ }\href
  {http://stacks.iop.org/0295-5075/92/i=1/a=13003} {\bibfield  {journal}
  {\bibinfo  {journal} {Eur. Phys. Lett.}\ }\textbf {\bibinfo {volume} {92}},\
  \bibinfo {pages} {13003} (\bibinfo {year} {2010})}\BibitemShut {NoStop}%
\bibitem [{\citenamefont {Ji}\ \emph {et~al.}(2012)\citenamefont {Ji},
  \citenamefont {Phillips},\ and\ \citenamefont {Platter}}]{Ji2011}%
  \BibitemOpen
  \bibfield  {author} {\bibinfo {author} {\bibfnamefont {C.}~\bibnamefont
  {Ji}}, \bibinfo {author} {\bibfnamefont {D.~R.}\ \bibnamefont {Phillips}}, \
  and\ \bibinfo {author} {\bibfnamefont {L.}~\bibnamefont {Platter}},\ }\href
  {\doibase http://dx.doi.org/10.1016/j.aop.2012.02.001} {\bibfield  {journal}
  {\bibinfo  {journal} {Annals of Physics}\ }\textbf {\bibinfo {volume}
  {327}},\ \bibinfo {pages} {1803 } (\bibinfo {year} {2012})},\ \bibinfo {note}
  {july 2012 Special Issue}\BibitemShut {NoStop}%
\bibitem [{\citenamefont {Ji}\ \emph {et~al.}(2015)\citenamefont {Ji},
  \citenamefont {Braaten}, \citenamefont {Phillips},\ and\ \citenamefont
  {Platter}}]{Ji2015}%
  \BibitemOpen
  \bibfield  {author} {\bibinfo {author} {\bibfnamefont {C.}~\bibnamefont
  {Ji}}, \bibinfo {author} {\bibfnamefont {E.}~\bibnamefont {Braaten}},
  \bibinfo {author} {\bibfnamefont {D.~R.}\ \bibnamefont {Phillips}}, \ and\
  \bibinfo {author} {\bibfnamefont {L.}~\bibnamefont {Platter}},\ }\href
  {\doibase 10.1103/PhysRevA.92.030702} {\bibfield  {journal} {\bibinfo
  {journal} {Phys. Rev. A}\ }\textbf {\bibinfo {volume} {92}},\ \bibinfo
  {pages} {030702} (\bibinfo {year} {2015})}\BibitemShut {NoStop}%
\bibitem [{\citenamefont {Blackley}\ \emph {et~al.}(2013)\citenamefont
  {Blackley}, \citenamefont {Le~Sueur}, \citenamefont {Hutson}, \citenamefont
  {McCarron}, \citenamefont {K\"oppinger}, \citenamefont {Cho}, \citenamefont
  {Jenkin},\ and\ \citenamefont {Cornish}}]{Blackley2013}%
  \BibitemOpen
  \bibfield  {author} {\bibinfo {author} {\bibfnamefont {C.~L.}\ \bibnamefont
  {Blackley}}, \bibinfo {author} {\bibfnamefont {C.~R.}\ \bibnamefont
  {Le~Sueur}}, \bibinfo {author} {\bibfnamefont {J.~M.}\ \bibnamefont
  {Hutson}}, \bibinfo {author} {\bibfnamefont {D.~J.}\ \bibnamefont
  {McCarron}}, \bibinfo {author} {\bibfnamefont {M.~P.}\ \bibnamefont
  {K\"oppinger}}, \bibinfo {author} {\bibfnamefont {H.-W.}\ \bibnamefont
  {Cho}}, \bibinfo {author} {\bibfnamefont {D.~L.}\ \bibnamefont {Jenkin}}, \
  and\ \bibinfo {author} {\bibfnamefont {S.~L.}\ \bibnamefont {Cornish}},\
  }\href {\doibase 10.1103/PhysRevA.87.033611} {\bibfield  {journal} {\bibinfo
  {journal} {Phys. Rev. A}\ }\textbf {\bibinfo {volume} {87}},\ \bibinfo
  {pages} {033611} (\bibinfo {year} {2013})}\BibitemShut {NoStop}%
\bibitem [{\citenamefont {Naidon}\ and\ \citenamefont
  {Endo}(2017)}]{Naidon2017}%
  \BibitemOpen
  \bibfield  {author} {\bibinfo {author} {\bibfnamefont {P.}~\bibnamefont
  {Naidon}}\ and\ \bibinfo {author} {\bibfnamefont {S.}~\bibnamefont {Endo}},\
  }\href {http://stacks.iop.org/0034-4885/80/i=5/a=056001} {\bibfield
  {journal} {\bibinfo  {journal} {Reports on Progress in Physics}\ }\textbf
  {\bibinfo {volume} {80}},\ \bibinfo {pages} {056001} (\bibinfo {year}
  {2017})}\BibitemShut {NoStop}%
\bibitem [{\citenamefont {Pethick}\ and\ \citenamefont
  {Smith}(2008)}]{Pethick2002}%
  \BibitemOpen
  \bibfield  {author} {\bibinfo {author} {\bibfnamefont {C.~J.}\ \bibnamefont
  {Pethick}}\ and\ \bibinfo {author} {\bibfnamefont {H.}~\bibnamefont
  {Smith}},\ }\href {\doibase 10.1017/CBO9780511802850} {\emph {\bibinfo
  {title} {Bose--Einstein Condensation in Dilute Gases}}},\ \bibinfo {edition}
  {2nd}\ ed.\ (\bibinfo  {publisher} {Cambridge University Press, Cambridge},\
  \bibinfo {year} {2008})\BibitemShut {NoStop}%
\bibitem [{\citenamefont {Lange}\ \emph {et~al.}(2009)\citenamefont {Lange},
  \citenamefont {Pilch}, \citenamefont {Prantner}, \citenamefont {Ferlaino},
  \citenamefont {Engeser}, \citenamefont {N\"agerl}, \citenamefont {Grimm},\
  and\ \citenamefont {Chin}}]{Lange2009}%
  \BibitemOpen
  \bibfield  {author} {\bibinfo {author} {\bibfnamefont {A.~D.}\ \bibnamefont
  {Lange}}, \bibinfo {author} {\bibfnamefont {K.}~\bibnamefont {Pilch}},
  \bibinfo {author} {\bibfnamefont {A.}~\bibnamefont {Prantner}}, \bibinfo
  {author} {\bibfnamefont {F.}~\bibnamefont {Ferlaino}}, \bibinfo {author}
  {\bibfnamefont {B.}~\bibnamefont {Engeser}}, \bibinfo {author} {\bibfnamefont
  {H.-C.}\ \bibnamefont {N\"agerl}}, \bibinfo {author} {\bibfnamefont
  {R.}~\bibnamefont {Grimm}}, \ and\ \bibinfo {author} {\bibfnamefont
  {C.}~\bibnamefont {Chin}},\ }\href {\doibase 10.1103/PhysRevA.79.013622}
  {\bibfield  {journal} {\bibinfo  {journal} {Phys. Rev. A}\ }\textbf {\bibinfo
  {volume} {79}},\ \bibinfo {pages} {013622} (\bibinfo {year}
  {2009})}\BibitemShut {NoStop}%
\bibitem [{\citenamefont {Berninger}\ \emph {et~al.}(2013)\citenamefont
  {Berninger}, \citenamefont {Zenesini}, \citenamefont {Huang}, \citenamefont
  {Harm}, \citenamefont {N\"agerl}, \citenamefont {Ferlaino}, \citenamefont
  {Grimm}, \citenamefont {Julienne},\ and\ \citenamefont
  {Hutson}}]{Berninger2013}%
  \BibitemOpen
  \bibfield  {author} {\bibinfo {author} {\bibfnamefont {M.}~\bibnamefont
  {Berninger}}, \bibinfo {author} {\bibfnamefont {A.}~\bibnamefont {Zenesini}},
  \bibinfo {author} {\bibfnamefont {B.}~\bibnamefont {Huang}}, \bibinfo
  {author} {\bibfnamefont {W.}~\bibnamefont {Harm}}, \bibinfo {author}
  {\bibfnamefont {H.-C.}\ \bibnamefont {N\"agerl}}, \bibinfo {author}
  {\bibfnamefont {F.}~\bibnamefont {Ferlaino}}, \bibinfo {author}
  {\bibfnamefont {R.}~\bibnamefont {Grimm}}, \bibinfo {author} {\bibfnamefont
  {P.~S.}\ \bibnamefont {Julienne}}, \ and\ \bibinfo {author} {\bibfnamefont
  {J.~M.}\ \bibnamefont {Hutson}},\ }\href {\doibase
  10.1103/PhysRevA.87.032517} {\bibfield  {journal} {\bibinfo  {journal} {Phys.
  Rev. A}\ }\textbf {\bibinfo {volume} {87}},\ \bibinfo {pages} {032517}
  (\bibinfo {year} {2013})}\BibitemShut {NoStop}%
\bibitem [{\citenamefont {D'Errico}\ \emph {et~al.}(2007)\citenamefont
  {D'Errico}, \citenamefont {Zaccanti}, \citenamefont {Fattori}, \citenamefont
  {Roati}, \citenamefont {Inguscio}, \citenamefont {Modugno},\ and\
  \citenamefont {Simoni}}]{DErrico2007}%
  \BibitemOpen
  \bibfield  {author} {\bibinfo {author} {\bibfnamefont {C.}~\bibnamefont
  {D'Errico}}, \bibinfo {author} {\bibfnamefont {M.}~\bibnamefont {Zaccanti}},
  \bibinfo {author} {\bibfnamefont {M.}~\bibnamefont {Fattori}}, \bibinfo
  {author} {\bibfnamefont {G.}~\bibnamefont {Roati}}, \bibinfo {author}
  {\bibfnamefont {M.}~\bibnamefont {Inguscio}}, \bibinfo {author}
  {\bibfnamefont {G.}~\bibnamefont {Modugno}}, \ and\ \bibinfo {author}
  {\bibfnamefont {A.}~\bibnamefont {Simoni}},\ }\href
  {http://stacks.iop.org/1367-2630/9/i=7/a=223} {\bibfield  {journal} {\bibinfo
   {journal} {New Journal of Physics}\ }\textbf {\bibinfo {volume} {9}},\
  \bibinfo {pages} {223} (\bibinfo {year} {2007})}\BibitemShut {NoStop}%
\bibitem [{\citenamefont {Dyke}\ \emph {et~al.}(2013)\citenamefont {Dyke},
  \citenamefont {Pollack},\ and\ \citenamefont {Hulet}}]{Dyke:2013}%
  \BibitemOpen
  \bibfield  {author} {\bibinfo {author} {\bibfnamefont {P.}~\bibnamefont
  {Dyke}}, \bibinfo {author} {\bibfnamefont {S.~E.}\ \bibnamefont {Pollack}}, \
  and\ \bibinfo {author} {\bibfnamefont {R.~G.}\ \bibnamefont {Hulet}},\ }\href
  {\doibase 10.1103/PhysRevA.88.023625} {\bibfield  {journal} {\bibinfo
  {journal} {Phys. Rev. A}\ }\textbf {\bibinfo {volume} {88}},\ \bibinfo
  {pages} {023625} (\bibinfo {year} {2013})}\BibitemShut {NoStop}%
\bibitem [{\citenamefont {Claussen}\ \emph {et~al.}(2003)\citenamefont
  {Claussen}, \citenamefont {Kokkelmans}, \citenamefont {Thompson},
  \citenamefont {Donley}, \citenamefont {Hodby},\ and\ \citenamefont
  {Wieman}}]{Claussen03}%
  \BibitemOpen
  \bibfield  {author} {\bibinfo {author} {\bibfnamefont {N.~R.}\ \bibnamefont
  {Claussen}}, \bibinfo {author} {\bibfnamefont {S.~J. J. M.~F.}\ \bibnamefont
  {Kokkelmans}}, \bibinfo {author} {\bibfnamefont {S.~T.}\ \bibnamefont
  {Thompson}}, \bibinfo {author} {\bibfnamefont {E.~A.}\ \bibnamefont
  {Donley}}, \bibinfo {author} {\bibfnamefont {E.}~\bibnamefont {Hodby}}, \
  and\ \bibinfo {author} {\bibfnamefont {C.~E.}\ \bibnamefont {Wieman}},\
  }\href {\doibase 10.1103/PhysRevA.67.060701} {\bibfield  {journal} {\bibinfo
  {journal} {Phys. Rev. A}\ }\textbf {\bibinfo {volume} {67}},\ \bibinfo
  {pages} {060701} (\bibinfo {year} {2003})}\BibitemShut {NoStop}%
\bibitem [{\citenamefont {Massignan}\ and\ \citenamefont
  {Stoof}(2008)}]{Massignan2008}%
  \BibitemOpen
  \bibfield  {author} {\bibinfo {author} {\bibfnamefont {P.}~\bibnamefont
  {Massignan}}\ and\ \bibinfo {author} {\bibfnamefont {H.~T.~C.}\ \bibnamefont
  {Stoof}},\ }\href {\doibase 10.1103/PhysRevA.78.030701} {\bibfield  {journal}
  {\bibinfo  {journal} {Phys. Rev. A}\ }\textbf {\bibinfo {volume} {78}},\
  \bibinfo {pages} {030701} (\bibinfo {year} {2008})}\BibitemShut {NoStop}%
\bibitem [{\citenamefont {Jona-Lasinio}\ and\ \citenamefont
  {Pricoupenko}(2010)}]{Pricoupenko2010}%
  \BibitemOpen
  \bibfield  {author} {\bibinfo {author} {\bibfnamefont {M.}~\bibnamefont
  {Jona-Lasinio}}\ and\ \bibinfo {author} {\bibfnamefont {L.}~\bibnamefont
  {Pricoupenko}},\ }\href {\doibase 10.1103/PhysRevLett.104.023201} {\bibfield
  {journal} {\bibinfo  {journal} {Phys. Rev. Lett.}\ }\textbf {\bibinfo
  {volume} {104}},\ \bibinfo {pages} {023201} (\bibinfo {year}
  {2010})}\BibitemShut {NoStop}%
\bibitem [{\citenamefont {Axilrod}\ and\ \citenamefont
  {Teller}(1943)}]{Axilrod1943}%
  \BibitemOpen
  \bibfield  {author} {\bibinfo {author} {\bibfnamefont {B.~M.}\ \bibnamefont
  {Axilrod}}\ and\ \bibinfo {author} {\bibfnamefont {E.}~\bibnamefont
  {Teller}},\ }\href@noop {} {\bibfield  {journal} {\bibinfo  {journal} {J.
  Chem. Phys.}\ ,\ \bibinfo {pages} {299}} (\bibinfo {year}
  {1943})}\BibitemShut {NoStop}%
\bibitem [{\citenamefont {Tang}\ \emph {et~al.}(2009)\citenamefont {Tang},
  \citenamefont {Yan}, \citenamefont {Shi},\ and\ \citenamefont
  {Babb}}]{Tang2009}%
  \BibitemOpen
  \bibfield  {author} {\bibinfo {author} {\bibfnamefont {L.-Y.}\ \bibnamefont
  {Tang}}, \bibinfo {author} {\bibfnamefont {Z.-C.}\ \bibnamefont {Yan}},
  \bibinfo {author} {\bibfnamefont {T.-Y.}\ \bibnamefont {Shi}}, \ and\
  \bibinfo {author} {\bibfnamefont {J.~F.}\ \bibnamefont {Babb}},\ }\href
  {\doibase 10.1103/PhysRevA.79.062712} {\bibfield  {journal} {\bibinfo
  {journal} {Phys. Rev. A}\ }\textbf {\bibinfo {volume} {79}},\ \bibinfo
  {pages} {062712} (\bibinfo {year} {2009})}\BibitemShut {NoStop}%
\bibitem [{\citenamefont {Babb}(2010)}]{Babb2010}%
  \BibitemOpen
  \bibfield  {author} {\bibinfo {author} {\bibfnamefont {J.~F.}\ \bibnamefont
  {Babb}},\ }in\ \href {\doibase 10.1016/S1049-250X(10)59001-3} {\emph
  {\bibinfo {booktitle} {Advances in Atomic, Molecular, and Optical
  Physics}}},\ \bibinfo {series} {Advances In Atomic, Molecular, and Optical
  Physics}, Vol.~\bibinfo {volume} {59},\ \bibinfo {editor} {edited by\
  \bibinfo {editor} {\bibfnamefont {P.~B.}\ \bibnamefont {E.~Arimondo}}\ and\
  \bibinfo {editor} {\bibfnamefont {C.}~\bibnamefont {Lin}}}\ (\bibinfo
  {publisher} {Academic Press, Cambridge, US},\ \bibinfo {year} {2010})\ pp.\
  \bibinfo {pages} {1 -- 20}\BibitemShut {NoStop}%
\bibitem [{\citenamefont {Tang}\ \emph {et~al.}(2012)\citenamefont {Tang},
  \citenamefont {Yan}, \citenamefont {Shi}, \citenamefont {Babb},\ and\
  \citenamefont {Mitroy}}]{Tang2012}%
  \BibitemOpen
  \bibfield  {author} {\bibinfo {author} {\bibfnamefont {L.-Y.}\ \bibnamefont
  {Tang}}, \bibinfo {author} {\bibfnamefont {Z.-C.}\ \bibnamefont {Yan}},
  \bibinfo {author} {\bibfnamefont {T.-Y.}\ \bibnamefont {Shi}}, \bibinfo
  {author} {\bibfnamefont {J.~F.}\ \bibnamefont {Babb}}, \ and\ \bibinfo
  {author} {\bibfnamefont {J.}~\bibnamefont {Mitroy}},\ }\href {\doibase
  10.1063/1.3691891} {\bibfield  {journal} {\bibinfo  {journal} {The Journal of
  Chemical Physics}\ }\textbf {\bibinfo {volume} {136}},\ \bibinfo {eid}
  {104104} (\bibinfo {year} {2012})}\BibitemShut {NoStop}%
\bibitem [{\citenamefont {Th\o{}gersen}\ \emph {et~al.}(2009)\citenamefont
  {Th\o{}gersen}, \citenamefont {Fedorov}, \citenamefont {Jensen},
  \citenamefont {Esry},\ and\ \citenamefont {Wang}}]{Thogersen2009}%
  \BibitemOpen
  \bibfield  {author} {\bibinfo {author} {\bibfnamefont {M.}~\bibnamefont
  {Th\o{}gersen}}, \bibinfo {author} {\bibfnamefont {D.~V.}\ \bibnamefont
  {Fedorov}}, \bibinfo {author} {\bibfnamefont {A.~S.}\ \bibnamefont {Jensen}},
  \bibinfo {author} {\bibfnamefont {B.~D.}\ \bibnamefont {Esry}}, \ and\
  \bibinfo {author} {\bibfnamefont {Y.}~\bibnamefont {Wang}},\ }\href {\doibase
  10.1103/PhysRevA.80.013608} {\bibfield  {journal} {\bibinfo  {journal} {Phys.
  Rev. A}\ }\textbf {\bibinfo {volume} {80}},\ \bibinfo {pages} {013608}
  (\bibinfo {year} {2009})}\BibitemShut {NoStop}%
\bibitem [{\citenamefont {Klauss}\ \emph {et~al.}(2017)\citenamefont {Klauss},
  \citenamefont {Xie}, \citenamefont {Lopez-Abadia}, \citenamefont {D'Incao},
  \citenamefont {Hadzibabic}, \citenamefont {Jin},\ and\ \citenamefont
  {Cornell}}]{Klauss2017}%
  \BibitemOpen
  \bibfield  {author} {\bibinfo {author} {\bibfnamefont {C.~E.}\ \bibnamefont
  {Klauss}}, \bibinfo {author} {\bibfnamefont {X.}~\bibnamefont {Xie}},
  \bibinfo {author} {\bibfnamefont {C.}~\bibnamefont {Lopez-Abadia}}, \bibinfo
  {author} {\bibfnamefont {J.~P.}\ \bibnamefont {D'Incao}}, \bibinfo {author}
  {\bibfnamefont {Z.}~\bibnamefont {Hadzibabic}}, \bibinfo {author}
  {\bibfnamefont {D.~S.}\ \bibnamefont {Jin}}, \ and\ \bibinfo {author}
  {\bibfnamefont {E.~A.}\ \bibnamefont {Cornell}},\ }\href {\doibase
  10.1103/PhysRevLett.119.143401} {\bibfield  {journal} {\bibinfo  {journal}
  {Phys. Rev. Lett.}\ }\textbf {\bibinfo {volume} {119}},\ \bibinfo {pages}
  {143401} (\bibinfo {year} {2017})}\BibitemShut {NoStop}%
\bibitem [{\citenamefont {Barth}\ and\ \citenamefont
  {Hofmann}(2015)}]{Barth2015}%
  \BibitemOpen
  \bibfield  {author} {\bibinfo {author} {\bibfnamefont {M.}~\bibnamefont
  {Barth}}\ and\ \bibinfo {author} {\bibfnamefont {J.}~\bibnamefont
  {Hofmann}},\ }\href {\doibase 10.1103/PhysRevA.92.062716} {\bibfield
  {journal} {\bibinfo  {journal} {Phys. Rev. A}\ }\textbf {\bibinfo {volume}
  {92}},\ \bibinfo {pages} {062716} (\bibinfo {year} {2015})}\BibitemShut
  {NoStop}%
\bibitem [{\citenamefont {Makotyn}\ \emph {et~al.}(2014)\citenamefont
  {Makotyn}, \citenamefont {Klauss}, \citenamefont {Goldberger}, \citenamefont
  {Cornell},\ and\ \citenamefont {Jin}}]{Makotyn2014}%
  \BibitemOpen
  \bibfield  {author} {\bibinfo {author} {\bibfnamefont {P.}~\bibnamefont
  {Makotyn}}, \bibinfo {author} {\bibfnamefont {C.~E.}\ \bibnamefont {Klauss}},
  \bibinfo {author} {\bibfnamefont {D.~L.}\ \bibnamefont {Goldberger}},
  \bibinfo {author} {\bibfnamefont {E.~A.}\ \bibnamefont {Cornell}}, \ and\
  \bibinfo {author} {\bibfnamefont {D.~S.}\ \bibnamefont {Jin}},\ }\href
  {http://dx.doi.org/10.1038/nphys2850} {\bibfield  {journal} {\bibinfo
  {journal} {Nat Phys}\ }\textbf {\bibinfo {volume} {10}},\ \bibinfo {pages}
  {116} (\bibinfo {year} {2014})}\BibitemShut {NoStop}%
\bibitem [{\citenamefont {Boer}(1948)}]{deboer1948}%
  \BibitemOpen
  \bibfield  {author} {\bibinfo {author} {\bibfnamefont {J.~D.}\ \bibnamefont
  {Boer}},\ }\href {\doibase http://dx.doi.org/10.1016/0031-8914(48)90032-9}
  {\bibfield  {journal} {\bibinfo  {journal} {Physica}\ }\textbf {\bibinfo
  {volume} {14}},\ \bibinfo {pages} {139 } (\bibinfo {year}
  {1948})}\BibitemShut {NoStop}%
\bibitem [{\citenamefont {Enss}\ \emph {et~al.}(2011)\citenamefont {Enss},
  \citenamefont {Haussmann},\ and\ \citenamefont {Zwerger}}]{Enss2011}%
  \BibitemOpen
  \bibfield  {author} {\bibinfo {author} {\bibfnamefont {T.}~\bibnamefont
  {Enss}}, \bibinfo {author} {\bibfnamefont {R.}~\bibnamefont {Haussmann}}, \
  and\ \bibinfo {author} {\bibfnamefont {W.}~\bibnamefont {Zwerger}},\ }\href
  {\doibase http://dx.doi.org/10.1016/j.aop.2010.10.002} {\bibfield  {journal}
  {\bibinfo  {journal} {Annals of Physics}\ }\textbf {\bibinfo {volume}
  {326}},\ \bibinfo {pages} {770 } (\bibinfo {year} {2011})}\BibitemShut
  {NoStop}%
\bibitem [{\citenamefont {S{\o}rensen}\ \emph {et~al.}(2013)\citenamefont
  {S{\o}rensen}, \citenamefont {Fedorov}, \citenamefont {Jensen},\ and\
  \citenamefont {Zinner}}]{Soerensen2013}%
  \BibitemOpen
  \bibfield  {author} {\bibinfo {author} {\bibfnamefont {P.~K.}\ \bibnamefont
  {S{\o}rensen}}, \bibinfo {author} {\bibfnamefont {D.~V.}\ \bibnamefont
  {Fedorov}}, \bibinfo {author} {\bibfnamefont {A.~S.}\ \bibnamefont {Jensen}},
  \ and\ \bibinfo {author} {\bibfnamefont {N.~T.}\ \bibnamefont {Zinner}},\
  }\href {http://stacks.iop.org/0953-4075/46/i=7/a=075301} {\bibfield
  {journal} {\bibinfo  {journal} {Journal of Physics B: Atomic, Molecular and
  Optical Physics}\ }\textbf {\bibinfo {volume} {46}},\ \bibinfo {pages}
  {075301} (\bibinfo {year} {2013})}\BibitemShut {NoStop}%
\end{thebibliography}

%

\end{document}